\documentclass[a4paper,11pt]{article}
\usepackage{heppub}
\pdfoutput=1
\usepackage{xspace}
\usepackage{placeins}
\usepackage{wasysym}
\usepackage{slashed}
\usepackage{empheq}
\usepackage[font={bf, footnotesize}]{subfig}
\usepackage[usenames,dvipsnames]{xcolor}

\newcommand{\abs}[1]{\lvert#1\rvert}
\newcommand{\Abs}[1]{\bigl\lvert#1\bigr\rvert}
\newcommand{\ord}[1]{\mathcal{O}(#1)}

\newcommand{\ORd}[1]{\mathcal{O}\Bigl(#1\Bigr)}

\newcommand{\mae}[3]{\langle#1\lvert#2\rvert#3\rangle}
\newcommand{\Mae}[3]{\big\langle#1\big\lvert#2\big\rvert#3\big\rangle}

\newcommand{\ket}[1]{\lvert#1\rangle}
\newcommand{\Ket}[1]{\big\lvert#1\big\rangle}

\newcommand{\bra}[1]{\langle#1\rvert}
\newcommand{\Bra}[1]{\big\langle#1\big\rvert}

\newcommand{\braket}[2]{\langle#1\vert#2\rangle}

\newcommand{\nn}{\nonumber}

\newcommand{\df}{\mathrm{d}}
\newcommand{\img}{\mathrm{i}}

\renewcommand{\Re}{\operatorname{Re}}

\newcommand{\tr}{\operatorname{tr}}
\newcommand{\Tr}{\operatorname{Tr}}

\newcommand{\Sl}[1]{\slashed{#1}}

\DeclareMathOperator*{\SumInt}{%
\mathchoice%
  {\ooalign{$\displaystyle\sum$\cr\hidewidth$\displaystyle\int$\hidewidth\cr}}
  {\ooalign{\raisebox{.14\height}{\scalebox{.7}{$\textstyle\sum$}}\cr\hidewidth$\textstyle\int$\hidewidth\cr}}
  {\ooalign{\raisebox{.2\height}{\scalebox{.6}{$\scriptstyle\sum$}}\cr$\scriptstyle\int$\cr}}
  {\ooalign{\raisebox{.2\height}{\scalebox{.6}{$\scriptstyle\sum$}}\cr$\scriptstyle\int$\cr}}
}

\newcommand{\eps}{\epsilon}

\newcommand{\bn}{{\bar n}}

\newcommand{\cF}{\mathcal{F}}
\newcommand{\cH}{\mathcal{H}}

\newcommand{\cL}{\mathcal{L}}

\newcommand{\cJ}{\mathcal{J}}

\newcommand{\as}{\alpha_s}
\newcommand{\aem}{\alpha_\mathrm{em}}

\newcommand{\lqcd}{\Lambda_\mathrm{QCD}}

\newcommand{\nbar}{\bar{n}}
\newcommand{\qTcut}{q_T^\mathrm{cut}}

\newcommand{\GeV}{\,\mathrm{GeV}}
\newcommand{\fb}{\,\mathrm{fb}}
\newcommand{\pb}{\,\mathrm{pb}}

\newcommand{\cut}{\mathrm{cut}}

\newcommand{\WidthTwoSubfigs}{0.5\textwidth}

\allowdisplaybreaks[2]

\title{\boldmath
Transverse Momentum Distributions\\ of Heavy Hadrons and Polarized Heavy Quarks
}

\author[a]{Rebecca von Kuk,\hspace{-0.2ex}}
\emailAdd{rebecca.von.kuk@desy.de}

\author[b]{Johannes K.~L.~Michel}
\emailAdd{jklmich@mit.edu}

\author[b]{and Zhiquan Sun}
\emailAdd{zqsun@mit.edu}

\affiliation[a]{Deutsches Elektronen-Synchrotron DESY, Notkestr. 85, 22607 Hamburg, Germany}
\affiliation[b]{Center for Theoretical Physics,\,Massachusetts Institute of Technology,\,Cambridge,\,MA\,02139,\,USA}

\abstract{%
We initiate the study of transverse momentum-dependent (TMD) fragmentation functions for heavy quarks,
demonstrate their factorization in terms
of novel nonperturbative matrix elements in heavy-quark effective theory (HQET),
and prove new TMD sum rules that arise from heavy-quark spin symmetry.
We discuss the phenomenology of heavy-quark TMD FFs at $B$ factories
and find that the Collins effect,
in contrast to claims in the literature,
is not parametrically suppressed by the heavy-quark mass.
We further calculate all TMD parton distribution functions
for the production of heavy quarks from polarized gluons within the nucleon
and use our results to demonstrate the potential of the future EIC
to resolve TMD heavy-quark fragmentation in semi-inclusive DIS,
complementing the planned EIC program to use heavy quarks as probes of gluon distributions.
}

\date{June 20, 2023}

\preprint{\vbox{%
\hbox{DESY 23-032}
\hbox{MIT-CTP 5537}
}
}

\begin{document}

\maketitle
\newpage

\section{Introduction}
\label{sec:intro}

Hadronization --- the nonperturbative mechanism
that confines quarks and gluons produced in high-energy collisions
into the experimentally observed color-singlet mesons and baryons ---
is a key aspect of virtually any process involving Quantum Chromodynamics (QCD),
but its fundamental description from first principles remains elusive~\cite{Webber:1999ui}.
In the quest for this fundamental understanding,
the fragmentation of bottom and charm quarks to heavy mesons
can play a vital role because the mass of the heavy quark imprints
as a perturbative scale on the otherwise nonperturbative dynamics of hadronization.
The unique properties of heavy quarks as color-charged, but perturbatively accessible objects
make them ideally suited as probes of the hadronization cascade,
effectively serving as a static color source coupling to the light degrees of freedom.
An improved field-theoretic understanding of heavy-quark fragmentation
will also benefit the description of heavy flavor in Monte-Carlo generators for the LHC~\cite{Campbell:2022qmc},
where many key searches and Higgs coupling measurements involve final-state charm or bottom quarks.

A rigorous field-theoretic framework in which hadronization can be studied in detail
is that of transverse momentum-dependent (TMD) fragmentation functions (FFs),
for which all-order factorization theorems have been established~\cite{Collins:1350496}.
Like collinear fragmentation functions,
TMD FFs depend on the longitudinal momentum fraction $z_H$
that the hadron retains from its parent quark.
In addition, they describe the transverse momentum
that the hadron picks up by recoiling against other fragmentation products,
including the full quantum correlations with the quark polarization,
which provides a three-dimensional picture of the fragmentation cascade.
For processes with initial-state hadrons,
TMD FFs are complemented by TMD parton distribution functions (PDFs)
describing the three-dimensional motion of quarks and gluons inside the nucleon.
The TMD dynamics of light quarks and gluons are a well-established field
of experimental study \cite{Alekseev:2008aa, Airapetian:2012ki, CLAS:2003qum, Aschenauer:2015eha, Adamczyk:2015gyk, Aghasyan:2017ctw, CLAS:2017yrm, Parsamyan:2018ovx, HERMES:2019zll, HERMES:2020ifk, CLAS:2021jhm},
phenomenological analysis (see e.g.\ \refscite{Scimemi:2019cmh,Bacchetta:2019sam,Bury:2021sue, Bacchetta:2022awv}),
and progress towards first-principle calculations
using lattice field-theory~\cite{Ebert:2018gzl,Ji:2019sxk,Shanahan:2020zxr,Shanahan:2021tst,Schlemmer:2021aij,Li:2021wvl,LPC:2022ibr};
for a recent comprehensive overview, see \refcite{Boussarie:2023izj}.
Precision TMD measurements are a key physics target of the future Electron-Ion Collider (EIC)~\cite{AbdulKhalek:2021gbh}.

In this paper we study, for the first time, the TMD FFs of heavy quarks to heavy hadrons.
Our theoretical tool to analyze the fragmentation of heavy quarks
is (boosted) Heavy-Quark Effective Theory (bHQET)~\cite{Eichten:1989zv, Isgur:1989vq, Isgur:1990yhj, Grinstein:1990mj, Georgi:1990um,Korner:1991kf,Mannel:1991mc,
Fleming:2007qr, Fleming:2007xt},
which has previously been applied to the well-understood
collinear (or longitudinal) heavy-quark FFs~\cite{Jaffe:1993ie, Falk:1993rf, Neubert:2007je, Fickinger:2016rfd}.
We demonstrate that applying bHQET to TMD FFs gives rise
to novel, universal matrix elements describing
the nonperturbative transverse dynamics of light QCD degrees of freedom in the presence
of a heavy quark (i.e., a static color source).
While a large part of this work is devoted to developing
this new theoretical formalism,
we will also consider the phenomenology
of heavy-quark TMD FFs in two distinct processes,
$e^{+}e^{-}$ collisions and semi-inclusive deep inelastic scattering (SIDIS),
which are illustrated in \fig{heavy_process}:

\begin{figure*}
\centering
\begin{minipage}{0.45\textwidth}
\subfloat[]{
   \includegraphics[width=\textwidth]{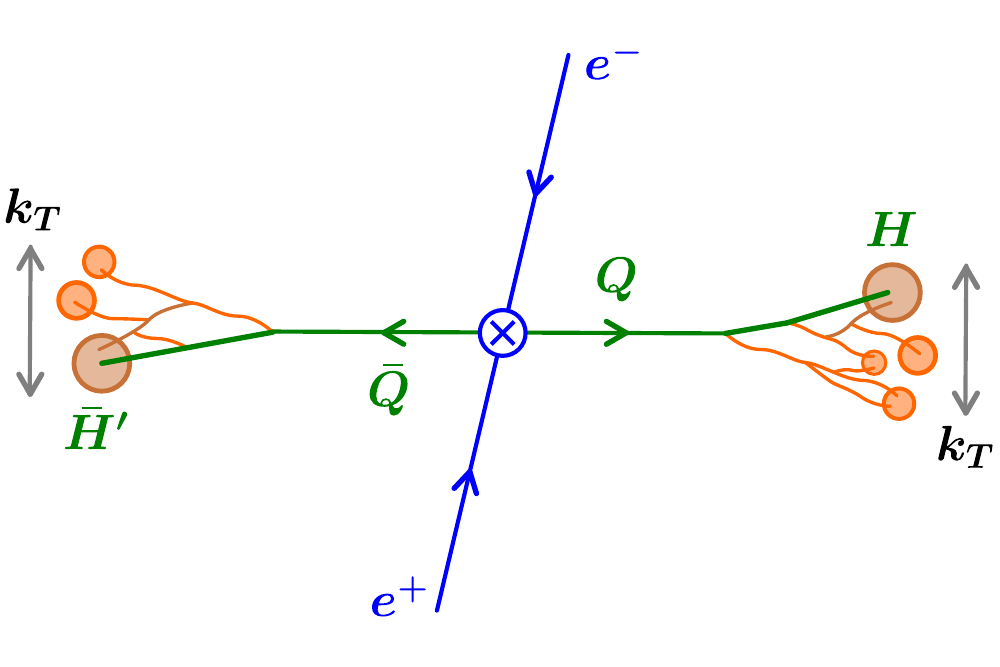}
}%
\\
\subfloat[]{
   \includegraphics[width=\textwidth]{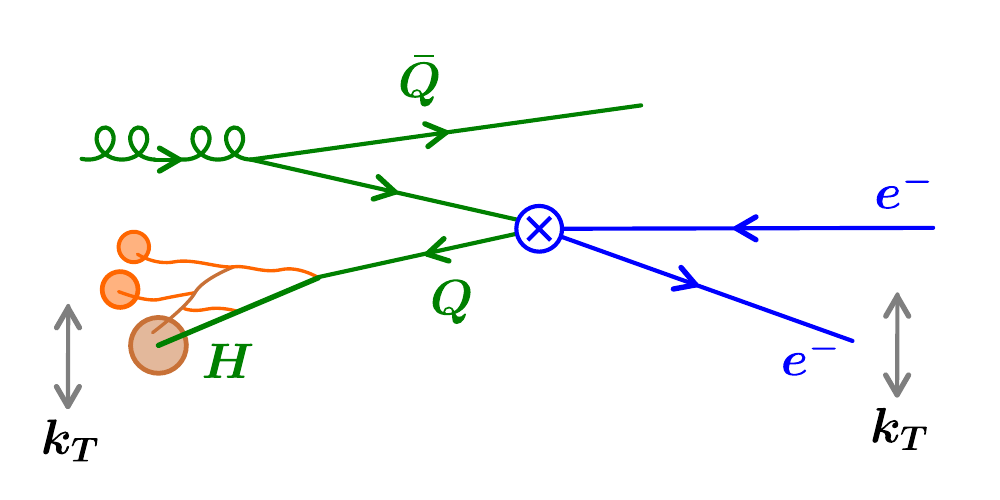}
}%
\end{minipage}
\hspace{0.25in}%
\raisebox{-1in}{\subfloat[]{
   \raisebox{0.1in}{
      \includegraphics[width=0.45\textwidth]{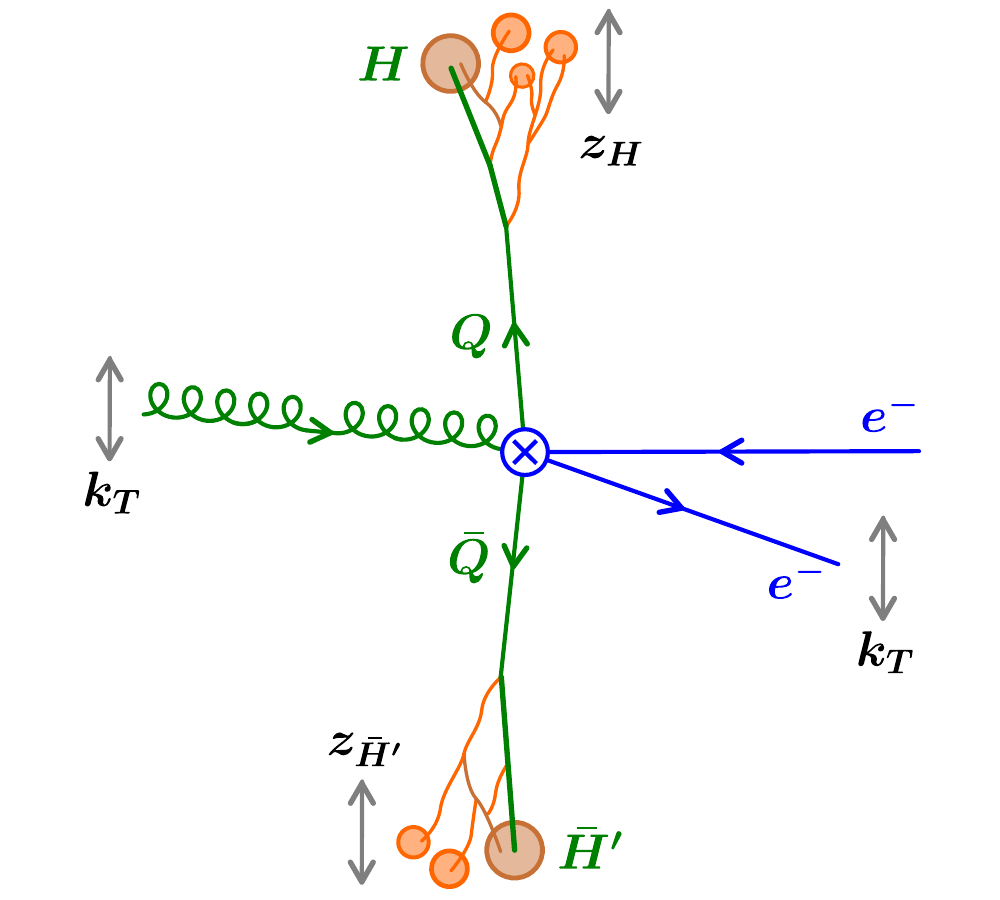}
   }
}}%
\caption{%
\textbf{(a)} Heavy-quark pair production in the back-to-back limit in $e^{+} e^{-}$ collisions,
which gives access to heavy-quark TMD FFs.
(Note that arrows indicate momentum flow, not fermion flow.)
\textbf{(b)} Semi-inclusive deep inelastic scattering with an identified heavy hadron in the final state.
This process gives access to individual heavy-quark TMD FFs
convolved with a heavy-quark TMD PDF that can be perturbatively computed in terms of the collinear gluon PDF.
\textbf{(c)} Heavy-quark pair production in electron-nucleon collisions,
which we do not consider in this work.
This process can probe the gluon TMD PDF,
but is only sensitive to longitudinal fragmentation dynamics.
}
\label{fig:heavy_process}
\end{figure*}

\begin{itemize}
   \item[(a)]
   Heavy quarks are copiously produced in $e^{+}e^{-}$ collisions.
   We are interested in the case where the quarks are produced relativistically,
   as is the case for charm quarks at existing $B$ factories,
   such that their fragmentation processes are independent.
   The production rate is largest in the back-to-back region,
   where the cross section differential in the small hadron transverse momenta
   factorizes in terms of two unpolarized TMD FFs $D_{1\, H/Q}(z_H, k_T)$.
   Furthermore, the heavy-quark pair is produced in an entangled transverse polarization state in central events.
   This entanglement imprints on the distribution of final-state hadrons as an azimuthal modulation known as the Collins effect,
   whose strength schematically is given by
   \begin{align} \label{eq:sketch_collins_effect}
   R_{\cos(2\phi_0)} \propto H^{\perp}_{1\, H/Q}(z_H, k_T) \otimes H^{\perp}_{1\, \bar H'/\,\bar Q}(z_{H'}, k_T)
   \,, \end{align}
   where $H^{\perp}_{1\, H/i}$ is called the Collins function.
   The light-quark Collins effect has been measured in detail in pion and kaon samples
   by the \textsc{Belle}~\cite{Belle:2005dmx, Belle:2008fdv,Belle:2019nve}
   and \textsc{BaBar} collaborations~\cite{BaBar:2013jdt, BaBar:2015mcn},
   but despite a proposal in \refcite{BaBar:2013jdt},
   a measurement (or search) of the heavy-quark Collins effect has not yet been performed.
   We will find that the heavy-quark Collins function encodes
   intricate nonperturbative physics,
   which motivates a dedicated measurement.%
   \footnote{It has been argued in the experimental literature~\cite{BaBar:2013jdt}
   that the Collins effect is suppressed by the quark mass.
   We contrast these claims with our findings in \sec{collins_comment}.}
   \item[(b)]
   A previously overlooked aspect of heavy-quark phenomenology
   in electron-nucleon collisions at the future EIC
   is that heavy quarks can be pair-produced in initial-state gluon splittings
   at small transverse momentum, with one quark e.g.\ going down the beam pipe
   and the other undergoing hard scattering and subsequent fragmentation
   into a heavy hadron, which in turn is reconstructed in a semi-inclusive measurement,
   as illustrated in \fig{heavy_process}~(b).
   Crucially, this process can be described by the standard TMD factorization
   for semi-inclusive deep inelastic scattering (SIDIS)
   when both the transverse momentum and the mass of the quark are small compared to the hard scattering energy $Q$.
   It thus retains the full sensitivity to the initial and final-state transverse momentum
   distribution (encoded in heavy-quark TMD PDFs and FFs) with respect to the photon direction.

   To make phenomenological predictions for heavy-quark SIDIS,
   we fill a gap in the literature and compute all polarized heavy-quark TMD PDFs
   by perturbatively matching them onto collinear twist-2 nucleon PDFs,
   extending the analysis of the unpolarized heavy-quark TMD PDF in \refscite{Nadolsky:2002jr, Pietrulewicz:2017gxc}.%
   \footnote{\Refcite{Pietrulewicz:2017gxc} also computed all secondary quark mass effects
   on light-quark distributions, including mass effects in the Collins-Soper kernel,
   which drive the renormalization of many of the objects we introduce here beyond leading-logarithmic order.
   The much more involved $\ord{\as^2}$ secondary quark mass effects
   in the gluon TMD PDF were recently calculated in \refcite{Pietrulewicz:2023dxt}.}
   Interestingly, while many polarized TMD PDFs are strongly suppressed for heavy quarks,
   we find a nonzero leading result for the so-called worm-gear~$L$ function $h_{1L}^{\perp}$,
   encoding the production of transversely polarized (heavy) quarks
   from linearly polarized light quarks and gluons, which is possible because the quark mass violates chirality.
   Indeed, the expectation that transverse quark polarization effects are suppressed by quark masses
   goes back to the early days of QCD~\cite{Kane:1978nd},
   and we find that heavy-quark TMDs provide an arena
   to make this statement precise within twist-2 collinear factorization.
   For future phenomenology at the EIC, this nonzero conversion rate
   provides an exciting avenue to observe the heavy-quark Collins function
   because the factorized SIDIS cross section contains an azimuthal spin asymmetry,
   \begin{align} \label{eq:sketch_sidis_sin2phih}
   A_{\sin(2\phi_{H})} \propto S_L h_{1L}^{\perp} \otimes H^{\perp}_{1\, H/i}
   \,.\end{align}
   Importantly, this asymmetry unlike \eq{sketch_collins_effect}
   involves only a single Collins function multiplying the perturbatively predicted worm-gear $L$ function,
   making it possible to extract its sign.
\end{itemize}

We point out that an extensive heavy-flavor physics program
is already being planned for the Electron Ion Collider (EIC),
which will leverage heavy-quark pair production
as a hard probe of gluon TMDs~\cite{Boer:2010zf, Zhu:2013yxa, Zhang:2017uiz, delCastillo:2020omr, Kang:2020xgk, delCastillo:2021znl}
and of cold nuclear matter~\cite{Li:2020sru, Li:2020zbk},
as illustrated in \fig{heavy_process}~(c).
We stress that this is \emph{not} the case we consider in this paper:
In case~(c), the transverse momentum imbalance, production rate, and distribution of the heavy-quark pair
are sensitive to the initial-state gluon TMD PDF (or the nuclear collinear gluon PDFs),
but on the fragmentation side are at most sensitive to the well-understood longitudinal momentum ($z_H$) distribution
at leading power~\cite{delCastillo:2020omr}.
In contrast to this, the TMD processes in \fig{heavy_process}~(a) and (b)
are directly sensitive to the nonperturbative transverse dynamics
of heavy-quark fragmentation.%
\footnote{
\label{ftn:groomed_heavy_jets}%
A very interesting middle ground is occupied by \refcite{Makris:2018npl},
which analyzed the transverse momentum spectrum of heavy headrons \emph{inside groomed jets}, also using bHQET.
The resulting bHQET matrix elements in \refcite{Makris:2018npl} are predominantly longitudinal:
Depending on the precise parametric regime,
the dominant contribution to the transverse momentum
either comes from perturbative collinear-soft modes stopping the soft-drop grooming algorithm~\cite{Larkoski:2014wba},
which factorize from the heavy-quark dynamics,
or from bHQET modes subject to a primary soft-drop criterion and a secondary measurement on perturbative transverse momenta.
(Compared to our results in this work for inclusive heavy-quark TMD FFs sensitive
to the full transverse structure of the fragmentation cascade,
this effect of the grooming is also responsible for the striking absence of Collins-Soper scaling reported in \refcite{Makris:2018npl}.)
Despite these differences in the experimental observable and theoretical structure,
we find that it should be possible to establish
a powerful connection between our results here and those of \refcite{Makris:2018npl}.
Specifically, if $\ord{\lqcd^2}$ power corrections from intrinsic transverse hadronic dynamics can be resolved even within groomed jets,
as outlined around their eq.~(5.9), this would access a second moment
of the unpolarized bHQET TMD fragmentation factor we will introduce in \sec{tmd_ffs_from_bhqet}.
If this connection can be made precise,
it would suggest that the bHQET fragmentation factors we consider in \sec{tmd_ffs_from_bhqet} can appear
also in scenarios where the mass has been integrated out
together with additional observables (i.e., the soft-drop criterion in this case).
}
We note that the TMD fragmentation of light quarks to quarkonia
has been studied in \refcite{Echevarria:2020qjk}, in that case by matching onto nonrelativistic QCD,
and similarly for light-quark TMD dynamics in hard quarkonium production and decay in \refcite{Echevarria:2019ynx, Fleming:2019pzj}.

The remainder of this paper is structured as follows:
In \Sec{heavy_quark_tmd_ffs}, we analyze heavy-quark TMD FFs
and identify the new bHQET matrix elements
and perturbative matching coefficients
that characterize the fragmentation dynamics.
In \Sec{polarized_heavy_quark_tmd_pdfs}, we discuss the all-order
structure of matching polarized heavy quark TMD PDFs onto collinear PDFs
and explicitly compute the $\ord{\as}$ matching onto gluon PDFs.
In \Sec{pheno}, we use our results from the previous two sections
to outline the prospects
for heavy-quark TMD phenomenology at $e^{+} e^{-}$ colliders
and the future EIC.

\paragraph{Important note on conventions:}
In an attempt to upset everybody equally,
we will use QCD fields for writing down TMD correlators in this paper
(thus hopefully making it accessible without background knowledge of soft-collinear effective theory),
but will consistently make use of lightcone coordinate conventions
as more commonly used in the SCET community (and also, for example, in bHQET).
Specifically, we decompose four-vectors $p^\mu$
in terms of lightlike vectors $n^\mu$, $\bn^\mu$ with $n^2 = \nbar^2 = 0$
and {\boldmath $n \cdot \bn = 2$},
\begin{align}
p^\mu
= \nbar \cdot p \, \frac{n^\mu}{2} + n \cdot p \, \frac{\nbar^\mu}{2} + p_\perp^\mu
\equiv p^- \frac{n^\mu}{2} + p^+ \frac{\nbar^\mu}{2} + p_\perp^\mu
\equiv (p^-, p^+, p_\perp)
\,,\end{align}
such that e.g.\ $p^2 = p^- p^+ + p_\perp^2$.
As another part of this convention, collinear momenta
near the mass shell will typically
have large components $p^- \gg p_\perp \gg p^+$.
We always take transverse vectors
with subscript $\perp$ to be Minkowskian, $p_\perp^2 \equiv p_\perp \cdot p_\perp < 0$,
and denote their magnitude by $p_T = \sqrt{- p_\perp^2}$.
Finally, we define the metric and antisymmetric tensor in transverse space as
\begin{align}
g^{\mu \nu}_\perp \equiv g^{\mu \nu} - \frac{n^\mu \bar{n}^\nu}{2} - \frac{\bar{n}^\mu {n}^\nu}{2}
\,, \qquad
\epsilon^{\mu \nu}_\perp \equiv \epsilon^{\mu \nu \rho \sigma} \frac{\bar{n}_\rho n_\sigma}{2}.
\end{align}
Our convention for the antisymmetric tensor is $\eps^{0123} = +1$.
\enlargethispage{1em}

\section{TMDs for heavy quark fragmentation into a heavy hadron}
\label{sec:heavy_quark_tmd_ffs}

\subsection{Calculational setup and parametric regimes}
\label{sec:tmd_ffs_setup}

We consider the fragmentation of a (possibly polarized) heavy quark $Q$
into a hadron $H$ that contains the heavy quark and carries momentum $P_H^\mu$.
For this paper, we assume that the heavy hadron polarization is not experimentally reconstructed.
We work in QCD with $n_f = n_\ell + 1$ flavors,
where the $n_\ell$ massless quark flavors are denoted by $q$
and the heavy quark $Q$ has a pole mass $m \equiv m_c, m_b \gg \lqcd$.
We decompose $P_H^\mu$ in terms of lightcone momenta as
\begin{align}
P_H^\mu = P_H^- \frac{n^\mu}{2} + \frac{M_H^2}{P_H^-} \frac{\bn^\mu}{2}
\,,\end{align}
where $P_H^- \gg P_H^+ = M_H^2/P_H^-$ is boosted in the frame of the hard scattering
and by definition $P_{H,\perp} = 0$,
coinciding with the ``hadron frame'' for fragmentation~\cite{Collins:1350496}.

We are interested in the dependence of the fragmentation process
on the total transverse momentum of additional hadronic radiation $X$ into the final state,
which is equal to the initial quark transverse momentum $k_\perp$ by momentum conservation,
and Fourier conjugate to the transverse spacetime separation $b_\perp$ between quark fields.
In position space,
the TMD quark-quark correlator
describing this fragmentation process is defined as
\begin{align} \label{eq:def_tmd_ff_correlator}
\Delta_{H/Q}^{\beta \beta'}(z_H, b_\perp)
&= \frac{1}{2 z_H N_c} \int \! \frac{\df b^+}{4\pi} \, e^{\img b^+ (P_H^-/z_H)/2}
\nn \\[0.4em] & \quad \times
\Tr \SumInt_{X}
\Mae{0}{W^\dagger(b) \, \psi_Q^\beta(b)}{H X}
\Mae{H X}{\bar\psi_Q^{\beta'}(0) \, W(0)}{0}
\,,\end{align}
where $z_H$ is the fraction of the quark's lightcone momentum retained by $H$,
$\beta, \beta'$ are the open spin indices of the quark fields,
$\Tr$ denotes a trace over fundamental color indices,
and $b \equiv (0, b^+, b_\perp)$.
We have kept a sum over the possible hadron helicities $h_H$,
which are not experimentally resolved, implicit in the constrained sum
over states, i.e.,
\begin{align} \label{eq:sum_over_hadron_helicities}
\SumInt_{X} \ket{H X} \bra{H X}
\equiv \SumInt_{X} \sum_{h_H} \ket{H, h_H; X} \bra{H, h_H; X}
\,.\end{align}
The Wilson line $W(x)$ is defined as an anti-path ordered exponential of gauge fields
extending to positive infinity along the lightcone direction $\nbar^\mu$,
\begin{align} \label{eq:def_wilson_line}
W(x) = \bar{P} \Bigl[ \exp \Bigl(
   - \img g \int_0^\infty \! \df s \, \bn \cdot A(x + \bn s)
\Bigr) \Bigr]
\,.\end{align}
For simplicity, we have suppressed the rapidity regulator,
the soft factor, and transverse gauge links at infinity in \eq{def_tmd_ff_correlator}.
The Wilson lines only depend on the direction of $\bn^\mu$
and are thus invariant under $\bn^\mu \mapsto e^{\alpha} \, \bn^\mu$.
Taking $P_H^\mu$ and $\bn^\mu$ to define $n^\mu$
and tracking the $\alpha$ dependence through the definition of $\Delta_{H/Q}(z, b_\perp)$
implies that the ``good components'' of $\Delta_{H/Q}$ \cite{PhysRevD.1.2901,Boussarie:2023izj},
by which we mean the components of the fermion fields
that are picked out by the projector $\Sl{n}\Sl{\bn}/4$ acting on $\psi_Q$
and that appear in leading-power factorization theorems,
transform as $\Delta_{H/Q} \mapsto e^{-\alpha} \, \Delta_{H/Q}$ under this relabeling.%
\footnote{In SCET this symmetry under the simultaneous relabeling $n^\mu \mapsto e^{-\alpha} n^\mu$
and $\bn^\mu \mapsto e^{\alpha} \bn^\mu$ is known as type-III reparameterization invariance~\cite{Manohar:2002fd, Marcantonini:2008qn}
and is a manifest symmetry of the entire correlator
because the bad components $(\Sl \nbar \Sl n/4) \, \psi_Q$ have been integrated out.}

In terms of the correlator in \eq{def_tmd_ff_correlator},
the bare unpolarized ($D_{1\,H/Q}$) and Collins fragmentation function ($H_{1\,H/Q}^{\perp (1)}$)
from the introduction are defined
in position space as%
\footnote{\label{ftn:same_symbol_momentum_position_space}%
We use the same symbol for transverse momentum distributions
in $k_T$ space and their Fourier transforms in $b_T$ space
throughout this paper, as the meaning will always be clear from the context.
Our conventions for Fourier transforms and the spin decomposition
of TMD correlators follow \refcite{Ebert:2021jhy}.
Note the superscript $(1)$ on the $b_T$-space Collins function indicating
a $b_T$ derivative that arises from integrating a term $\Sl k_\perp$
in the momentum-space correlator by parts, and that is specifically
required due to the conventional normalization to the hadron mass~\cite{Boer:2011xd}.
For reference, the stated definition of the Collins function in position space
is equivalent to a term
$
\mathrm{tr} [\tfrac{\img}{2} \sigma^{\alpha\beta}\bar{n}_{\beta}\gamma_{5} \,
\Delta_{H/Q}(z_{H},b_{\perp})]
\supset \img \eps_{\perp}^{\alpha\rho} b_{\perp\rho} M_{H} H_{1}^{\perp(1)} (z_{H},b_{\perp})
$
with $\sigma_{\mu \nu} = \tfrac{\img}{2} [\gamma_\mu, \gamma_\nu]$
appearing in the spin decomposition of the correlator.
}
\begin{align} \label{eq:def_tmd_ff_unpol_collins}
D_{1\,H/Q}(z_H, b_T) &= \tr \Bigl[ \frac{\Sl{\bn}}{2} \, \Delta_{H/Q}(z_H, b_\perp) \Bigr]
\,, \nn \\
H_{1\,H/Q}^{\perp (1)}(z_H, b_T) &= \tr \Bigl[ \frac{\Sl{\bn}}{2} \frac{\Sl{b}_\perp}{M_H b_T^2} \, \Delta_{H/Q}(z_H, b_\perp) \Bigr]
\,,\end{align}
where $\tr$ denotes a trace over spin indices.
The unpolarized TMD FF encodes the total rate for producing an unpolarized hadron
from an unpolarized quark, while the Collins TMD FF describes the strength of the correlation
between the quark's transverse polarization and the direction
of the hadron transverse momentum.
The leading TMD fragmentation functions have been proven
to be universal between processes~\cite{Collins:2004nx},
i.e., they are independent of whether the Wilson line
points to the future ($e^+e^- \to \text{hadrons}$) or the past (SIDIS).
Note that these scalar projections of the TMD fragmentation correlator
are invariant under $\bn^\mu \mapsto e^{\alpha} \bn^\mu$ by construction.

\begin{figure*}
\centering
\subfloat[]{
   \includegraphics[width=0.38\textwidth]{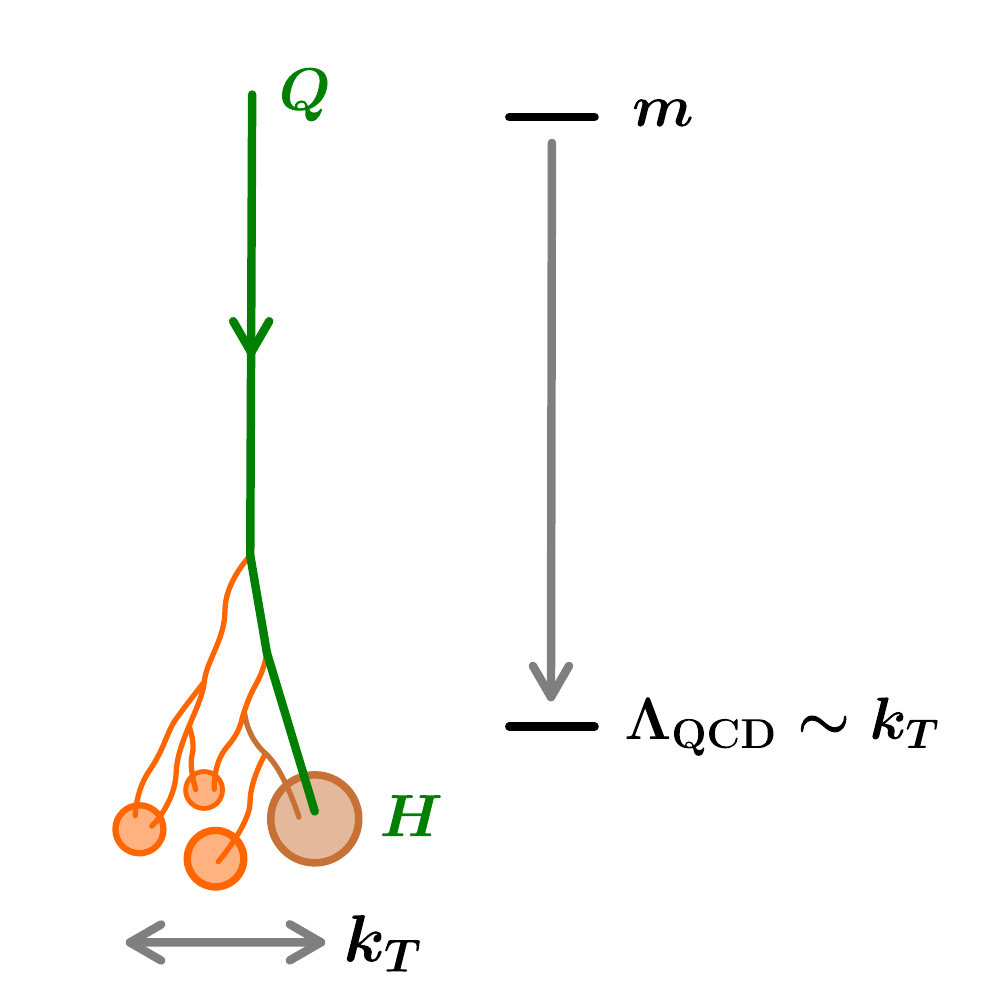}
}%
\hspace{2em}
\subfloat[]{
   \includegraphics[width=0.38\textwidth]{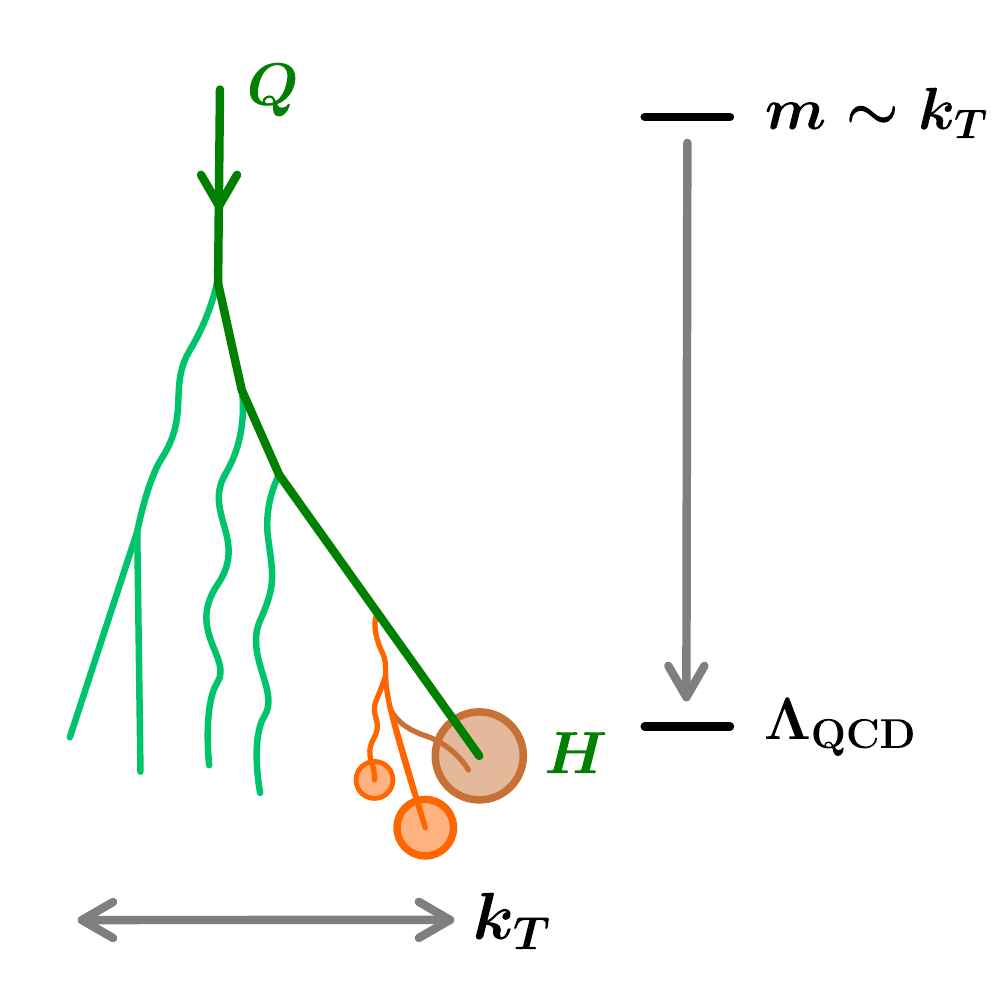}
}%
\caption{
Parametric regimes for the fragmentation of a heavy quark $Q$ (green)
into a single heavy hadron $H$ with a measurement on transverse momentum.
In case~\textbf{(a)}, the hadron picks up transverse momentum
relative to its parent quark as the ``brown muck'' (brown) coalesces around the quark
and splits from other soft hadronic radiation into the final state (orange).
In case~\textbf{(b)}, the transverse and longitudinal momentum distributions are dominated
by perturbative emissions at the scale of the heavy quark mass (teal).
In this regime, the nonperturbative hadronization process
is encoded in a normalization factor,
while its effect on the shape of the momentum distribution is subleading.
}
\label{fig:regimes_fragmentation}
\end{figure*}

Since $\lqcd \ll m$, the nonperturbative dynamics
in the fragmentation process are constrained
by heavy-quark symmetry in all cases, but differences arise depending
on the hierarchy between these two parametric scales
and the magnitude $k_T$ of the transverse momentum
or, equivalently, the inverse of the transverse distance $1/b_T \sim k_T$.
Broadly speaking, we will consider the two cases illustrated in \fig{regimes_fragmentation}.
In case~(a), which we analyze in \sec{tmd_ffs_from_bhqet},
$k_T \sim \lqcd$ is generated during the nonperturbative fragmentation process itself,
while perturbative emissions at the scale $m \gg k_T$ are suppressed.
In this case, the heavy hadron carries almost all the longitudinal momentum
provided by the initial heavy quark,
while the $k_T$ dependence is carried by universal nonperturbative functions
describing how the ``brown muck'' separates from other light hadronic final states.
In this regime, ``disfavored'' fragmentation functions
where the valence content of the identified heavy hadron does not match the
initial heavy quark, e.g.\ $Q \to \bar{H}$, $Q \to h$, or $q, g \to H$,
are forbidden by heavy-quark symmetry.
To simplify the analysis, we will count
\begin{align} \label{eq:away_from_endpoint}
1 - z_H \sim 1
\,,\end{align}
i.e., we will assume that the $z_H$ measurement does not probe
the precise longitudinal momentum distribution near the endpoint,
which is the case e.g.\ if $z_H$ is integrated over a sufficiently large bin
or when taking Mellin moments of the $z_H$ distribution.
While relaxing this modification on the fragmentation side would be straightforward
and leads to interesting fully-differential bHQET fragmentation shape functions,
this would also require consistently modifying the description
of the opposite collinear sector by reinstating
the transverse momentum dependence in the formalism of \refcite{Fickinger:2016rfd}
for $e^+ e^-$ collisions,
and by crossing the jet function in that reference
into the initial state using the methods of \refcite{Lustermans:2019cau} for SIDIS,
all of which we leave to future work.

In case~(b), which we consider in \sec{tmd_ffs_from_collinear_ffs},
the distributions in transverse and longitudinal momentum
are determined by perturbative dynamics at the scale $k_T \sim m \gg \lqcd$,
while the dynamics of the nonperturbative bound state
only contribute a normalization factor.
In the case of the unpolarized TMD FF,
this normalization factor admits an interpretation as the total probability for $Q$ to fragment into $H$,
as is well known for inclusive heavy hadron production cross sections~\cite{Mele:1990cw, Jaffe:1993ie, Falk:1993rf}.
In this case, the disfavored fragmentation functions for $Q \to \bar{H}$
or $q \to H$, and $Q \to h$ or $g \to H$,
are perturbatively suppressed by $\ord{\as^2}$ and $\ord{\as}$ at the scale $\mu \sim m$, respectively.
We stress that we continue to assume \eq{away_from_endpoint}
also in this regime, so longitudinal momentum distributions remain perturbative.

\subsection{Review of Boosted Heavy-Quark Effective Theory}
\label{sec:bhqet}

The appropriate theory that describes the dynamics at the scale $\lqcd$
in either case and makes the heavy-quark symmetry manifest
is boosted Heavy-Quark Effective Theory (bHQET)~\cite{Fleming:2007qr, Fleming:2007xt},
i.e., the application of HQET~\cite{Manohar:2000dt} to heavy quarks produced in an energetic collision.
The effective theory is constructed by integrating out the off-shell fluctuations of the heavy quark field
at the scale $m$;
these in particular include its antiquark component with energy gap $2m$.
The dynamic degrees of freedom are heavy-quark fields $h_v(x)$
that are labeled by the timelike direction $v^\mu$,
which we choose to be the velocity of the heavy hadron,
\begin{align}
v^\mu = \frac{P_H^\mu}{M_H}
\,, \qquad
v^2 = 1
\,,\end{align}
The tree-level matching of the massive QCD quark field onto $h_v$ at $\mu \sim m$ reads:
\begin{align} \label{eq:bhqet_tree_matching}
\psi_Q(x) = e^{-\img m v \cdot x} h_v(x)
\Bigl[ 1 + \ORd{\frac{1}{m}} \Bigr]
\,.\end{align}
The $h_v(x)$ are implemented as Dirac spinors satisfying the projection relations
\begin{align} \label{eq:hqet_projection_identities}
P_{\pm} \equiv \frac{1 \pm \Sl v}{2}
\,, \qquad
P_+ h_v = h_v
\,, \qquad
P_- h_v = 0
\,, \qquad
\bar{h}_v P_+ = \bar{h}_v
\,, \qquad
\bar{h}_v P_- = 0
\,.\end{align}
For external states, the matching reads $\ket{H, h_H; X} = \sqrt{m} \, \ket{H_v, h_H; X}$,
and we use a nonrelativistic normalization convention for the bHQET states.
In addition, the effective theory contains light-quark and gluon degrees of freedom
that have isotropic momentum $p^\mu \sim \lqcd$ in the rest frame of the heavy hadron.
The tree-level matching for these is trivial;
in particular, the Wilson line $W(x)$ takes the same form
in the effective theory, but consists of gluon fields
that only have support on a restricted set of modes.
For reference, the leading HQET Lagrangian is given by~\cite{Manohar:2000dt}
\begin{align}  \label{eq:hqet_lagrangian}
\cL = \bar{h}_v (\img v \cdot D) h_v + \cL_\mathrm{light} + \ORd{\frac{1}{m}}
\,,\end{align}
where $\cL_\mathrm{light}$ is a copy of the QCD Lagrangian
with $n_\ell$ massless quark flavors.

The spin degrees of freedom of the heavy-quark can be explicitly decoupled
from the light dynamics at leading power in $1/m$
by performing a field redefinition
involving static Wilson lines $Y_v(x)$~\cite{Korchemsky:1991zp, Bauer:2001yt},
\begin{align}
\label{eq:sterile_decoupling}
h_v(x) = Y_v(x) \, h_v^{(0)}(x)
\,, \qquad
Y_v(x) = P \Bigl[ \exp \Bigl(
   \img g \int_0^\infty \! \df s \, v \cdot A(x + v s)
\Bigr) \Bigr]
\,.\end{align}
In this way, the heavy-quark Lagrangian becomes that of a free theory,
$\cL^{(0)} = \bar h_v^{(0)} (\img v \cdot \partial) h_v^{(0)}$,
while the $Y_v(x)$ takes the place of $h_v(x)$ in all external operators,
acting as a static source of soft gluons.
Specifically, the action of $h_v(x)$ on a product state in the decoupled theory
is given by
\begin{align}
\label{eq:decoupling_outstate}
h_v(x) \, \ket{s_Q, h_Q; s_\ell, h_\ell, f_\ell; X}
= u(v, h_Q) \, Y_{v}(x) \, \ket{s_\ell, h_\ell, f_\ell ; X}
\,,\end{align}
where $s_Q = \tfrac{1}{2}$ and $h_Q = \pm \tfrac{1}{2}$ are the spin and helicity of the heavy quark,
$u(v, h_Q) = u(m v, h_Q)/\sqrt{m}$ is an HQET spinor,
and $s_\ell$, $h_\ell$, and $f_\ell$ are the total angular momentum,
helicity, and flavor content of the light degrees of freedom inside the hadron.
(We will specify a helicity axis in the following section.)
Note that the interpolating field for the light state on the right-hand side
formally contains a future-pointing Wilson line to form an overall color singlet,
which we suppress.
Physical hadron states of definite angular momentum $s_H$
and helicity $h_H$ also have definite $s_\ell$,
which is a good quantum number in the heavy-quark limit.
In general, they involve a coherent sum
over the helicity eigenstates in \eq{decoupling_outstate},
\begin{align} \label{eq:clebsch_gordan}
\ket{H_v, h_H}
\equiv \ket{s_H, h_H, s_Q, s_\ell}
= \sum_{h_Q}
\sum_{h_\ell}
\ket{s_Q, h_Q; s_\ell, h_\ell}
\braket{s_Q, h_Q; s_\ell, h_\ell}{s_H, h_H}
\,,\end{align}
where we suppressed the common $f_\ell$ and $X$,
and $\braket{s_Q, h_Q; s_\ell, h_\ell}{s_H, h_H}$
is a Clebsch-Gordan coefficient.
(We take the coefficient to vanish for $h_Q + h_\ell \neq h_H$,
i.e., one sum is always eliminated in practice by helicity conservation.)
For the case of inclusive fragmentation,
it has been known for a long time~\cite{Falk:1993rf, Manohar:2000dt}
that the factorized form of \eq{decoupling_outstate}
together with parity and \eq{clebsch_gordan}
implies relations between the fragmentation probabilities
to different hadron states within the same heavy-quark spin symmetry multiplet,
i.e., with the same $s_\ell = \tfrac{1}{2}, 1, \tfrac{3}{2}, \dots$.
As an example, at the strict leading order in $1/m$,
an unpolarized charm quark is exactly three times
as likely to fragment into an excited spin-$1$ vector meson ($D^*$)
than into the corresponding pseudoscalar state ($D$).
The physical reason for this is that the light dynamics
do not see the heavy-quark spin at leading power,
and thus the same nonperturbative matrix elements
with given $s_\ell, h_\ell$ appear in several cases.
This analysis is simplified by the fact that for unpolarized
or linearly polarized heavy quarks, light amplitudes for different helicities
cannot interfere.
One key goal of the next section will be to work out the consequences
of heavy-quark spin symmetry for \emph{transverse momentum-dependent} fragmentation functions,
where transverse quark polarization will let us access this interference for the first time.

\subsection{Calculating TMD FFs from bHQET for
\texorpdfstring{$\lqcd \lesssim k_T \ll m$}{LambdaQCD < kT << m} }
\label{sec:tmd_ffs_from_bhqet}

\subsubsection{Tree-level matching and discrete symmetries}
\label{sec:tmd_ffs_from_bhqet_tree_level}

Using the tree-level matching onto bHQET in \eq{bhqet_tree_matching}
at the leading order in $\lqcd/m \sim k_T/m$,
the correlator in \eq{def_tmd_ff_correlator} evaluates to
\begin{align} \label{eq:tmd_ff_correlator_in_hqet}
\Delta_{H/Q}(z_H, b_\perp)
&= \frac{m}{ z_H} \int \! \frac{\df b^+}{4\pi} \,
e^{\img b^+ ( P_H^-/z_H - m v^- )/2}
\Bigl\{
   F_H(b_\perp)  + \ord{\as} + \ORd{\frac{1}{m}}
\Bigr\}
\,,\end{align}
where $F_H$ is a bHQET bispinor defined by
\begin{align} \label{eq:def_f}
F_H(b_\perp)
&\equiv \frac{1}{2 N_c} \Tr
\SumInt_{X}
\Mae{0}{W^\dagger(b_\perp) \,
   h_v(b_\perp) \Ket{H_vX} \Bra{H_vX} \bar h_v(0) \,
W(0)}{0}
\,.\end{align}
Note that we have evaluated the matrix element at $b^+ = 0$,
which is justified at leading order in $1/m$.
This is easiest to see by using momentum conservation on the first matrix element
to translate the fields back to $b^+ = 0$,
\begin{align}
&\Mae{0}{W^\dagger(b) \, h_v(b)}{H_vX}
= \Mae{0}{W^\dagger(b_\perp) \, h_v(b_\perp)}{H_vX} \, e^{\img b^+ k^-/2}
\,,\end{align}
where $k^-$ is the total residual minus momentum of the final state,
and then expanding away $k^- \sim v^- \lqcd$ compared to the leading $\ord{P_H^-}$ term
$P_H^-/z_H - m v^-$ in the Fourier phase in \eq{tmd_ff_correlator_in_hqet}.%
\footnote{
Recall that $1 - z_H = \ord{1}$, see \eq{away_from_endpoint}.
If we instead count $m(1 - z_H) \sim \lqcd$,
this leading term is pushed down from $\ord{m v^-}$
to $\ord{\lqcd v^- }$
and the expansion cannot be performed.
In that case, we would instead obtain
a fully-differential bHQET fragmentation shape function.
}
Using $P_H^- = v^- [m + \ord{\lqcd}]$,
we can perform the $b^+$ integral,
\begin{align} \label{eq:tmd_ff_correlator_in_hqet_simplified}
\Delta_{H/Q}(z_H, b_\perp)
&= \frac{\delta(1-z_H)}{\nbar \cdot v} \, F_H(b_\perp) + \ORd{\frac{1}{m}}
\,.\end{align}
The leftover boost factor
$1/(\nbar \cdot v) = 1/v^- \sim m \, \df b^+$
is a consequence of the nonrelativistic normalization of the external state,
and ensures that projections of the above result onto good components
are indeed invariant under the rescaling transformation $\nbar \mapsto e^\alpha \, \nbar^\mu$
discussed below \eq{def_wilson_line}.

To analyze the spin structure of $F_H(b_\perp)$,
it is convenient to define the auxiliary vector
\begin{align} \label{eq:z_def}
z^\mu = v^\mu - \frac{\bn^\mu}{\bn \cdot v}
\qquad \text{with} \qquad
z^2 = -1
\,,\end{align}
which defines a unit $z$ axis oriented along the spatial component of $\bn$ in the rest frame.
Written out explicitly, $F_H(v, z, b_\perp)$ depends on the three four vectors $v^\mu$,
i.e., the label direction in bHQET, corresponding to $P_H^\mu$ in the full theory,
the spacelike vector $z^\mu$ parameterizing the Wilson line direction $\bn^\mu$ relative to $v^\mu$,
and the spatial separation $b_\perp^\mu$ of the fields
(with direction $x^\mu \equiv b_\perp^\mu/b_T$, $x^2 = -1$).
As these three are orthogonal, they define a unique fourth
unit direction $y^\rho = \eps^{\mu\nu\rho\sigma} v_\mu x_\nu z_\sigma$ with $y^2 = -1$.

There are three applicable symmetries constraining the form of $F_H$.
First, the correlator populates only the particle-particle components of the bispinor,
\begin{align} \label{eq:f_projections}
F_H = P_+ \, F_H = F_H \, P_+
\quad \Leftrightarrow \quad
P_- \, F_H = F_H \, P_- = 0
\,.\end{align}
Second, from its definition, the correlator transforms
under hermitian conjugation as
\begin{align} \label{eq:f_hermiticity}
F_H^\dagger(v, z, b_\perp)
&= \frac{1}{2N_c} \Tr
\SumInt_{X}
\Mae{0}{W^\dagger(0) \,
   h_v(0) \ket{H_vX} \bra{H_vX} \bar h_v(b_\perp) \,
W(b_\perp)}{0}
\\
&= \frac{1}{2N_c} \Tr
\SumInt_{X}
\Mae{0}{W^\dagger(-b_\perp) \,
   h_v(-b_\perp) \ket{H_vX} \bra{H_vX} \bar h_v(0) \,
W(0)}{0}
= F_H(v, z, -b_\perp)
\nn \,,\end{align}
where $h_v^\dagger = \bar h_v \, \Sl v = \bar h_v P_+ \Sl v = \bar h_v P_+ = \bar h_v$
simplifies for the particle components.
On the second identity we translated both matrix elements by $-b_\perp$,
exploiting that the resulting phase cancels between the states.
Third, since parity is conserved in QCD (and thus in bHQET),
the correlator satisfies
\begin{align} \label{eq:f_parity}
F_H(v, z, b_\perp)
= \mathbf{P} F_H(v, z, b_\perp) \mathbf{P}
= \Sl v \, F_H(v, -z, -b_\perp) \, \Sl v
= F_H(v, -z, -b_\perp)
\,,\end{align}
where $\mathbf{P}$ denotes the unitary operator implementing parity in the rest frame of the heavy meson.
Note that time reversal is broken by the presence of the out states,
and thus is not a good symmetry of fragmentation functions~\cite{Mulders:1995dh}.
Finally, charge conjugation relates the form of $F_H$ for $Q \to H$
to that of $\bar{Q} \to \bar{H}$, but does not constrain the form of $F_H$ for a given final state.

By applying \eqss{f_projections}{f_hermiticity}{f_parity}
to the most general form of $F_H(v, z, b_\perp)$ allowed by Lorentz covariance,
it is easy to show that
\begin{align} \label{eq:f_result}
F_H(v, z, b_\perp) = \chi_{1,H}(b_T) \, \frac{1 + \Sl v}{2} + \chi_{1,H}^\perp(b_T) \, \Sl z \frac{\Sl b_\perp}{b_T}\, \frac{1 + \Sl v}{2}
\end{align}
in terms of two real-valued scalar coefficient functions
$\chi_{1,H}(b_T)$ and $\chi_{1,H}^\perp(b_T)$
that can only depend on $v^2 = -z^2 = 1$ and $b_\perp^2 = b_T^2$.
By performing the traces in \eq{def_tmd_ff_unpol_collins},
we can identify these two functions with the unpolarized and Collins TMD FF, respectively,
\begin{align} \label{eq:tmdffs_bhqet_tree_level_final_results}
D_{1\,H/Q}(z_H, b_T)
&= \delta(1-z_H) \, \chi_{1,H}(b_T) + \ord{\as} + \ORd{\frac{1}{m}}
\,, \nn \\
b_T M_H \, H_{1\,H/Q}^{\perp (1)}(z_H, b_T)
&= \delta(1-z_H) \, \chi_{1,H}^\perp(b_T) + \ord{\as} + \ORd{\frac{1}{m}}
\,,\end{align}
These results hold at the leading order in the heavy-quark expansion%
\footnote{
While we could have expanded $M_H = m + \ord{\lqcd}$
in the conventional normalization factor for $H_{1\,H/Q}^{\perp(1)}$
at this order in the HQET expansion,
we choose to keep it exact as it cancels a corresponding term
in factorized cross sections.
}
and up to perturbative corrections at the scale $\mu \sim m$
(which we reinstate in \sec{all_order_matching_and_renormalization}),
but capture the exact nonperturbative dependence on $k_T \sim \lqcd$ within the $\chi_{1,H}$ and $\chi_{1,H}^\perp$.
For reference, we can also take suitable traces of \eq{f_result}
to obtain explicit definitions of $\chi_{1,H}$ and $\chi_{1,H}^\perp$
in terms of bHQET matrix elements,
\begin{align} \label{eq:def_chi_1h}
\chi_{1,H}(b_T)
= \frac{1}{2} \tr F_H(b_\perp)
\,, \qquad
\chi_{1,H}^\perp(b_T)
= \frac{1}{2} \tr \Bigl[ \frac{\Sl b_\perp}{b_T} \Sl z \, F_H(b_\perp) \Bigr]
\,.\end{align}
which we dub \emph{heavy-quark TMD fragmentation factors}.
Note that finding a nonzero result for the Collins fragmentation function
during this step crucially relies on the presence of the Wilson line,
which distinguished a nontrivial reference direction $z^\mu$ in the rest frame.

From our analysis so far, we can conclude
that both leading-power TMD fragmentation functions for unpolarized hadrons
are allowed by the discrete symmetries of QCD in the heavy-quark limit
at leading order in $1/m$.
Written as in \eq{def_chi_1h}, they are also manifestly independent
of the flavor and mass of the heavy quark.
However, the physical interpretation of $\chi_{1,H}(b_T)$ and $\chi_{1,H}^\perp(b_T)$
is still fairly unclear at this point, and in fact we have not yet made use
of heavy-quark spin symmetry.
We address these questions in the next section,
where we will derive an intuitive physical picture of the TMD fragmentation factors
in terms of the individual constituents of the heavy hadron,
and will derive powerful relations within spin symmetry multiplets.

\subsubsection{Heavy-quark spin symmetry}
\label{sec:tmd_ffs_heavy_quark_spin_symmetry}

We now return to the full correlator $F_H(b_\perp)$ defined in \eq{def_f}
and analyze its heavy-quark spin symmetry properties,
which are particularly transparent when working with sterile fields.
To do so, we first decompose the out states as in \eq{clebsch_gordan},
\begin{align}
\sum_{h_H} \ket{H_v, h_H; X}
\bra{H_v, h_H; X}
&= \sum_{h_H} \Bigl( \sum_{h_Q}
\sum_{h_\ell}
\ket{h_Q; s_\ell, h_\ell, f_\ell; X}
\braket{s_Q, h_Q; s_\ell, h_\ell}{s_H, h_H}
\Bigr)
\nn \\ & \quad \times
\Bigl( \sum_{h_Q'}
\sum_{h_\ell'}
\braket{s_H, h_H}{s_Q, h_Q'; s_\ell, h_\ell'}
\bra{h_Q'; s_\ell, h_\ell', f_\ell; X}
\Bigr)
\,.\end{align}
For definiteness, we take the magnetization axis defining the helicity eigenvalues
to be the spatial component $z^\mu$ of the Wilson line direction
in \eq{z_def}, which points back to the hard collision.
Crucially, different helicities of the quark and the light degrees of freedom
in the amplitude ($h_Q, h_\ell$) and the complex conjugate amplitude ($h_Q', h_\ell'$)
can interfere with each other, as only their sum is constrained to be equal
to a common $h_H$ by helicity conservation.
Note that we cannot use a completeness relation for the Clebsch-Gordan coefficients
because they are only summed over the hadron helicity $h_H$,
but are \emph{not} summed over the total hadron angular momentum $s_H$
because we assume that the experimental measurement
can e.g.\ tell apart $D$ and $D^*$ mesons.
Acting on these out states with sterile heavy-quark fields
as in \eq{decoupling_outstate} yields
\begin{align}
F_H(b_\perp)
&= \frac{1}{2} \sum_{h_H} \sum_{h_Q, h_Q'} \sum_{h_\ell, h_\ell'}
u(v, h_Q) \, \bar u(v, h_Q') \,
\braket{s_Q, h_Q; s_\ell, h_\ell}{s_H, h_H}
\braket{s_H, h_H}{s_Q, h_Q'; s_\ell, h_\ell'}
\nn \\ & \quad \times
\frac{1}{N_c} \Tr
\SumInt_{X}
\Mae{0}{W^\dagger(b_\perp) \,
   Y_v(b_\perp) \Ket{s_\ell, h_\ell, f_\ell ; X} \Bra{s_\ell, h_\ell', f_\ell ; X} Y^\dagger_v(0) \,
W(0)}{0}
\nn \\
&\equiv \frac{1}{2} \sum_{h_Q, h_Q'}
u(v, h_Q) \, \bar u(v, h_Q') \, \rho_{H,h_Q h_Q'}(b_\perp)
\,.\end{align}
On the last line we defined a generalized
spin-density matrix $\rho_{H,h_Qh_Q'}(b_\perp)$
for the heavy-quark helicities.
Its entries are determined by the soft dynamics, the experimentally
reconstructed values of $s_H, s_\ell$ (i.e., $H$), and angular momentum conservation.

To proceed, it is useful to express the outer product of spinors as
\begin{align}
u(v, h_Q) \, \bar u(v, h_Q')
= \frac{1 + \Sl v}{2} \bigl( \delta_{h_Q' h_Q} + 2 \gamma_5 {\Sl \Sigma_{Q,h_Q' h_Q}} \bigr)
\,,\end{align}
where ${\Sigma_Q^\mu}$ is the heavy-quark spin operator
acting on the nonrelativistic spin Hilbert space $\mathbb{C}^2$
spanned by $\ket{h_Q}, \ket{h_Q'} = \ket{\pm \tfrac{1}{2}}$,
with ${\Sigma_Q^\mu} = \tfrac{1}{2}(0, \sigma_x, \sigma_y, \sigma_z)$ in the rest frame,
and ${\Sigma^\mu_{Q,h_Q'h_Q}} = \mae{h_Q'}{{\Sigma_Q^\mu}}{h_Q}$ is its matrix representation.
Evaluating the traces in \eq{def_chi_1h} then yields
\begin{align} \label{eq:chi_1h_quark_spin_space}
\chi_{1,H}(b_T)
= \frac{1}{2} \tr_{\,2 \times 2} \bigl[ \rho_{H}(b_\perp) \bigr]
\,, \qquad
\chi_{1,H}^\perp(b_T)
= - \img \tr_{\, 2 \times 2} \bigl[ (y \cdot {\Sigma_Q}) \, \rho_{H}(b_\perp) \bigr]
\,,\end{align}
where $y_\mu$ with $y^2 = -1$ is orthogonal to both $b_\perp^\mu$
and the Wilson line direction $z^\mu$, see the discussion below \eq{z_def}.
As expected from its relation to the unpolarized TMD FF,
the Fourier transform of $\chi_{1,H}$
is simply the total conditional probability to produce $H$ at rest
given an initial quark momentum $k_T$ transverse
to the direction of the static color source provided by the Wilson line.
The bHQET analogue $\chi_{1,H}^\perp$ of the Collins function on the other hand
can be interpreted as a conditional density of quark spin ${\Sigma_Q}$ with respect
to a magnetization axis defined by the static color source
and the final-state transverse momentum.
These physical interpretations roughly correspond
to those of the associated relativistic fragmentation functions,
which are often written in a form similar to \eq{chi_1h_quark_spin_space}.
Crucially, however, the meaning of the spin space on which the density matrix is defined
is different in the heavy-quark case:
For light quarks, whose spin after hadronization is an ill-defined concept,
it only refers to the initial spin state in which the light quark is prepared.
Heavy quarks at leading power in $1/m$, by contrast, retain their initial spin state
throughout the fragmentation process,
and thus their spin density matrix together with the hadron spin measurement
instead probes the angular momentum distribution of the final-state
light constituents of the heavy hadron.%
\footnote{A related difference is that relativistic fragmentation functions
have to be interpreted as number densities rather than probabilities
due to the semi-inclusive measurement acting also on hadronic states at higher multiplicity.
By contrast, the unpolarized bHQET TMD fragmentation factor
upon Fourier transform, and up to renormalization effects~\cite{Ebert:2022cku},
has an interpretation as a probability density
because additional pair production of heavy quarks is power suppressed by $k_T \ll m$.}

To make this fully explicit, let us introduce a shorthand
for the spin-density matrix $\rho_\ell$
of the light degrees of freedom $\ell \equiv \{ s_\ell, f_\ell\}$,
\begin{align} \label{eq:def_rho}
\rho_{\ell, h_\ell h_\ell'}(b_\perp)
&\equiv \frac{1}{N_c} \Tr
\SumInt_{X}
\Mae{0}{W^\dagger(b_\perp) \,
   Y_v(b_\perp) \Ket{s_\ell, h_\ell, f_\ell ; X} \Bra{s_\ell, h_\ell', f_\ell ; X} Y^\dagger_v(0) \,
W(0)}{0}
\,, \nn \\
\Rightarrow \rho_{H, h_Q h_Q'}(b_\perp)
&= \sum_{h_H} \sum_{h_\ell, h_\ell'}
\braket{s_Q, h_Q; s_\ell, h_\ell}{s_H, h_H}
\braket{s_H, h_H}{s_Q, h_Q'; s_\ell, h_\ell'} \,
\rho_{\ell, h_\ell h_\ell'}(b_\perp)
\,.\end{align}
The fact that the same light spin density matrix $\rho_\ell$ appears
for all hadrons within the same spin symmetry multiplet
(same $s_\ell$ and $f_\ell$, but different $s_H$)
leads to relations between their TMD FFs in the heavy-quark limit.
While it is interesting to ask how many independent nonperturbative
functions the constraints on $\rho_\ell$ from parity and hermiticity
in \eqs{f_hermiticity}{f_parity} leave \emph{in principle},
and how many of them are observable when reconstructing
the hadron spin in addition,
we now push on towards the combinations
that are relevant for an unpolarized hadron
and that contribute to the two fragmentation factors at hand.

\paragraph{Unpolarized TMD FF:}
We begin with the unpolarized quark case
and perform the trace in \eq{chi_1h_quark_spin_space},
which sets $h_Q = h_Q'$ and thus $h_\ell = h_\ell'$,
\begin{align} \label{eq:chi1H_in_terms_of_rho_ell}
\chi_{1,H}(b_T)
&= \frac{1}{2} \sum_{h_H} \sum_{h_Q} \sum_{h_\ell} \,
\abs{ \braket{s_Q, h_Q; s_\ell, h_\ell}{s_H, h_H}}^2 \,
\rho_{\ell,h_\ell h_\ell}(b_\perp)
\,.\end{align}
To illustrate this, consider the pseudoscalar case, where
\begin{align} \label{eq:chi1H_pseudoscalar}
s_\ell = 1/2\,,~s_H = 0 \,: \quad
\chi_{1,H}(b_T)
&= \frac{1}{4} \bigl[ \rho_{\ell, ++}(b_\perp) + \rho_{\ell, --}(b_\perp) \bigr]
\,,\end{align}
and we have written helicities as $\pm \equiv \pm \tfrac{1}{2}$ for short.
We see that the unpolarized TMD FF encodes information
about the magnitude of the amplitude for producing a given light helicity state.
Summing over all hadrons $H$ within the same spin symmetry multiplet $M_\ell$
(i.e., all hadrons with identical light spin and flavor state $\ell$),
we further define
\begin{align} \label{eq:sum_H_in_Mell_chi1H}
\chi_{1,\ell}(b_T)
&\equiv \sum_{H \in M_\ell} \chi_{1,H}(b_T)
=  \sum_{h_\ell} \rho_{\ell, h_\ell h_\ell}(b_\perp)
\,,\end{align}
where we used the completeness relation of the Clebsch-Gordan coefficients.
Note that in the same way, this sum reduces the quark spin density matrix
and the correlator to
\begin{align} \label{eq:sum_H_in_Mell_rhoH_FH}
\sum_{H \in M_\ell} \rho_{H,h_Q h_Q'}(b_\perp) = \delta_{h_Q h_Q'} \, \chi_{1,\ell}(b_T)
\,, \qquad
\sum_{H \in M_\ell} F_H(b_\perp) = \frac{1 + \Sl v}{2} \chi_{1,\ell}(b_T)
\,.\end{align}
By evaluating the partial sums in \eq{chi1H_in_terms_of_rho_ell},
it is easy to see that in terms of this baseline,
the individual unpolarized TMD fragmentation factors are given by
\begin{alignat}{3} \label{eq:spin_symmetry_relations_chi1H}
s_\ell &= \tfrac{1}{2} \,: \qquad
&\chi_{1,H}(b_T) &= \frac{1}{4} \chi_{1,\ell}(b_T)
\,,  \qquad
&\chi_{1,H^*}(b_T) &= \frac{3}{4} \chi_{1,\ell}(b_T)
\,, \nn \\[0.4em]
s_\ell &= 1 \,: \qquad
&\chi_{1,\Sigma_Q}(b_T) &= \frac{1}{3} \chi_{1,\ell}(b_T)
\,, \qquad
&\chi_{1,\Sigma_Q^*}(b_T) &= \frac{2}{3} \chi_{1,\ell}(b_T)
\,, \nn \\[0.4em]
s_\ell &= \tfrac{3}{2} \,: \qquad
&\chi_{1,H_1}(b_T) &= \frac{3}{8} \chi_{1,\ell}(b_T)
\,, \qquad
&\chi_{1,H_2^*}(b_T) &= \frac{5}{8} \chi_{1,\ell}(b_T)
\,,\end{alignat}
where for the purpose of this equation
we used $H$ as a shorthand for $D$ ($\bar{B}$) when $Q = c$ ($b$),
and similarly for the excited and higher-spin states.
These relations are textbook knowledge
in the inclusive fragmentation case ($b_T = 0$)~\cite{Manohar:2000dt}.
Our analysis shows, for the first time, that they hold without modification
and point by point in the distribution when resolving the hadron transverse momentum.
We anticipate that a transverse momentum-dependent version of the Falk-Peskin parameter~\cite{Falk:1993rf}
appears when resolving individual (linear) hadron polarizations $h_H$
of hadrons with $s_\ell = 3/2$.
Generalizations of the above to higher multiplets are trivial.

Given that the sum over fragmentation factors within a multiplet
encodes the full information on each individual one,
it is interesting to ask what form the \emph{total} fragmentation factor takes.
Performing the complete sum over states, we have
\begin{align} \label{eq:def_chi1}
\chi_1(b_T)
\equiv \sum_{H} \chi_{1,H}(b_T)
= \frac{1}{N_c} \Tr
\Mae{0}{W^\dagger(b_\perp) \,
   Y_v(b_\perp) \, Y^\dagger_v(0) \,
W(0)}{0}
\,.\end{align}
We see that the total unpolarized heavy-quark TMD FF
reduces to a vacuum matrix element
of two staple-shaped Wilson-line configurations
along the lightlike and timelike direction, respectively.
(Recall that we have suppressed transverse gauge links at infinity;
the Wilson lines in the interpolating fields for the out states cancel.)
This is reminiscent of the heavy quark-antiquark form factor
proposed in \refcite{Ji:2019sxk} to extract the TMD soft function on the lattice,
but in contrast to that proposal directly relates to a physical observable,
i.e., the total TMD cross section for producing heavy hadrons in $e^+e^-$ collisions.
We contend that this makes \eq{def_chi1}
the theoretically simplest TMD observable possible in QCD,
since it is entirely given in terms of vacuum matrix elements
of Wilson loops.

\begin{figure*}
\centering
\subfloat[]{
   \includegraphics[width=0.38\textwidth]{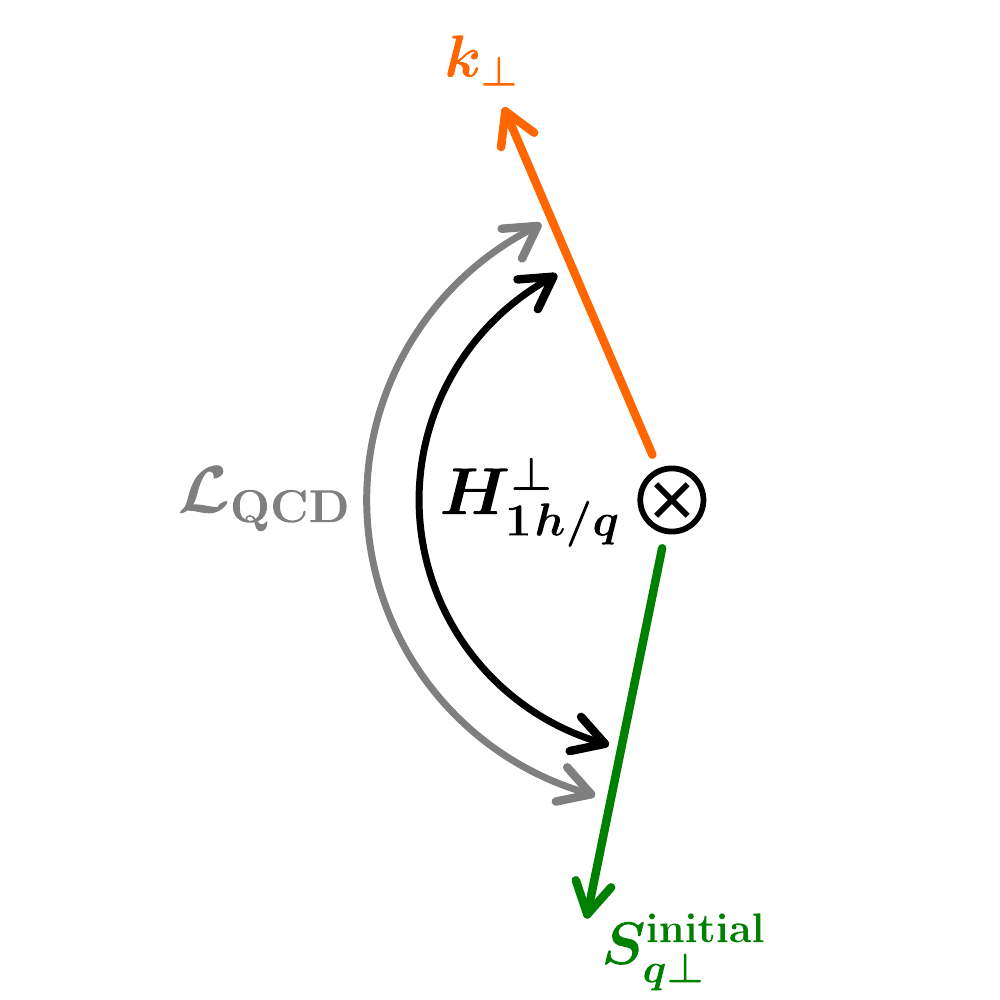}
}%
\hspace{2em}
\subfloat[]{
   \includegraphics[width=0.38\textwidth]{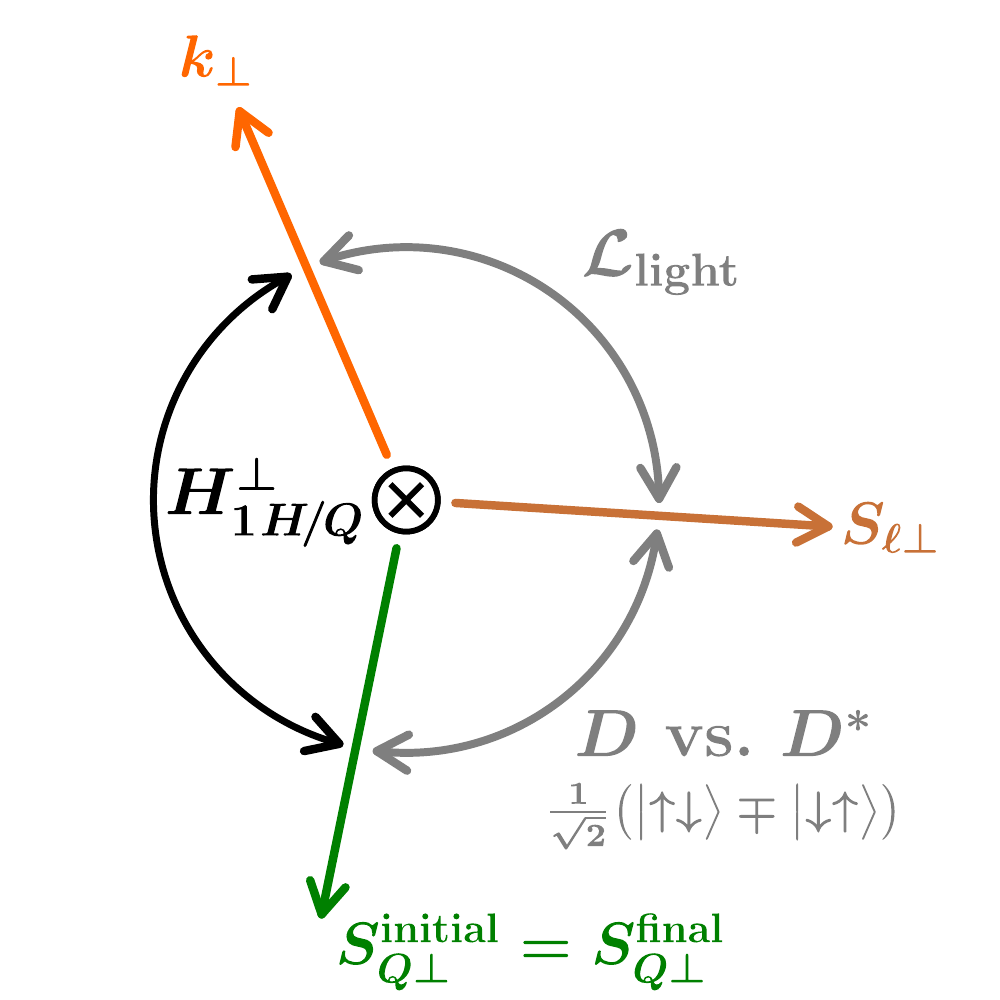}
}%
\caption{
Origin of the Collins TMD FF $H_1^{\perp}$
for \textbf{(a)}~light and \textbf{(b)}~heavy quarks.
Here $k_\perp$ is the transverse momentum of additional hadronic radiation
into the final state, $S_{q\perp}$ (or $S_{Q\perp}$)
is the quark transverse polarization vector,
and $S_{\ell\perp}$ is the transverse polarization vector
of the light hadron constituents.
}
\label{fig:illustration_collins_tmd_ff}
\end{figure*}

\paragraph{Collins TMD FF:}
A naive expectation from heavy-quark spin symmetry
might be that the Collins FF should be suppressed by $1/m$
because it encodes a correlation between
the initial quark transverse polarization vector
and the transverse momentum of hadronic final-state radiation.
In the case of light quarks, this correlation arises
directly from the nonperturbative dynamics of the QCD Lagrangian,
as illustrated in \fig{illustration_collins_tmd_ff}~(a),
but in the heavy-quark case
it naively seems to require a suppressed magnetic interaction
with the heavy-quark spin.
We will now see that this is not the case.
As illustrated in \fig{illustration_collins_tmd_ff}~(b),
the angle between the final-state heavy-quark
and light transverse polarization vectors
(i.e, the relative phase between their helicity states)
determines which hadron in the spin symmetry multiplet is produced,
even without a dynamical heavy-quark spin interaction taking place.
Reconstructing this information experimentally thus induces a correlation
between the heavy-quark and the light spin state.
Crucially, spin symmetry ensures that the final-state
heavy-quark spin state is identical to the one it was prepared in.
The light spin state in turn \emph{is} in general correlated
with the transverse momentum $k_\perp$ of hadronic final-state radiation,
since they both arise from the same nonperturbative dynamics
of the light degrees of freedom, leading to a Collins effect at the leading order in $1/m$.
To illustrate this, it is again instructive to consider
the case of the pseudoscalar meson,
\begin{align} \label{eq:chi1Hperp_pseudoscalar}
s_\ell = 1/2\,,~s_H = 0 \,: \quad
\chi_{1,H}^\perp(b_T)
= \frac{1}{4} 
   \bigl[ \rho_{\ell,-+}(b_\perp) - \rho_{\ell,+-}(b_\perp) \bigr]
\end{align}
As expected, the Collins FF in the heavy-quark limit contains information
about the strength of the interference, and hence the relative nonperturbative phases,
of amplitudes for different light helicities.

As a corollary, we conclude that the Collins FF must vanish
at leading order in $1/m$
when summing over all the hadrons in the spin symmetry multiplet,
\begin{align} \label{eq:sum_H_in_Mell_chi1Hperp}
\sum_{H \in M_\ell} \chi_{1,H}^\perp(b_T) = 0
\,.\end{align}
This is immediate to see from the diagonal form
of the summed quark spin density matrix or the full correlator in \eq{sum_H_in_Mell_rhoH_FH}.
Concretely, this means that the Collins FF vanishes altogether
for $s_\ell = 0$ baryons, $\chi_{1,\Lambda_Q} = 0$.
For the next few multiplets and using the same notation as in \eq{spin_symmetry_relations_chi1H},
the explicit relations are
\begin{alignat}{2} \label{eq:spin_symmetry_relations_chi1Hperp}
s_\ell &= \tfrac{1}{2} \,: \qquad
&\chi_{1,H}^\perp(b_T) + \chi_{1,H^*}^\perp(b_T)
&= 0
\,, \nn \\[0.4em]
s_\ell &= 1 \,: \qquad
&\chi_{1,\Sigma_Q}^\perp(b_T) + \chi_{1,\Sigma_Q^*}^\perp(b_T)
&= 0
\,, \nn \\[0.4em]
s_\ell &= \tfrac{3}{2} \,: \qquad
&\chi_{1,H_1}^\perp(b_T) + \chi_{1,H_2^*}^\perp(b_T)
&= 0
\,.\end{alignat}

\paragraph{Discussion:}
The spin symmetry relations in \eqs{spin_symmetry_relations_chi1H}{spin_symmetry_relations_chi1Hperp}
are the main results of this section.
They hold for all values of $b_T$,
which means that they also hold point by point in $k_T$ upon Fourier transform.
Furthermore, they are unaffected by renormalization,
as we discuss in the next section.
This makes them substantially stronger than
the known sum rules for relativistic TMD fragmentation functions.
For the light-quark Collins function in particular,
the Sch{\"a}fer-Teryaev sum rule~\cite{Schafer:1999kn}
has only been rigorously proven~\cite{Meissner:2010cc} in the bare case,
and requires a sum over all possible hadrons, an integral over $z_h$,
and a weighted integral over $k_T$.
This is in contrast to the novel sum rule in \eq{sum_H_in_Mell_chi1Hperp}
that we have derived for the heavy-quark limit,
which only requires summing the Collins function
over a subset of hadrons and holds at any value of $k_T$ and $z_H$.
(We will explicitly extend these spin symmetry relations
to $k_T \sim m$ in \sec{tmd_ffs_from_collinear_ffs}.)
It thus implies the Sch{\"a}fer-Teryaev sum rule
for heavy quarks as a corollary.

We caution that as in the inclusive case,
the validity of the spin symmetry relations rests on the assumption
that spin symmetry violation is negligible during the entire fragmentation process.
This assumptions breaks down e.g.\ for those $H, H^*$ produced from the decays of $H_1, H_2^*$
whose spin symmetry-violating mass splitting is comparable to their widths~\cite{Falk:1993rf}.

\subsubsection{All-order matching and renormalization}
\label{sec:all_order_matching_and_renormalization}

Using known results~\cite{Hoang:2015vua, Hoang:2019fze}
for the perturbative matching of SCET onto bHQET,
it is in fact straightforward to generalize \eq{tmdffs_bhqet_tree_level_final_results}
to all orders in perturbation theory,
\begin{align} \label{eq:tmdffs_bhqet_all_order_final_results}
D_{1\,H/Q}(z_H, b_T, \mu, \zeta)
&= \delta(1-z_H) \, C_m(m, \mu, {\zeta}) \, \chi_{1,H}\Bigl(b_T, \mu, {\frac{\sqrt{\zeta}}{m}}\Bigr) + \ORd{\frac{1}{m}}
\,, \nn \\
b_T M_H \, H_{1\,H/Q}^{\perp (1)}(z_H, b_T, \mu, \zeta)
&= \delta(1-z_H) \, C_m(m, \mu,{\zeta}) \, \chi_{1,H}^\perp\Bigl(b_T, \mu, {\frac{\sqrt{\zeta}}{m}}\Bigr) + \ORd{\frac{1}{m}}
\end{align}
Here the matching coefficient $C_m = 1 + \ord{\as}$ arises from separately
matching the collinear (``unsubtracted'') and soft contributions
to the TMD FFs onto bHQET and QCD with $n_\ell$ light flavors, respectively.

Importantly, the matching is diagonal in spin and forces $z_H = 1$
because real radiation is parametrically forbidden at the scale $\mu \sim m$
due to $k_T \ll m$, and thus we can separately match the collinear fields
in the two matrix elements in \eq{def_tmd_ff_correlator} onto bHQET.
Starting at two loops, the matching coefficient features rapidity logarithms
of the Collins-Soper scale $\zeta$ over the mass
as a consequence of the large boost separating
the heavy hadron rest frame and the frame
where the soft radiation is isotropic.
The appearance of the Collins-Soper scale can most transparently be understood
by organizing the matching of the collinear sector
onto bHQET in terms of gauge-invariant building blocks $W^\dagger \psi_Q$
with definite large lightcone momentum $\omega = \sqrt{\zeta}$,
as commonly done in SCET.
The soft matching is nontrivial because starting at $\ord{\as^2}$,
vacuum polarization diagrams involving the heavy quark contribute
to the expectation values of soft Wilson line operators in the $n_\ell + 1$ theory.
The two-loop result for $C_m$ was obtained in \refcite{Hoang:2015vua},
and our notation relates to theirs as
\begin{align}
C_m(m, \mu, {\zeta}) =
   H_{m,n}\Bigl(m, \mu, \frac{\nu}{\sqrt{\zeta}} \Bigr)
   \Bigl[ H_{m,s}\Bigl(m, \mu, \frac{\nu}{m} \Bigr) \Bigr]^{1/2}
\,.\end{align}
Here the dependence on the rapidity scale $\nu$ cancels
between the individual matching coefficients on the right-hand side,
leaving behind the dependence on $\zeta/m^2$.

The renormalization properties of $\chi_{1,H}$ and $\chi_{1,H}^\perp$
follow from \eq{tmdffs_bhqet_all_order_final_results}
by consistency with the bHQET matching.
Making the renormalization explicit requires
introducing a rapidity regulator into the bHQET matrix element definitions,
e.g.\ by modifying the lightlike Wilson lines in a standard fashion,
and canceling rapidity divergences using the known TMD soft factor
in a theory with $n_\ell$ light quarks,
which leaves behind an anomalous dependence
on the boost factor $\sqrt{\zeta}/m = \nbar \cdot v$
governed by the Collins-Soper kernel.
(In the following we find it convenient to use
the shorthand $\rho = \nbar \cdot v$ for the third argument
of the TMD fragmentation factors,
which physically is given by the exponential
of the hadron's rapidity in the frame of the hard collision.)
This proceeds in close analogy to the standard relativistic case,
so we leave an explicit check to a future perturbative calculation.
Note that the Wilson lines in \eq{def_f} are in fact still lightlike
up to possible off-lightcone regulators
(and thus feature standard rapidity divergences) despite the presence of the mass
because the opposite collinear sector is boosted close to the lightcone
from the point of view of the bHQET rest frame.
Conversely, the bHQET dynamics inside either collinear sector
are also boosted from the point of view of the central soft modes.
Since TMD renormalization is multiplicative in $b_T$
and independent of the hadronic state, it acts in the same way
on all terms in the spin symmetry relations in \eqs{spin_symmetry_relations_chi1H}{spin_symmetry_relations_chi1Hperp},
which implies that they also hold for the renormalized objects
point by point in $b_T$ (or $k_T$).

\subsubsection{Relation to bHQET fragmentation probabilities for
\texorpdfstring{$\lqcd \ll k_T$}{LQCD << kT}}
\label{sec:relation_to_bhqet_fragmentation_probabilities}

An important property of the TMD fragmentation factors
we defined above is their limiting behavior as $k_T \gg \lqcd$
or, equivalently, $b_T \to 0$.
In this limit,
the unpolarized TMD fragmentation factor $\chi_{1,H}$ is related
to the total probability $\chi_H$ for the quark to fragment into $H$,
which has previously been analyzed in HQET~\cite{Jaffe:1993ie, Falk:1993rf, Neubert:2007je},
\begin{align} \label{eq:chi1h_ope}
\chi_{1,H}(b_T, \mu, {\rho}) = C_1(b_T, \mu, {\rho}) \, \chi_H + \ord{\lqcd^2 b_T^2}
\,.\end{align}
where the matrix-element definition of $\chi_H$~\cite{Fickinger:2016rfd}
is equal to $\chi_{1,H}(b_T = 0)$ at the bare level,
\begin{align} \label{eq:def_chiH}
\chi_H \equiv \frac{1}{4 N_c} \Tr \tr
\SumInt_{X}
\Mae{0}{W^\dagger
   h_v \Ket{H_vX} \Bra{H_vX} \bar h_v
W}{0}
\,.\end{align}
Because $\chi_H$ is not renormalized~\cite{Mele:1990cw},
we generally expect a perturbative Wilson coefficient
$C_1(b_T, \mu, \zeta) = 1 + \ord{\as}$ to appear
in \eq{chi1h_ope} at the renormalized level from integrating out partonic physics
at the scale $\mu \sim k_T$.
The Wilson coefficient can be calculated in perturbation theory
by replacing the heavy hadron state $\ket{H_v X}$ in \eqs{def_f}{def_chi_1h}
with partonic states $\ket{Q_v}$, $\ket{Q_v \,g}$, $\ket{Q_v \, q \bar q}$ etc.
\Eq{chi1h_ope} can be considered the leading term of a twist-like expansion
of $\chi_{1,H}(b_T)$ in $\lqcd b_T$ at leading order in the strong coupling,%
\footnote{
It is well known that the formal OPE of relativistic fragmentation functions
is ambiguous due to an unconstrained choice of boundary condition at lightcone infinity~\cite{Balitsky:1990ck, Moos:2020wvd}.
While this fundamental issue remains present here,
it is interesting to ask whether the case of bHQET TMD fragmentation factors,
which are Wilson loops, can provide additional insight into this issue.
}
and is analogous to the behavior of the usual TMD PDF as $b_T \to 0$
where it approaches its total momentum-space integral
given by the collinear PDF, up to renormalization effects and radiative corrections~\cite{Ebert:2022cku}.
Here we also assumed without detailed proof that corrections
to this relation are quadratic in $b_T$ based on the azimuthal symmetry of $\chi_{1,H}(b_T)$.
From $\sum_H \chi_H = 1$,
it follows that the total TMD fragmentation factor $\chi_{1}(b_T)$
defined in \eq{def_chi1} is purely perturbative in this regime,
\begin{align} \label{eq:chi1_perturbative}
\chi_1(b_T, \mu, {\rho}) = C_1(b_T, \mu, {\rho}) + \ord{\lqcd^2 b_T^2}
\,.\end{align}

In contrast to \eq{chi1h_ope}, the Collins TMD fragmentation factor $\chi_{1,H}^\perp$
must vanish at least linearly as $b_T \to 0$
because there is no leading bHQET matrix element
it could match onto in this limit.
This is easiest to see by repeating the symmetry analysis
of the bHQET correlator $F_H$ in \sec{tmd_ffs_from_bhqet_tree_level} at $b_T = 0$,
which only admits $F_H(v, z, b_\perp = 0) = \chi_H (1 + \Sl v)/2$.
As we will see by comparing to the limit $\lqcd \ll k_T \lesssim m$ in \sec{tmd_ffs_consistency},
the expansion indeed starts at the linear order,
and the matrix-element definition of the relevant nonperturbative
parameter at $\ord{\lqcd b_T}$, as well as its tree-level Wilson coefficient,
can all be inferred from consistency.

\subsection{Matching TMD FFs onto bHQET for
\texorpdfstring{$\lqcd \ll m \lesssim k_T$}{LambdaQCD << m < kT} }
\label{sec:tmd_ffs_from_collinear_ffs}

We next consider case~(b) in \fig{regimes_fragmentation}.
In this regime, the transverse and longitudinal momentum distributions
are determined by dynamics at the scale $\mu \sim m \sim k_T$ and are fully perturbative.
The nonperturbative dynamics in this case are encoded in bHQET matrix elements
that involve additional gluon fields or derivatives
and that can be nonlocal along the lightcone,
but in contrast to the previous section are local in the transverse direction.
Similar to a standard twist expansion,
these bHQET matrix elements are categorized by their mass dimension,
which determines their scaling as $\lqcd \ll m, k_T$,
i.e., their mass dimension $\sim \lqcd^n$ is compensated
by powers of $b_T$ or $1/m$ in the Wilson coefficient.
This story plays out differently
for the unpolarized vs.\ the Collins TMD FF,
which scale as $\ord{1}$ and $\ord{\lqcd \, b_T}$, respectively,
so we will go through the two cases separately in the following.
We note that the expansion of TMD FFs in terms of bHQET operators
differs from a standard twist expansion insofar
as the HQET field $h_v$ encoding the interactions with the heavy valence quark
remains present in all low-energy matrix elements.

\subsubsection{Unpolarized TMD FF}

For the unpolarized TMD FF, the unique bHQET matrix element
that can arise in the infrared at the leading order in $1/m$
is the total fragmentation probability $\chi_H$ as defined in \eq{def_chiH},
which follows from symmetry arguments similar to those below \eq{chi1_perturbative}:
\begin{align} \label{eq:tmd_ff_unpol_bhqet}
D_{1\,H/Q}(z_H, b_T, \mu, \zeta)
= d_{1\,Q/Q}(z_H, b_T, \mu, \zeta) \, \chi_H
+ \ORd{\frac{\lqcd}{m}}
+ \ord{\lqcd b_T}
\,.\end{align}
Importantly, we have again made use of the assumption in \eq{away_from_endpoint}
that we are sufficiently far away from (or have fully integrated over)
the endpoint regime $z_H \to 1$,
as otherwise there would be a nontrivial bHQET shape function
on the right-hand side~\cite{Neubert:2007je, Fickinger:2016rfd}.
The unique matching coefficient of $\chi_H$, which we dub
the \emph{partonic heavy-quark TMD FF} $d_{1\,Q/Q}(z, b_T, \mu, \zeta)$,
is a new object that, to our knowledge,
appears in our analysis for the first time.%
\footnote{
Curiously, the perturbative transverse dynamics of heavy-quark fragmentation
$\bar b \to B_c$ have previously been evaluated
in \refscite{Cheung:1994xx, Yuan:1994iv}.
The complete tree-level result given in the first reference,
which starts at $\ord{\as^2}$,
can be considered a very specific subset of the NNLO corrections
to the TMD FF we define here if we sum over final states.
If we tag on the charm instead, their result corresponds
to a different perturbative TMD FF $d_{1\,b\bar{c}/b} \sim \as^2$
whose renormalization, by our analysis, is governed by standard (massive) TMD evolution.
}
It is independent of the precise hadronic final state,
carries the exact dependence on $b_T m \sim 1$,
and can be calculated perturbatively
by evaluating \eqs{def_tmd_ff_correlator}{def_tmd_ff_unpol_collins}
for partonic final states including at least one heavy quark, i.e.,
\begin{align} \label{eq:def_tmd_ff_unpol_partonic}
d_{1\,Q/Q}(z_H, b_T)
&= \tr \Bigl[ \frac{\Sl{\bn}}{2} \, \Delta_{Q/Q}(z_H, b_\perp) \Bigr]
= \delta(1 - z_H) + \ord{\as}
\,.\end{align}
Its rapidity renormalization is governed by the Collins-Soper kernel
of a theory with $n_\ell$ massless and one massive quark degree of freedom~\cite{Pietrulewicz:2017gxc}.
We leave a dedicated NLO calculation of $d_{1\,Q/Q}(z, b_T, \mu, \zeta)$ to future work.
Since the dependence on the hadronic final state is purely encoded in $\chi_H$,
which satisfies the same spin symmetry relations as in \eq{spin_symmetry_relations_chi1H},
we conclude that the unpolarized heavy-quark TMD FF satisfies
\begin{alignat}{3} \label{eq:spin_symmetry_relations_d1hq}
D_{1\,H/Q} = \frac{1}{3} D_{1\,H^*/Q}
\,, \qquad
D_{1\,\Sigma_Q/Q} = \frac{1}{2} D_{1\,\Sigma_Q^*/Q}
\,, \qquad
D_{1\,H_1/Q} = \frac{3}{5} D_{1\,H_2^*/Q}
\,,\end{alignat}
for \emph{all} values of $b_T$ (or $k_T$), including $1/b_T \gtrsim m$,
up to corrections of $\ord{\lqcd/m}$.

\Eq{tmd_ff_unpol_bhqet} continues to be valid for $k_T \gg m$,
but features large perturbative logarithms of $b_T m \ll 1$ in this limit.
Their resummation is enabled by further factorizing the physics at those two scales.
To do so, we can first match the heavy-quark TMD FF
onto twist-2 heavy-quark collinear FFs at the scale $\mu \sim m$~\cite{Nadolsky:2002jr},
\begin{align} \label{eq:light_tmd_ff_unpol_leading_twist}
D_{1\,H/Q}(z_H, b_T, \mu, \zeta)
= \frac{1}{z_H^2} \sum_i \int \! \frac{\df z}{z} \,
\cJ_{i/q}(z, b_T, \mu, \zeta) \,
D_{H/i}\Bigl( \frac{z_H}{z}, \mu \Bigr)
+ \ord{m^2 b_T^2}
\,,\end{align}
where the sum runs over $i = q, \bar q, g$.
This matching takes the same form as the standard matching of light-quark TMD FFs
onto twist-2 FFs at $\mu \sim \lqcd$,
except that the highest IR scale here is given by $m$.
The Wilson coefficients $\cJ_{i/q}(z, b_T, \mu, \zeta)$
encode the perturbative process $q \to i$
in a theory with $n_\ell + 1$ massless flavors at the scale $\mu \sim k_T$,
with the quark retaining a fraction $z$ of the parent's lightcone momentum,
and are known to N$^3$LO~\cite{Luo:2020epw, Ebert:2020qef}.
In a second step, we perform the well-known~\cite{Jaffe:1993ie, Falk:1993rf, Neubert:2007je, Fickinger:2016rfd} matching
of the collinear FF of a massive quark onto bHQET to separate $\lqcd \ll m$,
\begin{align} \label{eq:coll_ff_bhqet}
D_{H/i}(z_H, \mu)
= d_{Q/i}(z_H, \mu) \,
\chi_H + \ORd{\frac{\lqcd}{m}}
\,,\end{align}
where $d_{Q/i}(z_H, \mu)$ is
the perturbative collinear heavy-quark FF for $i \to Q$~\cite{Mele:1990cw},
which is known to NNLO~\cite{Melnikov:2004bm, Mitov:2004du}.
Combining \eqs{light_tmd_ff_unpol_leading_twist}{coll_ff_bhqet}
and comparing to \eq{tmd_ff_unpol_bhqet},
we conclude that the perturbative ingredients are related by
\begin{align} \label{eq:d1qq_refactorization_large_kT}
d_{1\,Q/Q}(z_H, b_T, \mu, \zeta)
= \frac{1}{z_H^2} \sum_i \int \! \frac{\df z}{z} \,
\cJ_{i/Q}(z, b_T, \mu, \zeta) \,
d_{Q/i}\Bigl( \frac{z_H}{z}, \mu \Bigr)
+ \ord{m^2 b_T^2}
\,.\end{align}
This refactorization condition for $d_{1\,Q/Q}$
can serve as a cross check on future perturbative calculations,
and in addition enables resumming logarithms of $k_T/m \gg 1$.

\subsubsection{Collins TMD FF}

To identify the low-energy bHQET matrix element
that the Collins TMD FF matches onto in the limit $\lqcd \ll m \sim k_T$,
we use a two-step matching procedure formally valid
for the hierarchy $\lqcd \ll m \ll k_T$.
(We will later show that the result is correct for either hierarchy.)
As for the unpolarized TMD FF above, this lets us make
use of well-known results for the matching of \emph{light-quark} TMD FFs
onto collinear FFs, which we can then further match onto bHQET.

We start from the diagrammatic small-$b_T$ expansion
of the bare Collins TMD FF for light quarks,
which is valid for $\lqcd \ll k_T$
and given by~\cite{Mulders:1995dh, Boer:1997nt, Yuan:2009dw}%
\footnote{
In the literature, this relation is more commonly given
as a tree-level equality between $\hat{H}_{h/q}$
and a weighted $k_T$ integral over the bare momentum-space Collins FF.
Using \eq{bessel_integrals_ffs} and integrating by parts,
it is easy to see that this reduces to the derivative
of the $b_T$-space Collins FF at $b_T = 0$.
The $\ord{\as}$ corrections to \eq{light_tmd_ff_collins_twist3}
were evaluated at finite $k_T > 0$ in \refcite{Yuan:2009dw}
and involve twist-3 matrix elements
that depend on two independent momentum fractions
and reduce to $\hat H_{h/q}$ in certain limits by use of the equation of motion.
We anticipate that matching these more general matrix elements onto bHQET
will reduce the number of independent (residual) momenta to one
because the heavy-quark momentum is fixed.
}
\begin{align} \label{eq:light_tmd_ff_collins_twist3}
b_T M_h \, H_{1\,h/q}^{\perp(1)}(z_h, b_T)
&= b_T \hat{H}_{h/q}(z_h) + \ord{\as} + \ord{\lqcd^2 b_T^2}
\,,\end{align}
where $\hat{H}_{h/q}$ is a twist-3 collinear fragmentation matrix element
at the scale $\mu \sim \lqcd$,
\begin{align} \label{eq:def_hat_h}
\hat{H}_{h/q}(z_h)
&\equiv \frac{z_h^2}{2N_c} \int \! \frac{\df x^+}{4\pi} \, e^{\img x^+(P_h^-/z_h)/2}
\Tr \tr \SumInt_X \biggl\{
\Mae{0}{
   W^\dagger(x)
\nn \\ & \quad \times
   \sigma_{\alpha -} \bigl[
      \img D_{\perp}^\alpha(x)
      + g \mathcal{G}^{\alpha}_\perp(x)
   \bigr] \psi_q(x)
}{hX}
\Mae{hX}{\bar \psi_q(0) W(0)}{0}
+ \text{h.c.}
\biggr\}
\,,\end{align}
where $\sigma_{\mu -} = \tfrac{\img}{2} [\gamma_\mu, \Sl \nbar]$
and we have defined
\begin{align}
\mathcal{G}^{\alpha}_\perp(x)
\equiv \int_0^{+\infty} \!\!\! \df s \, W(x, x + \nbar s) \, G^{\alpha-}(x + \nbar s) \, W(x + \nbar s, x)
\end{align}
as a shorthand for the insertion of a gluon field strength tensor $G^{\mu\nu}$
anywhere along the lightcone, with $W(x, y)$ a straight Wilson-line segment connecting $x$ and $y$.

We now consider the heavy-quark Collins FF
and at first assume the hierarchy $\lqcd \ll m \ll k_T$.
For the matching at the scale $\mu \sim k_T$,
the mass is an infrared scale, and thus the twist expansion in \eq{light_tmd_ff_collins_twist3}
immediately carries over. The collinear matrix element $\hat{H}_{H/Q}$
takes the same form as \eq{def_hat_h}, but is now defined at the scale $\mu \sim m$.
To implement the separation of scales $\lqcd \ll m$,
we match $\hat{H}_{H/Q}$ onto bHQET.
At tree level, this amounts to a replacement
of the quark fields as in \eq{bhqet_tree_matching},
and after expanding the momentum-conserving phase results in
\begin{align} \label{eq:bhqet_matching_hat_h}
\hat{H}_{H/Q}(z_H)
&= \delta(1-z_H) \, \chi_{H,G} + \ord{\as} +  \ORd{\frac{1}{m}}
\,,\end{align}
where $\chi_{H,G} \sim \lqcd$ is a novel subleading bHQET matrix element defined by
\begin{align} \label{eq:def_chi_h_g}
\chi_{H,G} &\equiv \frac{1}{2 N_c}
\Tr \tr \SumInt_X \Bigl\{
\Mae{0}{ W^\dagger \, \sigma_{\beta \alpha} z^\beta \,
\bigl[
   \img D_{\perp}^\alpha
   + g \mathcal{G}_\perp^\alpha
\bigr]
h_v}{H_vX}
\Mae{H_vX}{\bar h_v W}{0}
+ \text{h.c.}
\Bigr\}
\,.\end{align}
Similar to the total fragmentation probability $\chi_H$ defined in \eq{def_chiH},
$\chi_{H,G}$ no longer depends on $b_{\perp}$,
but is simply a constant that depends on the identified hadron $H$.
Note that a nonzero value of $\chi_{H,G}$
is compatible with all the symmetries of bHQET:
Its defining spin correlator $X(v, z)$ (dropping the spin trace)
is hermitian by construction, satisfies $P_+ X = X P_+ = X$,
and under parity transforms as $\mathbf{P} X(v, z) \mathbf{P} = X(v, -z)$.
Repeating the analysis in \sec{tmd_ffs_from_bhqet_tree_level},
it is therefore proportional to $1 + \Sl v$, which has nonzero trace.

In the last step, we combine \eqs{light_tmd_ff_collins_twist3}{bhqet_matching_hat_h}
to arrive at our final result for the tree-level matching of the heavy-quark
Collins TMD FF onto bHQET:
\begin{align} \label{eq:twist3_bhqet_matching_collins_ff}
b_T M_H \, H_{1\,H/Q}^{\perp(1)}(z_H, b_T)
= \delta(1-z_H) \, b_T \chi_{H,G}
+ \ord{\as} + \ord{\lqcd^2}
\,.\end{align}
Because this derivation assumed $\lqcd \ll m \ll k_T$,
\eq{twist3_bhqet_matching_collins_ff} a priori is only valid up to power corrections in $m b_T$.
However, since we found a nonzero result at our tree-level working order
and power corrections in $m b_T$ can only arise from real radiation
in the calculation of the Wilson coefficient,
\eq{twist3_bhqet_matching_collins_ff} as written
also holds when integrating out both scales simultaneously.
We note that additional low-energy matrix elements
will in general be generated when performing the matching at higher orders in $\as$,
but leave a dedicated construction of the basis of bHQET operators
at this order in $\lqcd$ to future work.

We point out that an observation of the heavy-quark Collins function
in this regime would provide interesting insight
into novel gluon correlations in the heavy-quark fragmentation process
that are encoded in $\chi_{H,G}$.
More specifically, $\chi_{H,G}$ encodes a correlation between
the gluon field polarization and the transverse polarization
of the light constituents of the heavy hadron in the final state,
which as in \sec{tmd_ffs_heavy_quark_spin_symmetry} is indirectly resolved
by reconstructing the total hadron spin,
e.g.\ by distinguishing $D$ and $D^*$ mesons.
Conversely, $\chi_{H,G}$ must vanish
when summing over all hadrons in the spin symmetry multiplet $M_\ell$,
\begin{align} \label{eq:sum_rule_chi_h_g}
\sum_{H \in M_\ell} \chi_{H,G} = 0
\,.\end{align}
This result is straightforward to prove along the lines of \sec{tmd_ffs_heavy_quark_spin_symmetry}
by decoupling the heavy quark fields in \eq{def_chi_h_g}
and exploiting the completeness relation
of the Clebsch-Gordan coefficients, which leaves a trace of the form
$\tr \bigl[ \sigma_{\beta \alpha} (1 + \Sl v)/2 \bigr] = 0$.

Combining these results at large $k_T \sim m$
with those in \eq{spin_symmetry_relations_chi1Hperp}
we conclude that the Collins TMD FF
satisfies the following relations for all values of $b_T$ (or $k_T$),
\begin{alignat}{3} \label{eq:spin_symmetry_relations_h1perphq}
H^{\perp}_{1\,H/Q} = -H^{\perp}_{1\,H^*/Q}
\,, \qquad
H^{\perp}_{1\,\Sigma_Q/Q} = -H^{\perp}_{1\,\Sigma_Q^*/Q}
\,, \qquad
H^{\perp}_{1\,H_1/Q} = -H^{\perp}_{1\,H_2^*/Q}
\,,\end{alignat}
which we have proven here up to corrections of $\ord{\lqcd/m}$
and up to radiative corrections at the scale $\mu \sim k_T \sim m$
for large $k_T$.
We conjecture that the additional bHQET matrix elements generated
by the matching at higher orders in $\as$
will involve the same Dirac structure as \eq{def_chi_h_g},
i.e., an additional insertion of $\gamma_\perp^\mu$,
and thus will also satisfy \eq{sum_rule_chi_h_g},
but leave a detailed all-order analysis in this regime to future work.
(We recall that for $k_T \ll m$ \eq{spin_symmetry_relations_h1perphq} holds to all orders
in the strong coupling, see \sec{tmd_ffs_heavy_quark_spin_symmetry}.)

\subsection{Consistency between regimes for
\texorpdfstring{$\lqcd \ll k_T \ll m$}{LambdaQCD << kT << m}}
\label{sec:tmd_ffs_consistency}

Our results in the previous two sections share a common
domain of validity when the transverse dynamics are already perturbative,
$\lqcd \ll k_T$, but still subject to heavy-quark symmetry, $k_T \ll m$.
In this section we analyze the consistency relations
that arise from this overlap and relate the perturbative bHQET
fragmentation factors to the partonic heavy-quark TMD FFs.

We start with the unpolarized case.
Comparing \eqs{tmdffs_bhqet_all_order_final_results}{chi1h_ope},
which are valid for $\lqcd \lesssim k_T$,
to \eq{tmd_ff_unpol_bhqet}, valid for $k_T \lesssim m$,
we find the following all-order refactorization relation
for the partonic heavy-quark TMD FF
in the limit $k_T \ll m$,
\begin{align} \label{eq:tmd_ff_unpol_consistency}
d_{1\,Q/Q}(z, b_T, \mu, \zeta)
= \delta(1-z) \,
C_m(m, \mu, {\zeta}) \,
C_1\Bigl(b_T, \mu, {\frac{\sqrt \zeta}{m}} \Bigr)
+ \ORd{\frac{1}{b_T m}}
\,.\end{align}
Here we have canceled off the common nonperturbative factors of $\chi_H$.
To interpret the $z$ dependence,
\eq{tmd_ff_unpol_consistency} says that counting $1-z \sim 1$,
$d_{1\,Q/Q}$ must approach $\delta(1-z)$ up to an overall factor
at the distributional level for $b_T m \to \infty$,
i.e., all Mellin moments of $d_{1 \, Q/Q}$ must become equal in this limit.
We expect that \eq{tmd_ff_unpol_consistency} will provide
a powerful consistency check of future perturbative calculations of $d_{1\,Q/Q}$.
It also enables the resummation of large perturbative logarithms of $k_T/m \ll 1$,
complementing the factorized result
in \eq{d1qq_refactorization_large_kT} for the opposite limit.

For the Collins TMD FF we compare
\eq{tmdffs_bhqet_all_order_final_results}
to \eq{twist3_bhqet_matching_collins_ff} and use $C_m = 1 + \ord{\as}$.
Canceling off the $z$ dependence, which is trivial at tree level, this yields
\begin{align} \label{eq:chi1h_perp_ope}
\chi_{1,H}^\perp(b_T, \mu, {\rho}) = \chi_{H,G} \, b_T + \ord{\as} + \ord{\lqcd^2 b_T^2}
\,,\end{align}
which can be interpreted as the leading linear term
in a small-$b_T$ expansion of $\chi_{1,H}^\perp$,
as anticipated in \sec{relation_to_bhqet_fragmentation_probabilities}.
As for the Collins function at $k_T \sim m$,
we leave a dedicated higher-order matching calculation to future work,
which will involve nontrivial Wilson coefficients
integrated against at least one additional $\ord{\lqcd}$ bHQET matrix element.

\subsection{Model functions and numerical results}
\label{sec:tmd_ffs_models}

For our numerical results we will assume a simple Gaussian model
for the unpolarized TMD fragmentation factor,
\begin{align} \label{eq:chi_1h_model}
\chi_{1,H}(b_T, \mu_0, {\rho_0})
= \chi_H \exp\Bigl(- \kappa_H^2 b_T^2 \Bigr)
\,,\end{align}
where $\kappa_H \sim \lqcd$ has units of $\GeV$.
\Eq{chi_1h_model} is valid at initial scales $\mu_0 \sim {m \, \rho_0} \sim 1/b_T$
of the TMD evolution and satisfies \eq{chi1h_ope}
up to corrections in $\as(\mu_0)$.
To be specific, we apply a $\mu^*$ prescription~\cite{Lustermans:2019plv, Billis:2023xxx}
(also known as a ``local'' $b^*$ prescription)
starting at $\ord{b_T^4}$ to ensure that $\mu_0$ stays perturbative
without polluting nonperturbative corrections at $\ord{\lqcd^2 b_T^2}$~\cite{Ebert:2022cku},
\begin{align} \label{eq:mu_star_prescription}
\mu_0
= \Bigl( \frac{b_0^4}{b_T^4} + \mu_\mathrm{min}^4 \Bigr)^{1/4}
= \frac{b_0}{b_T} \Bigl[ 1 + \ord{\mu_\mathrm{min}^4 b_T^4} \Bigr]
\,,\end{align}
where $b_0 = 2 e^{-\gamma_E} \approx 1.12292$
and we take $\mu_\mathrm{min} = 1 \GeV$.
We take $\zeta_{0} \equiv m \rho_{0}$ to always be equal to its canonical value,
$\zeta_0 = (b_0/b_T)^2$.
We then use leading-logarithmic (LL) perturbative TMD evolution $U_q(\mu_0, \zeta_0, \mu, \zeta)$
to evolve \eq{chi_1h_model} to the overall scales
$\mu \sim \sqrt{\zeta} \sim Q$, with $Q$ the hard scattering energy.%
\footnote{We refer to \refcite{Ebert:2022cku} for the details of our evolution setup.}
This order is sufficient for the exploratory phenomenology we have in mind,
and in particular lets us use TMD evolution and $\beta$ functions
in QCD with $n_f = 5$ massless flavors at all scales
since the quark decoupling only induces next-to-leading logarithms of $b_T m$.
Specifically, we ignore the decoupling relations
and NNLL power-like secondary quark mass corrections to the Collins-Soper kernel
$\gamma_\zeta^q(b_T, \mu)$ that were determined in \refcite{Pietrulewicz:2017gxc}.
We also ignore nonperturbative contributions to the Collins-Soper kernel,
since they are orthogonal to the effects we are interested in here.
Overall, this results in the following expression
for the evolved unpolarized heavy-quark TMD FF,
\begin{align} \label{eq:tmd_ff_unpol_model}
\int_{z_\cut} \! \df z_H \, D_{1\,H/Q}(z_H, b_T, \mu, \zeta)
&= \chi_H \exp\Bigl(- \kappa_H^2 b_T^2 \Bigr) \, U_q(\mu_0, \zeta_0, \mu, \zeta)
\,, \\[0.4em]
U_q(\mu_0, \zeta_0, \mu, \zeta)
&= \exp \biggl[ \frac{1}{2} \gamma^q_\zeta(b_T, \mu_0) \ln \frac{\zeta}{\zeta_0} \biggr]
\exp \biggl[ \int_{\mu_0}^\mu \! \frac{\df \mu'}{\mu'} \, \gamma^q_\mu(\mu', \zeta) \biggr] \,
\nn \,,\end{align}
where for definiteness we considered the integral over $z_\cut \leq z_H \leq 1$.
To our working order, the right-hand side of \eq{tmd_ff_unpol_model}
is independent of $z_\cut$
as long as $1 - z_\cut \sim 1$ in order to satisfy \eq{away_from_endpoint},
and also holds for any truncated $z_H$ moment of the TMD FF.
Note that the single-parameter model in \eq{tmd_ff_unpol_model} is also accurate
at large $k_T \gtrsim m$, cf.\ \eq{tmd_ff_unpol_bhqet},
where it reduces to $\chi_H$ and thus is correct up to radiative corrections.

We assume a similar model for the Collins TMD fragmentation factor,
but have to account for the suppression at small $b_T$
by modifying the Gaussian, see \eq{chi1h_perp_ope},
\begin{align} \label{eq:chi_1h_perp_model}
\chi_{1,H}^\perp(b_T, \mu_0, {\rho_0})
= \chi_H \, \lambda_{H\perp} b_T \,
\exp\Bigl(- \kappa_{H\perp}^2 b_T^2 \Bigr)
\,,\end{align}
where we find it convenient
to express the overall effect strength
in terms of $\lambda_{H\perp} = \chi_{H,G}/\chi_H \sim \lqcd$,
i.e., relative to the total fragmentation probability $\chi_H$.
The parameter $\kappa_{H\perp} \sim \lqcd$ controls the relative impact of higher power corrections
and is in general distinct from $\kappa_H$ in \eq{chi_1h_model}.
Combining this with NLL $n_f = 5$ TMD evolution as above,
we find, for the evolved heavy-quark Collins function
in position space,
\begin{align} \label{eq:tmd_ff_collins_model}
b_T M_H \int_{z_\cut} \! \df z_H \, H_{1\,H/Q}^{\perp(1)}(z_H, b_T, \mu, \zeta)
=
\chi_H \, \lambda_{H\perp} b_T \,
\exp\Bigl(- \kappa_{H\perp}^2 b_T^2 \Bigr) \,
\, U_q(\mu_0, \zeta_0, \mu, \zeta)
\,.\end{align}

\begin{figure*}
\centering
\includegraphics[width=\WidthTwoSubfigs]{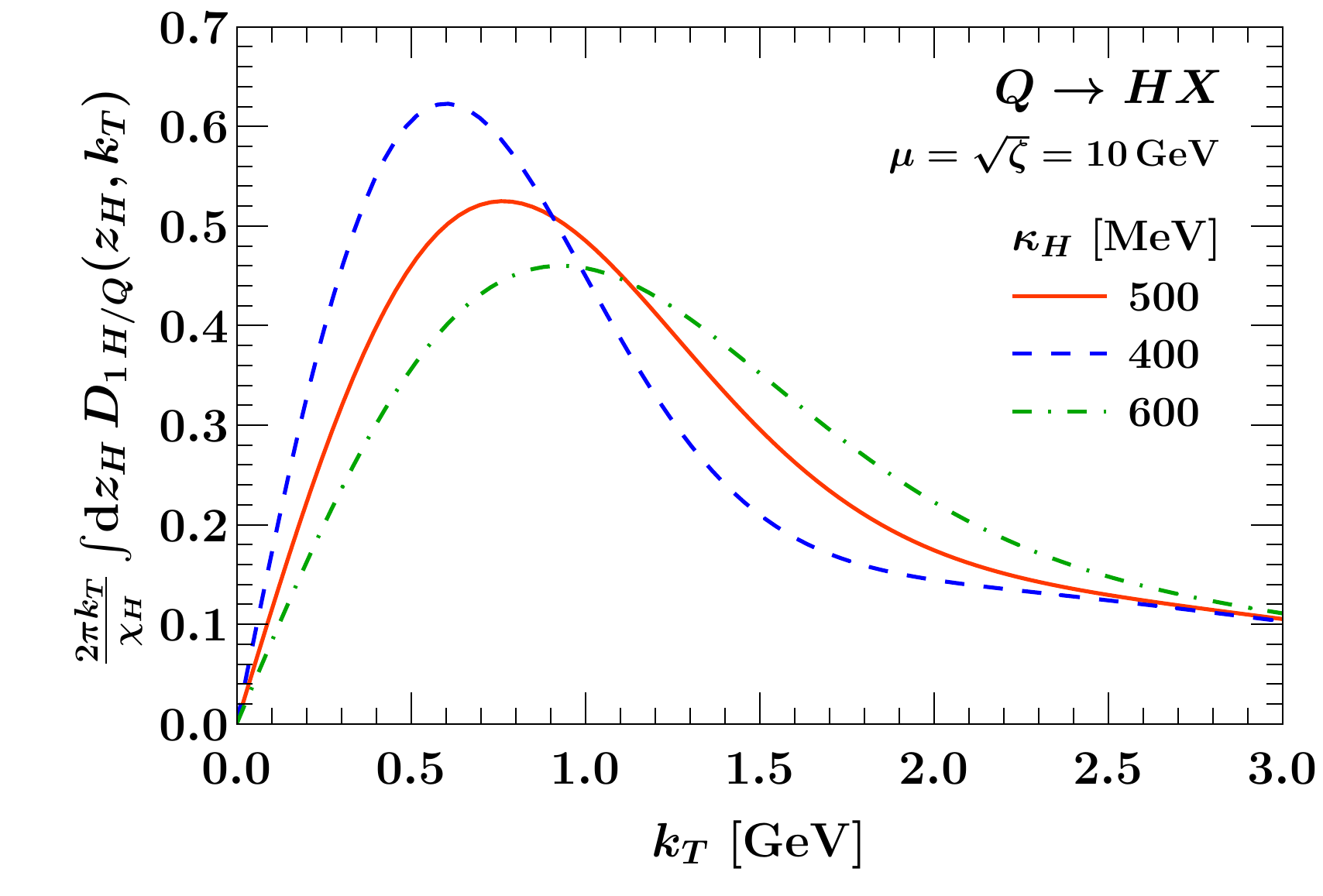}%
\hfill%
\includegraphics[width=\WidthTwoSubfigs]{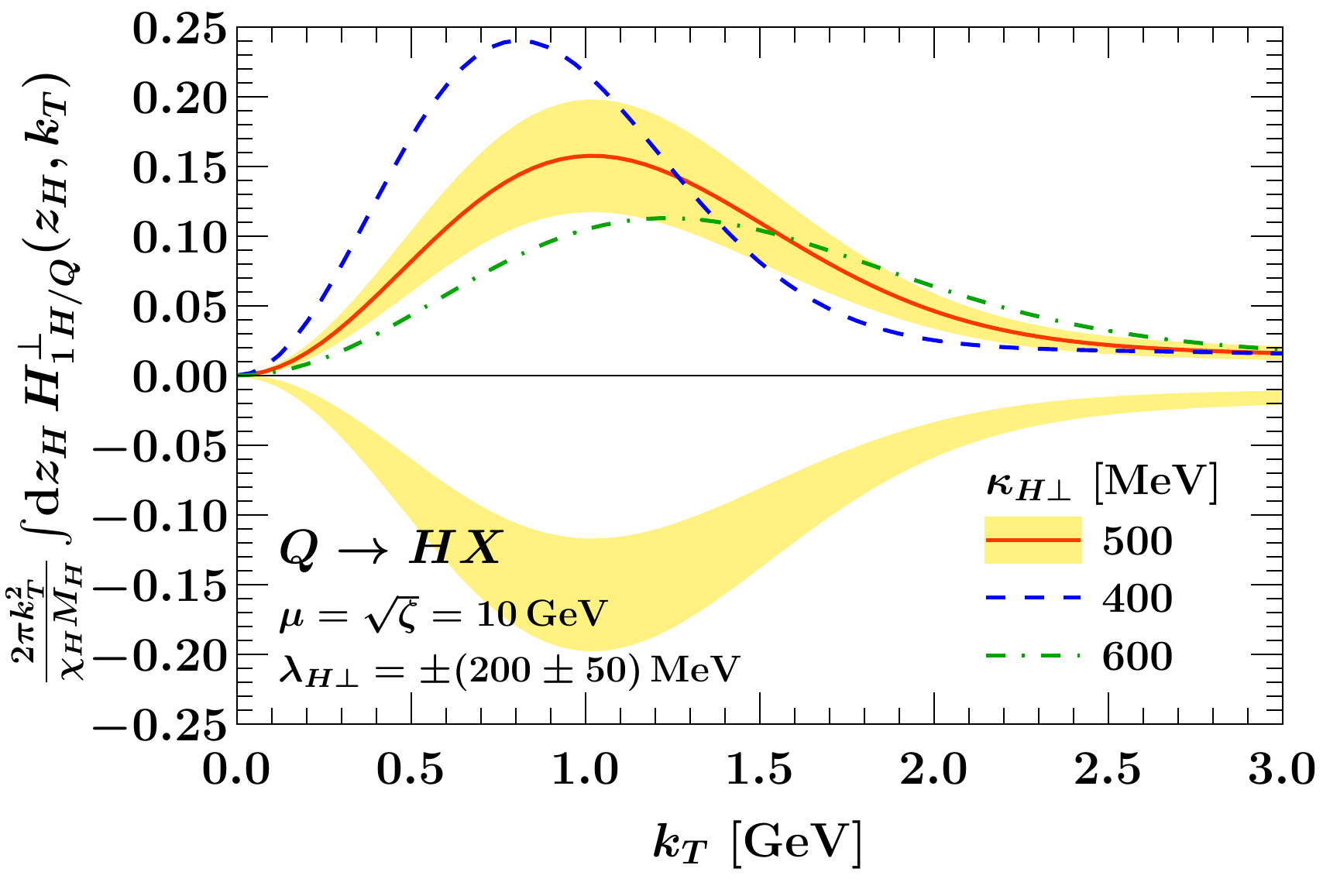}%
\caption{%
Unpolarized heavy-quark (left) and Collins TMD fragmentation functions (right)
as a function of $k_T$ and integrated over $z_H$.
All results are normalized to the total fragmentation probability $\chi_H$.
The yellow band in the case of the Collins function
corresponds to the indicated variations of the sign and magnitude of $\lambda_{H\perp}$.
These results are universal for charm and bottom quarks;
see the text for details.
}
\label{fig:tmd_ffs}
\end{figure*}

Taking appropriate Bessel integrals~\cite{Boer:2011xd},
we finally transition to momentum space,
\begin{align} \label{eq:bessel_integrals_ffs}
D_{1\,H/Q}(z_H, k_T, \mu, \zeta)
= \frac{1}{2\pi} \int_0^\infty \! \df b_T \, b_T \,  J_0(b_T k_T) \,
D_{1\,H/Q}(z_H, b_T, \mu, \zeta)
\,, \nn \\
\frac{k_T}{M_H} H_{1\,H/Q}^\perp(z_H, k_T, \mu, \zeta)
= \frac{M_H}{2\pi} \int_0^\infty \! \df b_T \, b_T^2 J_1(b_T k_T) \,
H_{1\,H/Q}^{\perp(1)}(z_H, b_T, \mu, \zeta)
\,.\end{align}
To evaluate the TMD evolution and the Bessel integrals,
we use the numerical implementation of TMD anomalous dimensions,
QCD renormalization-group solutions,
and double-exponential oscillatory integration in \texttt{SCETlib}~\cite{scetlib}.

Our results for the $z_H$-integrated heavy-quark TMD FFs
are shown as a function of $k_T$ for different values
of the model parameters in \fig{tmd_ffs}.
We use $\as(m_Z) = 0.118 \GeV$ as the input value for the strong coupling.
We note that due to heavy quark flavor symmetry,
the charm and bottom-quark TMD FFs are exactly equal at small $k_T \ll m$.
In other words, they only depend on the universal Gaussian parameters
$\kappa_{H}$ (for the unpolarized TMD FF), $\kappa_{H\perp}$ (for the Collins TMD FF),
and the Collins effect strength $\lambda_{H\perp}$.
At large $k_T \sim m$, the TMD FFs remain independent
of the heavy quark mass up to radiative corrections of $\ord{\as}$,
which we ignore at our LL working order.
These plots are thus identical for both flavors we consider.
We point out that the Collins function can in general take any sign,
as indicated by the yellow band scanning various values of the effect strength $\lambda_{H\perp}$.
The effect of varying the size of higher-power corrections ($\kappa_H$, $\kappa_{H\perp}$)
decreases as $k_T$ increases for both TMD FFs, as expected.

\section{Polarized heavy-quark TMD PDFs}
\label{sec:polarized_heavy_quark_tmd_pdfs}

\subsection{Calculational setup}
\label{sec:tmd_pdfs_setup}

In this section we consider the production of a heavy quark $Q$
with pole mass $m = m_c, m_b \gg \lqcd$
from light partons within a polarized nucleon $N$.
The nucleon has momentum
\begin{align}
P_N^\mu = P_N^- \frac{n^\mu}{2} + \frac{M_N^2}{P_N^-} \frac{\nbar^\mu}{2}
\,,\end{align}
with $P_N^- \gg P_N^+ = M_N^2/P_N^-$ in the rest frame of the hard scattering
that the heavy quark participates in.
Note that we again take the large component of the hadron (nucleon) momentum
to be along the $n^\mu$ direction to make this section self contained,
but the case of an $\nbar$-collinear incoming hadron
(as would be consistent with the $n$-collinear outgoing hadron
we considered in \sec{heavy_quark_tmd_ffs})
follows from $n^\mu \leftrightarrow \nbar^\mu$.

This time, we are interested in the transverse momentum $k_\perp$ of the heavy quark
with respect to the nucleon beam axis, which is again Fourier conjugate
to the transverse spacetime separation $b_\perp$ between quark fields.
The bare TMD quark-quark correlator between forward nucleon states
that describes this process is
\begin{align} \label{eq:def_tmd_pdf_correlator}
\Phi_{Q/N}^{\beta \beta '} (x, b_{\perp}) =
\int \frac{\df b^{+}}{4 \pi} e^{-\img b^{+} (x P_{N}^{-}) / 2}
\Mae{N}{\bar \psi^{\beta'}_{Q} (b) W(b) W^\dagger(0) \psi^{\beta }_{Q}(0)}{N}
\,,\end{align}
where $b \equiv (0, b^+, b_\perp)$, $W$ was defined in \eq{def_wilson_line},
$x$ is the lightcone momentum fraction carried by the heavy quark,
and we have suppressed the rapidity regulator, the soft factor,
and transverse gauge links at infinity for simplicity.
For the explicit perturbative calculations in this section,
it will also be useful to define the momentum-space version of the above correlator,
\begin{align} \label{eq:fourier_transform_tmd_pdf_correlator}
\Phi_{Q/N}^{\beta \beta'}(x, k_\perp)
= \int \! \frac{\df^2 b_\perp}{(2 \pi)^2} \, e^{-\img k_\perp \cdot \, b_\perp}
\Phi_{Q/N}^{\beta \beta '}(x, b_\perp)
\,.\end{align}
The spin decomposition of \eq{fourier_transform_tmd_pdf_correlator}
in terms of scalar TMD PDFs is well known~\cite{Bacchetta:2006tn, Ebert:2021jhy},
\begin{align} \label{eq:tmd_pdf_correlator_decomposition}
\Phi_{Q/N}(x > 0, k_\perp)
&= \Bigl\{
   f_{1\,Q/N}(x, k_T)
   + g_{1L\,Q/N}(x, k_T) \, S_{L} \gamma_{5}
   + h_{1L\,Q/N}^{\perp}(x, k_T) \, S_{L} \gamma_{5} \frac{\Sl k_\perp}{M_{N}}
\nn \\ & \qquad \qquad
   + \img h_{1\,Q/N}^{\perp}(x, k_T) \, \frac{\Sl k_\perp}{M_{N}}
   + \bigl( \text{terms $\propto S_\perp$} \bigr)
\Bigr\} \frac{\Sl n}{4}
\,,\end{align}
where $M_N$ is the nucleon mass,
$S_L$ is the longitudinal nucleon polarization in the Trento frame~\cite{Bacchetta:2004jz},
and in our convention $\Phi_{Q/N}(x < 0, k_\perp)$ decomposes
in the same way in terms of the antiquark TMD PDFs
$f_{1\,\bar Q/N}(\abs{x}, k_T)$, etc.
We have suppressed terms (``bad components'') that do not contribute
to leading-power TMD factorization theorems.
As we will see in the next section, the terms proportional
to the transverse nucleon polarization $S_\perp$
vanish for heavy quarks to all orders in the strong coupling
when matched onto the leading (twist-2) collinear PDFs.
We will also find that the twist-2 matching
for the Boer-Mulders function $h_{1}^{\perp}$ vanishes
at $\ord{\as}$.
The remaining TMD PDFs on the first line,
for which we will find nonzero results at $\ord{\as}$,
are the unpolarized TMD PDF $f_{1}$,
the helicity TMD PDF $g_{1L}$,
and the so-called worm-gear~$L$ function $h_{1L}^{\perp}$;
the latter will be of particular significance,
and encodes the production of a transversely polarized quark
from a linearly polarized nucleon.
For reference, the explicit Hankel transforms relating
scalar TMDs in $b_T$ and $k_T$ space read%
\footnote{
We continue to distinguish momentum and position-space functions by their argument,
see \ftn{same_symbol_momentum_position_space}.
For the meaning of the superscript $(1)$, see also there.
}
\begin{align} \label{eq:tmd_pdfs_hankel_transforms}
f_{1}(x, k_T) &= \int \! \frac{\df b_T}{2\pi} \, b_TJ_{0}(k_T b_T) \,
f_{1}(x, b_T)
\,, \nn \\
\frac{k_T}{M_{N}} h_{1}^{\perp} (x, k_T)
&= M_{N} \int \! \frac{\df b_T}{2\pi} \, b_T^{2} J_{1} (b_T k_T) \,
h_{1}^{\perp (1)} (x, b_T)
\,, \nn \\
g_{1L}(x, k_T) &= \int \! \frac{\df b_T}{2\pi} \, b_TJ_{0}(k_T b_T) \,
g_{1L}(x, b_T)
\,, \nn \\
\frac{k_T}{M_{N}} h_{1L}^{\perp} (x, k_T)
&= M_{N} \int \! \frac{\df b_T}{2\pi} \, b_T^{2} J_{1} (b_T k_T) \,
h_{1L}^{\perp (1)} (x, b_T)
\,.\end{align}

\subsection{Matching onto twist-2 collinear PDFs}
\label{sec:tmd_pdfs_matching}

Heavy-quark TMD PDFs are different from their TMD FF counterparts
because the heavy quark cannot be part of the
initial-state nucleon wave function at the scale $\mu \sim \lqcd$
at leading power in $\lqcd/m$,%
\footnote{
Power corrections of this kind,
which are known as ``intrinsic charm''
and have received substantial recent interest
on the collinear PDF side~\cite{Ball:2022qks, Guzzi:2022rca},
would be an interesting subject to explore in the TMD case in the future.
Very recently, the TMD PDFs for charm quarks within $\Lambda_c$ baryons,
which are leading valence contributions and do not have to be produced from gluons,
have been evaluated in a lightfront Hamiltonian model in \refcite{Zhu:2023nhl};
while these are phenomenologically inaccessible,
it would be interesting to analyze these valence dynamics
in the heavy-quark limit as we did for TMD FFs in \sec{heavy_quark_tmd_ffs}.
}
whereas in the fragmentation case the heavy quark is always part
of the final-state heavy hadron until its eventual weak decay.
This means that heavy quarks must be pair-produced
in initial-state gluon splittings at the scale $\mu \sim m$ instead.
In particular, this means there is at least one perturbative emission
with transverse momentum $\gtrsim m$ setting the scale of $k_T \gtrsim m$,
while the region of $k_T \ll m$ can only be populated
by several emissions with small net recoil,
which is a power-suppressed configuration.
In field theory terms, this means that heavy-quark TMD PDFs can be computed
by perturbatively matching them onto collinear twist-2 nucleon PDFs
in a theory with $n_\ell$ light flavors,
which are the only nonperturbative piece of information in this case.
The matching onto twist-2 collinear PDFs is well developed
for light quark and gluon TMDs, with notable results including
all unpolarized quark matching coefficients
through $\ord{\as^3}$~\cite{Luo:2019szz, Ebert:2020yqt}
and results for polarized TMDs through $\ord{\as^2}$~\cite{Bacchetta:2013pqa, Gutierrez-Reyes:2018iod},
and many of the following steps are standard, see e.g.~\cite{Collins:1350496}.
Likewise, the $\ord{\as}$ matching of the unpolarized heavy-quark TMD PDF
onto gluon collinear PDFs has been given in \refscite{Nadolsky:2002jr, Pietrulewicz:2017gxc}.
We nevertheless aim for a self-contained description,
giving us the opportunity to point out the ways in which (polarized)
heavy-quark TMD PDFs behave differently.

The bare light-quark and gluon twist-2 collinear correlators are defined as
\begin{align} \label{eq:def_collinear_pdf_correlators}
\Phi_{q/N}^{\alpha \alpha'}(x)
&= \int \frac{\df b^{+}}{4\pi} e^{-\img b^{+} (x P_{N}^{-}) / 2}
\Mae{N}{\bar \psi^{\alpha'}_{q}(b) W(b, 0) \psi^{\alpha }_{q}(0)}{N}
\,, \nn \\
\Phi_{g/N}^{\mu\nu} (x)
&= \int \frac{\df b^{+}}{4\pi} e^{-\img b^{+} (x P_{N}^{-}) / 2}
\Mae{N}{F^{-\mu}(b) W(b,0)  F^{-\nu}(0)}{N}
\,,\end{align}
where $b \equiv (0, b^+, 0)$ in this case
and $W(b,0)$ denotes a straight Wilson line segment.
The collinear correlators are conventionally decomposed as \cite{Collins:1350496}
\begin{align} \label{eq:collinear_pdf_correlators_decomposition}
\Phi_{q/N}(x > 0)
&= \Bigl\{
   f_{q/N}(x)
   + g_{q/N}(x) \, S_{L} \gamma_{5}
   + h_{q/N}(x) \, \gamma_5 \Sl S_\perp
\Bigr\} \frac{\Sl n}{4}
\,, \nn \\
\Phi_{g/N}^{\mu\nu} (x > 0)
& = -\frac{g_{\perp}^{\mu\nu}}{2} \, f_{g/N}(x)
+ \frac{\img \eps_{\perp}^{\mu\nu}}{2}  \, g_{g/N}(x) \, S_{L}
\,.\end{align}
in terms of the unpolarized (helicity) quark and gluon PDFs $f_{i/N}$ ($g_{i/N}$)
and the transversity quark PDF $h_{q/N}$.
The contribution $\propto S_\perp$ to the gluon correlator
(i.e., the transversity gluon PDF) vanishes identically
for spin-$0$ and spin-$1/2$ hadrons in the initial state
due to helicity conservation~\cite{Jaffe:1989xy}.

The matching relation between heavy-quark TMD PDFs
and twist-2 collinear PDFs holds at the operator level,
and constitutes the leading term in the OPE of the former.
Taking nucleon matrix elements of the bare operators,
the relation for general spin indices reads
\begin{align} \label{eq:tmd_pdf_matching_general_spin}
\Phi_{Q/N}^{\beta \beta'}(x, k_\perp)
&= \int \! \frac{\df p^-}{p^-} \,
   C_{Q/g , \mu \nu}^{\beta \beta'}(x P_N^-, p^-, k_\perp, m) \,
   \Phi_{g/N}^{\mu\nu}\Bigl(\frac{p^-}{P_N^-}\Bigr)
\nn \\ & \quad
   + \sum_q \int \! \frac{\df p^-}{p^-} \,
   C_{Q/q , \alpha \alpha'}^{\beta \beta'}(x P_N^-, p^-, k_\perp, m) \,
   \Phi_{q/N}^{\alpha \alpha'}\Bigl(\frac{p^-}{P_N^-}\Bigr)
\,,\end{align}
where $p^-$ is the lightcone momentum carried
by the light parton extracted from the collinear PDF
and the sum runs over the $n_\ell$ light quark flavors.
In pure dimensional regularization,
the bare matching coefficients are given by the partonic diagrams
\begin{align} \label{eq:tmd_pdf_matching_all_order_diagrams}
C_{Q/g , \mu \nu}^{\beta \beta'}(zp^-, p^-, k_\perp, m)
~= \raisebox{-0.45\height} {
   \includegraphics[width=0.25\textwidth]{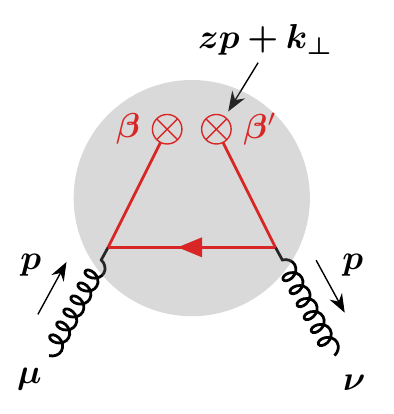}
}
\,, \nn \\
C_{Q/q , \alpha \alpha'}^{\beta \beta'}(zp^-, p^-, k_\perp, m)
~= \raisebox{-0.45\height} {
   \includegraphics[width=0.25\textwidth]{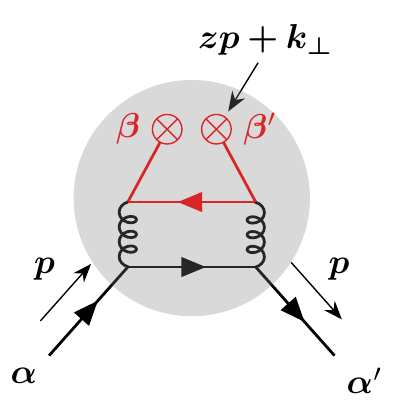}
}
\,,\end{align}
where $z = xP_N^-/p^-$ is the fraction of $p^-$ injected into the hard scattering process
and we have indicated the heavy quark lines in red.
The gray-shaded circles denote the sum of all possible QCD diagrams with these external legs,
including gluon attachments to the Wilson lines
that are part of the operators indicated by $\textcolor{red}{\otimes}$.
We have included the respective lowest-order diagram for illustration.
As is standard, matching relations between individual scalar TMD and collinear PDFs
follow by inserting \eq{collinear_pdf_correlators_decomposition}
into \eq{tmd_pdf_matching_general_spin}
and tracing the resulting Dirac bispinors $(\dots)^{\beta \beta'}$ against
the relevant Dirac structures.

Flavor conservation in QCD implies
that a single fermion line has to connect the external light-quark states
in \eq{tmd_pdf_matching_all_order_diagrams}.
It follows that contractions with the quark transversity PDF
involve an odd number of Dirac matrices on the light-quark line
and vanish to all orders,
i.e., flavor conservation and chirality for light quark flavors
imply that all terms $\propto S_\perp$ vanish
at twist-2 level in \eq{tmd_pdf_correlator_decomposition}.
This is distinct from e.g.\ the light-quark transversity TMD PDF,
which receives a tree-level contribution from the transversity collinear PDF of the same flavor.
As in the case of light-quark TMD PDFs, Lorentz covariance further implies
that only unpolarized (helicity) collinear PDFs
can contribute to the unpolarized and Boer-Mulders (helicity and worm-gear~$L$) TMD PDFs,
matching the dependence on $S_L$ in the spin decomposition.
These conclusions are not modified by the inclusion of the soft factor,
the rapidity renormalization,
and the UV renormalization of the TMD PDFs,
all of which are orthogonal to the spin structure.
They are likewise unaffected by the renormalization of the collinear PDFs,
which acts autonomously on the unpolarized and longitudinally polarized sectors.
Passing to renormalized objects, this altogether leaves us
with the following four nontrivial matching relations
for heavy-quark TMD PDFs onto collinear PDFs,
\begin{align} \label{eq:tmd_pdf_scalar_matching_relations_all_orders}
f_{1\,Q/N}(x, k_T, \mu, \zeta)
&= \sum_j \int \! \frac{\df z}{z} \,
C_{Q/j}(z, k_T, m, \mu, \zeta) \,
f_{j/N} \Bigl(\frac{x}{z}, \mu \Bigr)
\, , \nn \\
\frac{k_T}{M_N} \, h_{1\,Q/N}^\perp(x, k_T, \mu, \zeta)
&= \sum_j \int \! \frac{\df z}{z} \,
C_{Q_\perp/j}(z, k_T, m, \mu, \zeta) \,
f_{j/N} \Bigl(\frac{x}{z}, \mu \Bigr)
\, , \nn \\
g_{1L\,Q/N}(x, k_T, \mu, \zeta)
&= \sum_j \int \! \frac{\df z}{z} \,
C_{Q_\parallel/j_\parallel}(z, k_T, m, \mu, \zeta) \,
g_{j/N} \Bigl(\frac{x}{z}, \mu \Bigr)
\, , \nn \\
\frac{k_T}{M_N} \, h_{1L\,Q/N}^{\perp}(x, k_T, \mu, \zeta)
&= \sum_j \int \! \frac{\df z}{z} \,
C_{Q_\perp/j_\parallel}(z, k_T, m, \mu, \zeta) \,
g_{j/N}  \Bigl(\frac{x}{z}, \mu \Bigr)
\,.\end{align}
These relations are the key result of this section,
and hold up to power corrections in $\ord{\lqcd/k_T}$ and $\ord{\lqcd/m}$.
Here the subscripts $\lambda, \lambda' = \varnothing, \parallel, \perp$
on $C_{Q_\lambda/j_{\lambda'}}(z, k_T, \mu, \zeta)$
label the polarization of the heavy quark and the light parton $j$,
the sum runs over gluons and the $n_\ell$ flavors of light quarks and antiquarks,
and we have included a factor of $k_T/M_N$ on the left-hand side as needed
to ensure that the matching coefficient is independent of the hadronic state.
We have also changed integration variables from $p^-$ in \eq{tmd_pdf_scalar_matching_relations_all_orders}
to $z$, exploiting the fact that projections of the matching coefficients
onto good components can only depend on $z$ by reparameterization invariance.

Note that in a crucial difference to the light-quark case,
the heavy-quark worm-gear~$L$ TMD PDF,
which involves an odd number of Dirac matrices on the \emph{heavy-quark line}
in \eq{tmd_pdf_matching_all_order_diagrams}
is \emph{allowed} at twist-2 level because the quark mass breaks chirality.
The same is true for the Boer-Mulders function.
In both cases, the original argument of \refcite{Bacchetta:2008xw}
why the twist-2 matching for these functions vanishes to all orders in the light-quark case
critically relied on chirality.%
\footnote{An earlier version of this manuscript incorrectly
stated that the twist-2 matching for the heavy-quark Boer-Mulders function
should vanish to all orders based on its transformation behavior under time reversal,
which however only constrains the leading $\ord{\as}$ diagram
we consider in the next section.
We thank Markus Diehl for pointing this out to us.}
Conversely, the respective matching coefficients must vanish linearly as $m \to 0$
to afford the helicity flip,
\begin{align} \label{eq:worm_gear_l_small_mass_kT_space}
\lambda = \varnothing, \parallel \,: \quad
C_{Q_\perp/j_\lambda}(z, k_T, m, \mu, \zeta)
\propto \frac{m}{k_T^3} + \ord{m^2}
\,.\end{align}
Lastly, note that to all orders it is only the gluon PDF $f_g(x)$
and the quark singlet PDF $\sum_{i = q, \bar{q}} f_i(x)$
that contribute to the sum $f_{1\,Q/N} + f_{1\,\bar{Q}/N}$
due to the invariance of \eq{tmd_pdf_matching_all_order_diagrams}
under the $n_\ell$ light flavor symmetry,
and similarly for the two polarized cases.
The difference $f_{1\,Q/N} - f_{1\,\bar{Q}/N}$ of heavy quark and antiquark TMD PDFs
receives a nonzero contribution proportional
to $\sum_{i = q} f_i(x) - \sum_{i = \bar{q}} f_i(x)$
starting at $\ord{\as^3}$ due to the relative orientation of the color flow
along the fermion lines in \eq{tmd_pdf_matching_all_order_diagrams},
as in the light-quark case~\cite{Luo:2019szz, Ebert:2020qef, Luo:2020epw}.

Inverting the Hankel transforms in \eq{tmd_pdfs_hankel_transforms},
we find the $b_T$-space matching relations
\begin{align} \label{eq:tmd_pdf_scalar_matching_relations_bT}
f_{1\,Q/N}(x, b_T, \mu, \zeta)
&= \sum_j \int \! \frac{\df z}{z} \,
C_{Q/j}(z, b_T, m, \mu, \zeta) \,
f_{j/N} \Bigl(\frac{x}{z}, \mu \Bigr)
\, , \nn \\
b_T M_N \, h_{1\,Q/N}^{\perp (1)}(x, b_T, \mu, \zeta)
&= \sum_j \int \! \frac{\df z}{z} \,
C_{Q_\perp/j}(z, b_T, m, \mu, \zeta) \,
f_{j/N}  \Bigl(\frac{x}{z}, \mu \Bigr)
\, , \nn \\
g_{1L\,Q/N}(x, b_T, \mu, \zeta)
&= \sum_j \int \! \frac{\df z}{z} \,
C_{Q_\parallel/j_\parallel}(z, b_T, m, \mu, \zeta) \,
g_{j/N} \Bigl(\frac{x}{z}, \mu \Bigr)
\, , \nn \\
b_T M_N \, h_{1L\,Q/N}^{\perp (1)}(x, b_T, \mu, \zeta)
&= \sum_j \int \! \frac{\df z}{z} \,
C_{Q_\perp/j_\parallel}(z, b_T, m, \mu, \zeta) \,
g_{j/N}  \Bigl(\frac{x}{z}, \mu \Bigr)
\,,\end{align}
where the matching coefficients are given by
($n = 1$ for $\lambda = \perp$ and $n = 0$ otherwise)
\begin{align} \label{eq:matching_coeffs_hankel_transforms}
C_{Q_\lambda/j_{\lambda'}}(z, b_T, m, \mu, \zeta)
&= 2 \pi \int \! \df k_T \, k_T J_n(k_T b_T) \,
C_{Q_\lambda/j_{\lambda'}}(z, k_T, m, \mu, \zeta)
\,.\end{align}
For the dimensionless $b_T$-space matching coefficients,
\eq{worm_gear_l_small_mass_kT_space} simply reads
\begin{align} \label{eq:worm_gear_l_small_mass_bT_space}
\lambda = \varnothing, \parallel \,: \quad
C_{Q_\perp/j_\lambda}(z, b_T, m, \mu, \zeta)
\propto m b_T + \ord{m^2}
\,.\end{align}

\subsection{One-loop evaluation of matching coefficients}
\label{sec:tmd_pdfs_one_loop}

At $\ord{\alpha_s}$, only the gluon diagram
in \eq{tmd_pdf_matching_all_order_diagrams} is nonzero.
Using standard QCD Feynman rules, we find the leading-order result
\begin{align}\label{eq:matching_coeff}
&C_{Q/g\,\mu \nu}^{\beta \beta '}(z p^-, p^-, k_\perp, m) ~=
\raisebox{-0.45\height}{
   \includegraphics[scale=0.9]{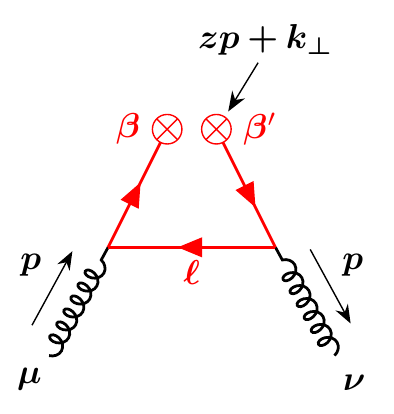}
}
\\[-0.4em]
&=-\frac{\img g^{2}}{2} \int \!\! \frac{\df ^4 \ell}{ (2\pi)^{4}} \,
\delta(zp^- - p^- - \ell^-) \, \delta^{(2)}(k_\perp - \ell_\perp)
\frac{ [ \, (\Sl p + \Sl \ell + m) \gamma_{\mu} (\Sl \ell + m)
	\gamma_{\nu} (\Sl p + \Sl \ell + m)]^{\beta \beta '}}
{[(p+\ell)^{2} - m^{2} + \img 0]^2(\ell^{2} - m^{2} + \img 0)}
\nn \,,\end{align}
where $p = (p^-, 0, 0)$ is the momentum of the external gluon
and $\ell$ is defined as indicated (in the direction of fermion flow).
The $\ell^+$ integral is straightforward
to evaluate by contours, which amounts to setting
an emitted antiquark on shell, $\ell^2 = m^2$.
Note that the diagram is finite in four dimensions and without a rapidity regulator
because the quark mass cuts off infrared singularities.
This is expected, as the UV and rapidity renormalization
only become nontrivial at the next order.

Dotting \eq{matching_coeff} into the gluon collinear PDF correlator
in \eq{collinear_pdf_correlators_decomposition}
and projecting onto quark spin structures,
we find individual momentum-space matching coefficients
\begin{align}
C_{Q_\lambda/g_{\lambda'}}(z, k_T, m, \mu, \zeta)
= \frac{\as(\mu)}{4\pi} \, C^{(1)}_{Q_\lambda/g_{\lambda'}}(z, k_T, m) + \ord{\as^2}
\,,\end{align}
with leading-order coefficient functions given by
\begin{align} \label{eq:matching_momentum}
C^{(1)}_{Q/g}(z, k_T, m)
&= T_F \, \Theta(z) \Theta(1-z) \, \frac{2}{\pi} \, \frac{k_{T}^{2}(1-2z + 2z^{2}) + m^{2} }{(k_T^2 + m^2)^2}
\,, \nn \\
C^{(1)}_{Q_\perp/g}(z, k_T, m)
&= 0
\,, \nn \\
C^{(1)}_{Q_\parallel/g_\parallel}(z, k_T, m)
&= T_F \, \Theta(z) \Theta(1-z) \,  \frac{2}{\pi} \, \frac{k_{T}^{2}(2z-1) + m^{2} }{(k_T^2 + m^2)^2}
\,, \nn \\
C^{(1)}_{Q_\perp/g_\parallel}(z, k_T, m)
&= T_F  \, \Theta(z) \Theta(1-z) \,  \frac{4}{\pi} \frac{m k_{T} (z-1)}{(k_T^2 + m^2)^2}
\,.\end{align}
As a nontrivial check, we have confirmed
that using massive SCET Feynman rules~\cite{Leibovich:2003jd}
results in the same expressions after performing the spin traces and integrating over the loop momentum.
Note that the projection of the $\ord{\as}$ twist-2 matching diagram
onto the Boer-Mulders function remains zero even for finite quark masses.
This is expected because the Boer-Mulders function
is odd under time reversal~\cite{Collins:2002kn},
i.e., it changes sign depending
on whether the Wilson lines in the operator point to the future (SIDIS)
or the past (Drell-Yan).
The diagram in \eq{matching_coeff} does not yet feature
gluon attachments to the Wilson lines that could resolve their direction,
and thus its projection onto the Boer-Mulders function has to vanish.
Starting at $\ord{\as^2}$, the matching coefficient can in general
receive nonzero contributions from the absorptive part of real-virtual diagrams
because chirality is broken by the quark mass,
and it would be interesting to investigate these contributions further.

Evaluating the inverse Hankel transforms in \eq{matching_coeffs_hankel_transforms},
we find the position space matching coefficients
\begin{align}
C_{Q_\lambda/g_{\lambda'}}(z, b_T, m, \mu, \zeta)
= \frac{\as(\mu)}{4\pi} \, C^{(1)}_{Q_\lambda/g_{\lambda'}}(z, b_T m) + \ord{\as^2}
\,,\end{align}
which at this order only depend on the dimensionless combination $b_T m$ and are given by
\begin{align} \label{eq:matching_position}
C^{(1)}_{Q/g}(z, b_T m)
&= T_F \, \Theta(z) \Theta(1-z) \, 4 \Bigl[ ( 1-2z + 2z^{2} ) \, K_{0}(b_Tm) + z (1-z)\, b_Tm \, K_{1}(b_Tm) \Bigr]
\,, \nn \\
C^{(1)}_{Q_\perp/g}(z, b_T m)
&= 0
\,, \nn \\
C^{(1)}_{Q_\parallel/g_\parallel}(z, b_T m)
&= T_F \, \Theta(z) \Theta(1-z) \, 4  \Bigl[ (2z -1) \, K_{0}(b_Tm) + (1-z) \, b_Tm \, K_{1}(b_Tm) \Bigr]
\,, \nn \\
C^{(1)}_{Q_\perp/g_\parallel}(z, b_T m)
&= T_F \, \Theta(z) \Theta(1-z) \,4 (z-1) \, b_T m \, K_{0}(b_T m)
\, ,\end{align}
where $K_0$ and $K_1$ are modified Bessel functions of the second kind.
These are the main analytic results of this section.
The unpolarized matching coefficient $C^{(1)}_{Q/g}$
has been computed long ago~\cite{Nadolsky:2002jr},
and we agree with the $b_T$-space expression given in that reference
as well as with the $k_T$-space result in \refcite{Pietrulewicz:2017gxc}.
The results for the polarization-dependent matching coefficients are new.

\subsection{Consistency with the light-quark limit}
\label{sec:tmd_pdfs_consistency}

For $\lqcd \ll m \ll k_T$, heavy-quark TMD PDFs can be determined
using a two-step matching~\cite{Pietrulewicz:2017gxc}.
First, the TMD operators at the scale $\mu \sim k_T$ are matched
onto collinear PDFs in a theory with $n_\ell + 1$ massless quark flavors,
which results in the standard massless TMD matching coefficients.
In a second step, the $n_\ell + 1$-flavor PDFs
are matched onto those in a theory with $n_\ell$ flavors at the scale $\mu \sim m$.
At fixed order, this implies the following consistency relation
for the unpolarized and linearly polarized massive TMD matching coefficients,
\begin{align} \label{eq:tmd_pdfs_consistency}
C_{Q / k}(z, b_T, m, \mu, \zeta)
&= \sum_{j}\int \! \frac{\df z'}{z'} \,
C_{Q / j}(z', b_T, \mu, \zeta) \, \mathcal{M}_{j / k} \Bigl(\frac{z}{z'}, m, \mu\Bigr)
+ \ord{m^2 b_T^2}
\,,  \\
C_{Q_\parallel / k_\parallel}(z, b_T, m, \mu, \zeta)
&= \sum_{j}\int \! \frac{\df z'}{z'}  \,
C_{Q_\parallel / j_\parallel}(z', b_T, \mu, \zeta) \,
\mathcal{M}_{j_\parallel /k_\parallel}\Bigl(\frac{z}{z'}, m, \mu\Bigr)
+ \ord{m^2 b_T^2}
\,, \nn \end{align}
where $\mathcal{M}_{j_\lambda /k_\lambda}$ denotes the PDF matching function,
the sum runs over all light degrees of freedom,
and the subscript $\lambda = \varnothing, \parallel$ again
labels the polarization of the heavy quark and the light partons $j$ and $k$.
Perturbatively expanding the matching functions as
\begin{align}
C_{i_\lambda/ j_\lambda}(z', b_T, \mu, \zeta)
= \delta_{ij} \, \delta(1-z)
+ \sum_{n=1}^{\infty} \Bigl(\frac{\alpha_s (\mu)}{4 \pi} \Bigr)^n
C_{i_\lambda/ j_\lambda}^{(n)}(z, b_T, \mu, \zeta)
\,, \nn \\
\mathcal{M}_{j_\lambda /k_\lambda}(z, m, \mu)
= \delta_{jk} \, \delta(1-z) + \sum_{n=1}^{\infty} \Bigl(\frac{\alpha_s (\mu)}{4 \pi} \Bigr)^n
\mathcal{M}_{j_\lambda /k_\lambda}^{(n)}(z, m,\mu)
\,,\end{align}
these relations simplify
for our dimensionless $\ord{\as}$ coefficient functions in $b_T$ space,
\begin{align} \label{eq:tmd_pdfs_consistency_one_loop}
C^{(1)}_{Q_\lambda/ g_\lambda}(z, b_T m)
= C_{q_\lambda/ g_\lambda}^{(1)}(z, b_T, \mu)
+ \mathcal{M}_{q_\lambda / g_\lambda}^{(1)}(z, m, \mu)
\,,\end{align}
where the $\mu$ dependence has to cancel within the matching coefficient.
For the unpolarized case, this relation has previously
been verified in \refscite{Nadolsky:2002jr, Pietrulewicz:2017gxc}.
At NLO, the polarized PDF matching function relevant
for our case is given by~\cite{Diehl:2022dia}
\begin{align}
\mathcal{M}_{Q_\parallel/ g_\parallel} ^{(1)}(z , m ,\mu) =
T_F (2z -1)\ln \frac{\mu^2}{m^2}
\,.\end{align}
The massless matching coefficient for the quark helicity TMD PDF
onto the collinear gluon helicity PDF was calculated
in \refcite{Bacchetta:2013pqa},
\begin{align}
C^{(1)}_{q_\parallel / g_\parallel}(z, b_T, \mu, \zeta)
= 4T_F \Bigl[ (2z-1) \ln  \frac{2 e ^{-\gamma_E}}{\mu b_T}  + (1-z)\Bigr]
\,.\end{align}
Using $K_0(x) = -\ln \frac{2e^{-\gamma_E}}{x} + \ord{x}$, it is straightforward
to see that our result in \eq{matching_position} indeed satisfies
\eq{tmd_pdfs_consistency_one_loop}.

By contrast, the worm-gear $L$ matching coefficient
is suppressed by one power of the mass, see \eq{worm_gear_l_small_mass_bT_space},
and therefore cannot be reproduced by a leading-power PDF matching
at the scale $\mu \sim m$.
Interestingly, it contains a logarithm of $m b_T$ at subleading power,
\begin{align} \label{eq:c1_qperp_gparallel_expanded}
C^{(1)}_{Q_\perp / g_\parallel}(z, b_T m)
= 4 T_F \, b_T m (z-1)
\ln \frac{2 e ^{-\gamma_E}}{m b_T} + \ord{m^3 b_T^3}
\,.\end{align}
Unlike the leading-power logarithms
in $C^{(1)}_{Q/ g}$ and $C^{(1)}_{Q_\parallel / g_\parallel}$,
this logarithm cannot be resummed by the evolution of $n_\ell+1$-flavor PDFs
between $\mu \sim m$ and $\mu \sim 1/b_T$.
Subleading-power mass logarithms in amplitudes that require a helicity flip
have received much attention in the context of Higgs boson production
through a bottom-quark loop, see e.g.\ \refscite{Liu:2019oav, Liu:2020tzd, Liu:2020wbn, Liu:2021chn, Liu:2022ajh},
and it would be interesting to understand whether the logarithm
in \eq{c1_qperp_gparallel_expanded} might be amenable to similar techniques.

\subsection{Numerical results for TMD PDFs}

For numerics, we evaluate \eq{tmd_pdf_scalar_matching_relations_bT}
at the boundary scales $\mu_0 \sim \sqrt{\zeta_0} \sim 1/b_T$
given in and below \eq{mu_star_prescription}, perform the TMD evolution back to $\mu = \sqrt{\zeta} = Q$
as described around \eq{tmd_ff_unpol_model},
and finally take a numerical Fourier transform as in \eq{tmd_pdfs_hankel_transforms}.
E.g., we have
\begin{align}
f_1(x, b_T, \mu, \zeta)
&= U_q(\mu_0, \zeta_0, \mu, \zeta) \,
\frac{\as(\mu_0)}{4\pi} \int \! \frac{\df z}{z} \, C^{(1)}_{Q/g}(z, b_T m) \,
f_{g} \Bigl(\frac{x}{z}, \mu_0 \Bigr)
\end{align}
for the evolved unpolarized heavy-quark TMD PDF,
and similarly for the other cases.
For the input collinear gluon PDFs
we use the \texttt{NNPDF31\_nnlo\_as\_0118} unpolarized proton PDF set~\cite{NNPDF:2017mvq}
together with the \texttt{NNPDFpol11\_100} set for the polarized case~\cite{Nocera:2014gqa}.
Our input values for the strong coupling and the quark pole masses
were given in \sec{tmd_ffs_models}.

\begin{figure*}
\centering
\includegraphics[width=\WidthTwoSubfigs]{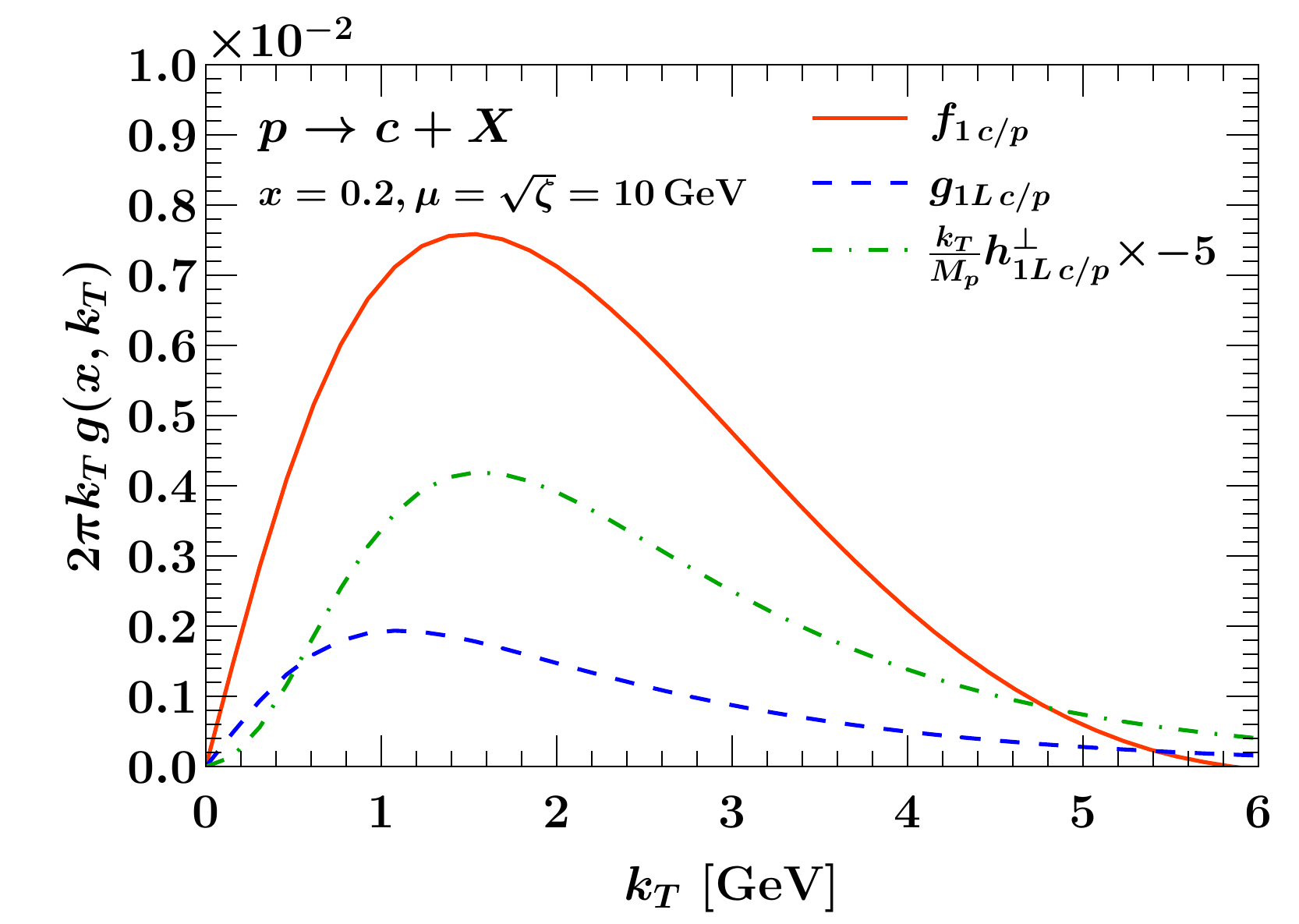}%
\hfill%
\includegraphics[width=\WidthTwoSubfigs]{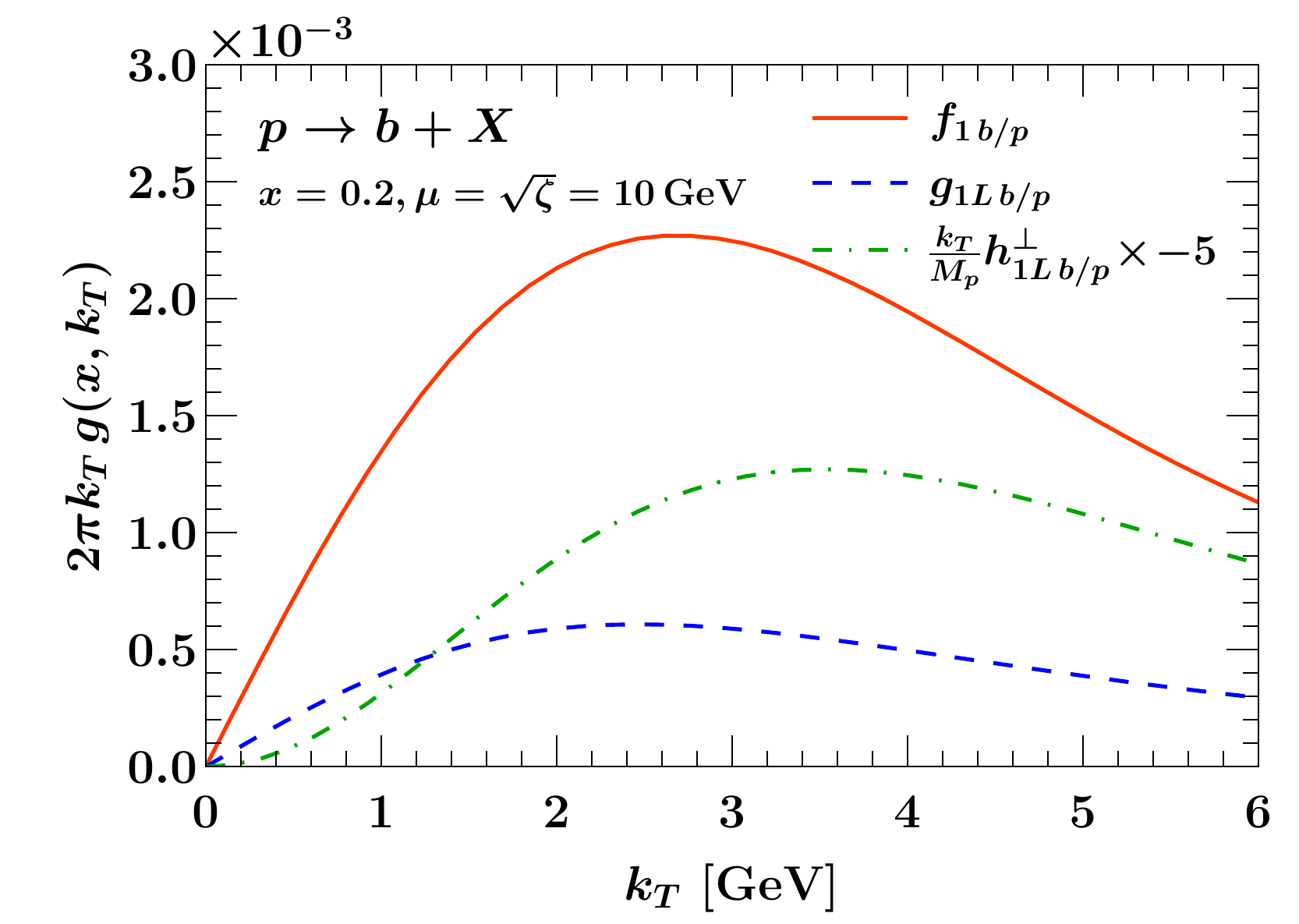}%
\\
\includegraphics[width=\WidthTwoSubfigs]{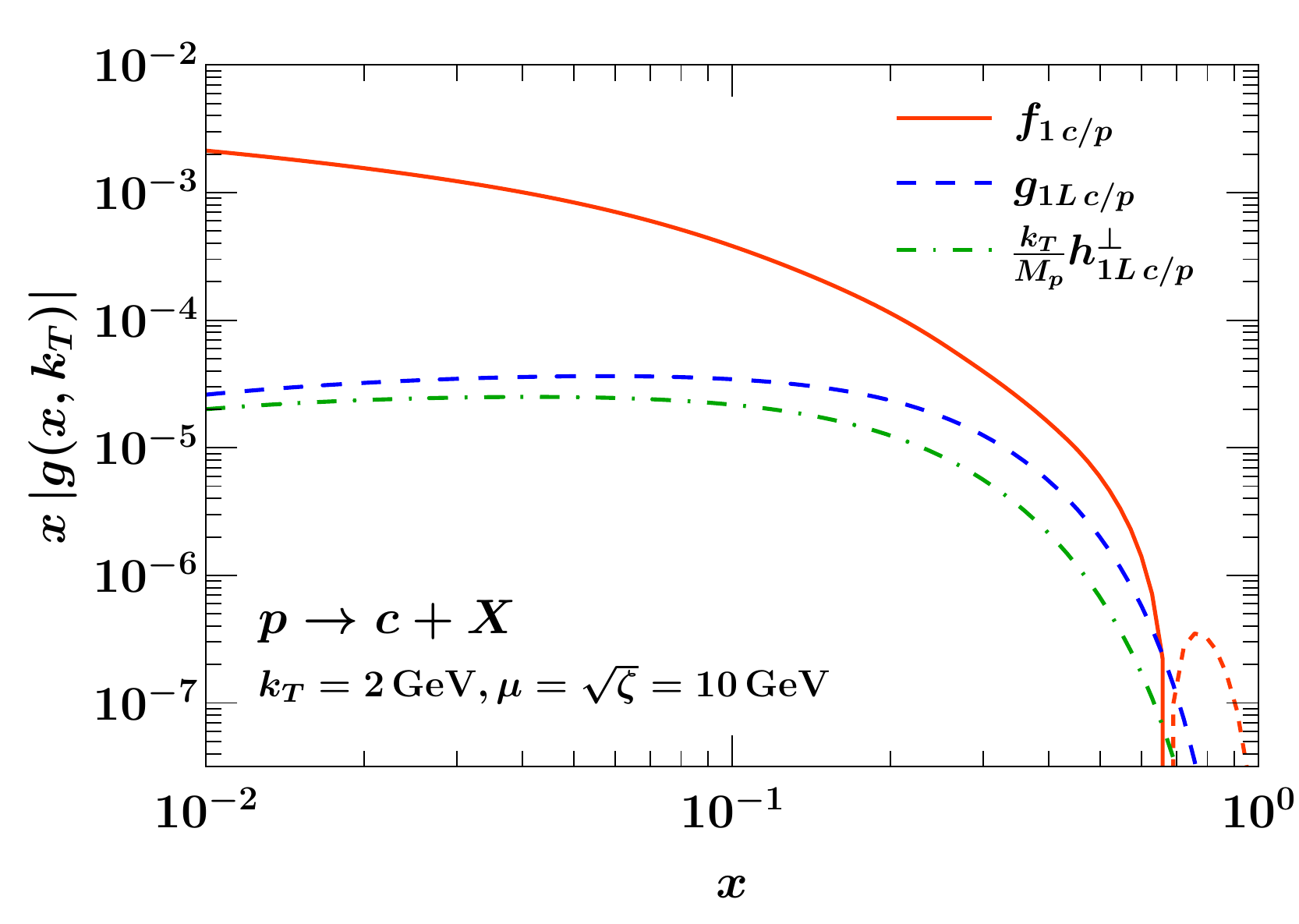}%
\hfill%
\includegraphics[width=\WidthTwoSubfigs]{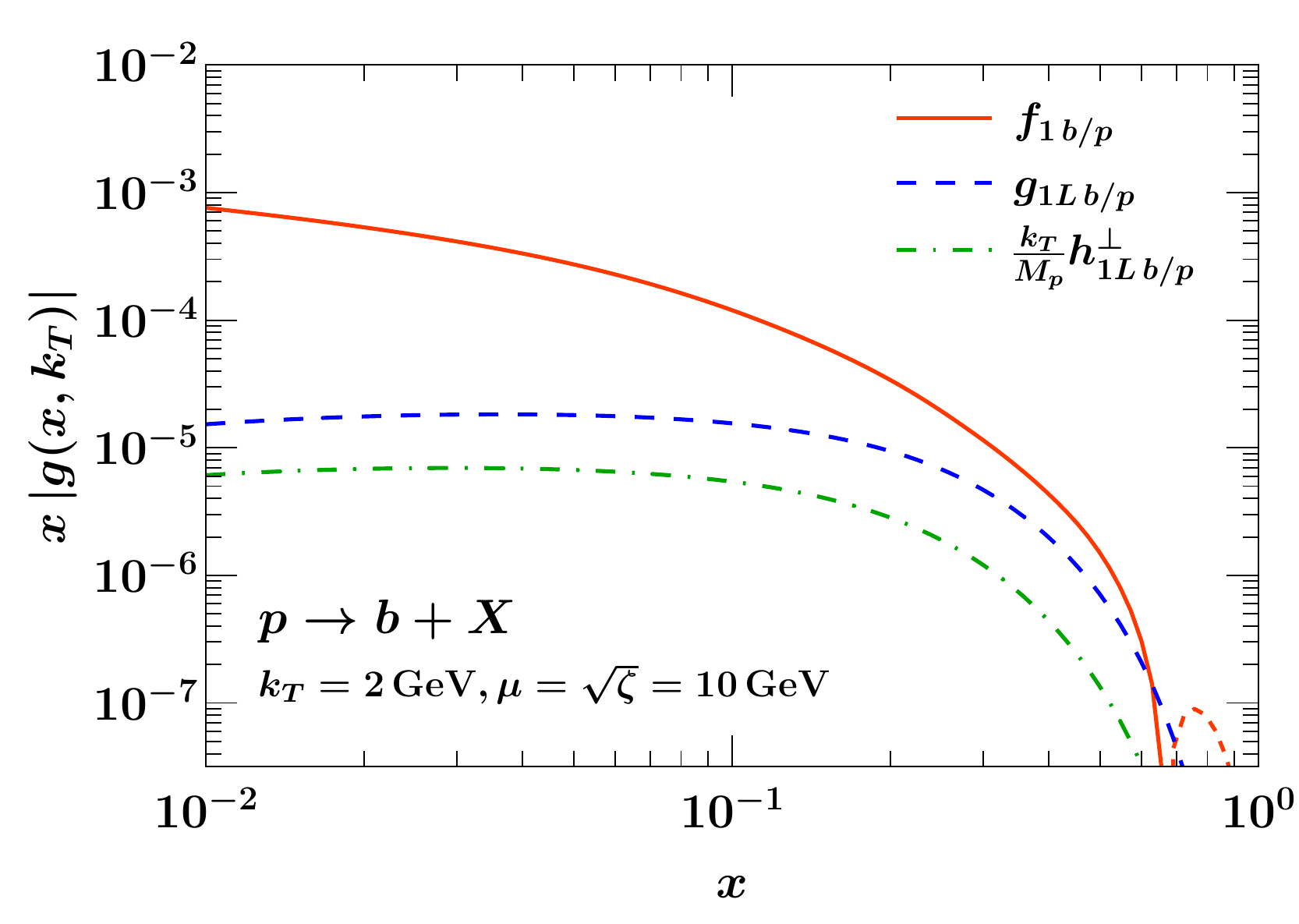}%
\caption{%
TMD PDFs for producing a charm (left) or bottom quark (right)
from gluons within a longitudinally polarized proton
as a function of $k_T$ at fixed $x$ (top) or vice versa (bottom).
Dashed red lines in the bottom two panels indicate negative sign.
}
\label{fig:tmd_pdfs}
\end{figure*}

In \fig{tmd_pdfs}, we show our numerical results for the heavy quark TMD PDFs
for producing a charm or bottom quark from a longitudinally polarized proton
as a function of $k_{T}$ and $x$, respectively.
The bottom quark TMD PDFs have a wider peak in $k_{T}$
compared to the charm because of its larger mass,
as can be understood from the fact that the expressions in \eq{matching_position}
only depend on $m b_T$ up to RG effects.
Note also that the worm-gear $L$ function
(after including a Jacobian $2 \pi k_{T}$)
is quadratic in the small $k_T$ region with a coefficient proportional to $1/m^3$,
whereas the unpolarized and helicity TMD PDFs are linear in $k_T$.
As this approximation is valid to higher $k_T$ in the case of the bottom quark
than that of the charm, the bottom-quark TMD PDF
has a numerically smaller value over a wide range.
As $x$ decreases, the unpolarized heavy-quark TMD PDF
rises much more rapidly than the polarized ones,
as expected from the smaller gluon polarization fraction at smaller $x$.
We point out that the unpolarized TMD PDF changes sign
at very high $x \geq 0.6$, indicating a need
for resumming subleading-power threshold logarithms of $1-x$
using e.g.\ the tools of \refscite{Beneke:2020ibj}.

\section{Towards phenomenology with heavy-quark TMDs}
\label{sec:pheno}

\subsection{Accessing heavy-quark TMDs in \texorpdfstring{$e^+ e^-$}{ee} collisions}
\label{app:tmds_ee_hadrons}

In $e^+ e^-$ collisions, TMD fragmentation functions may be accessed
from double-inclusive measurements with two identified hadrons,
$e^+ e^- \to H_a H_b X$.
For instance,
the six-fold differential cross section for this process in the TMD limit
$P_{a,T}, M_{a,b} \ll Q$ is given by \cite{Boer:1997mf, Metz:2016swz}
\begin{align} \label{eq:tmd_factorization_ee_hadrons}
\frac{\df \sigma_{e^+e^- \to H_a H_b X}}{\df \cos \theta \, \df \phi \, \df z_a \, \df z_b \, \df^2 \vec{P}_{a,T}}
&= \frac{3 \aem^2}{Q^2} \Bigl[
   \Bigl( \frac{1}{2} - y + y^2 \Bigr) W_\mathrm{incl}(Q^2, z_a, z_b, P_{a,T}/z_a)
\nn \\ & \quad
+ y (1-y) \cos(2 \phi_0) \, W_{\cos(2 \phi_0)}(Q^2, z_a, z_b, P_{a,T}/z_a)
\Bigr]
\nn \\ & \quad
+ (\text{odd under $y \leftrightarrow 1-y$})
\,,\end{align}
where $\cos \theta$ and $\phi$ are the spherical coordinates of hadron $H_b$
with respect to the incoming beams in the center-of-mass frame,
$z_a$ and $z_b$ are the lightcone momentum fractions of the two hadrons,
and $\vec{P}_{a,T}$ is the transverse momentum of hadron $H_a$.
On the right-hand side, $\aem$ is the fine-structure constant,
$Q$ is the center-of-mass energy of the collision,
$y = (1 + \cos \theta)/2$,
and $\phi_0$ is the azimuthal angle of $\vec P_{a,T}$
measured relative to the plane spanned by $H_b$ and the beams.
The hadronic structure functions factorize into TMD FFs,
\begin{align} \label{eq:structure_functions_ee_hadrons}
W_\mathrm{incl}(Q^2, z_a, z_b, q_T)
&= \cF_{ee} \Bigl[
   \cH \, D_{1} \, D_{1}
\Bigr]
\,, \nn \\
W_{\cos(2 \phi_0)}(Q^2, z_a, z_b, q_T)
&= \cF_{ee} \Bigl[
   \cH \, H_{1}^{\perp(1)} \, H_{1}^{\perp(1)}
\Bigr]
\,,\end{align}
where $\cF_{ee}$ denotes a weighted sum over flavors
and a convolution of two TMD FFs (i.e., a product in $b_T$ space)
at total partonic transverse momentum $q_T = P_{a,T}/z_a$,
\begin{align} \label{eq:def_f_ee_hadrons}
\cF_{ee}\Bigl[ \cH \, {D}^{(n)} \, {D}^{(m)} \Bigr]
&= z_a^2 z_b^2 \int_0^\infty \! \frac{\df b_T\,b_T}{2\pi}
(M_{a} b_T)^n (M_{b} b_T)^m
J_{n+m}(b_T q_T)
\nn \\ & \quad \times
\sum_{i, j} \cH_{ee \to ij}(Q^2, \mu) \,
{D}_{H_a/i}(z_a, b_T, \mu, Q^2) \,
{D}_{H_b/j}(z_b, b_T, \mu, Q^2)
\,,\end{align}
and the hard function describing the pair production of quarks is given by
\begin{align} \label{eq:hard_ee_hadrons_tree}
\cH_{ee \to ij}(Q^2, \mu)
&= \delta_{i \bar{j}} \,
\Bigl\{
   e_i^2
   - 2 v_e v_i e_i \Re \bigl[ P_Z(Q^2) \bigr]
   + (v_e^2 + a_e^2)(v_i^2 + a_i^2) \Abs{P_Z(Q^2)}^2
\Bigr\}
\,.\end{align}
Here we have kept the contribution from $Z$ boson exchange
and $Z$-photon interference, as relevant for measurements on the $Z$ pole,
where $P_Z(Q^2) = Q^2/(Q^2 - m_Z^2 + \img \Gamma_Z m_Z)$ is the reduced $Z$ propagator
and $e_f$ ($v_f$, $a_f$) are the electromagnetic charge (vector, axial coupling to the $Z$)
of a fermion $f$.
We may assume that the experimental measurement involves an integral over symmetric ranges in $\cos \theta$
such that the forward-backward asymmetry and an associated odd Collins effect
in \eq{tmd_factorization_ee_hadrons} drop out.

Crucially, the TMD factorization theorems
in \eqs{tmd_factorization_ee_hadrons}{structure_functions_ee_hadrons}
only assume that the hard scale $Q \sim z_a Q \sim z_b Q$
is large compared to all other scales,
i.e., all masses and transverse momenta,
and therefore hold for \emph{both} light-quark and heavy-quark fragmentation
at $z_{a,b} \sim 1$ without modification.
In particular, the heavy quarks are approximately massless at the scale $\mu \sim Q$
at which they are produced,
and their polarization states are thus fully entangled.
The hard function in \eq{hard_ee_hadrons_tree} could be modified to account for the effect
of perturbative spin flips,
but this amounts to retaining power corrections
in $m/Q$ further suppressed by powers of $\as$.
Importantly, this means that a characteristic $\cos(2\phi_0)$ modulation
(the Collins effect) is present both for light and for heavy quarks
at leading power and at tree level.
As is commonly done for light quarks, the Collins effect strength
\begin{align}
R_{\cos(2\phi_0)}(Q^2, q_T)
\equiv \frac{
   \int \! \df z_a \, \int \! \df z_b \,
   W_{\cos(2 \phi_0)}(Q^2, z_a, z_b, q_T)
}{
   \int \! \df z_a \, \int \! \df z_b \,
   W_\mathrm{incl}(Q^2, z_a, z_b, q_T)
}
\end{align}
can be accessed by taking suitable ratios of
weighted cross sections, which we here take to be integrated over $z_a$ and $z_b$
as likely relevant for an initial study of the heavy-quark Collins effect.

\begin{figure*}
\centering
\includegraphics[width=\WidthTwoSubfigs]{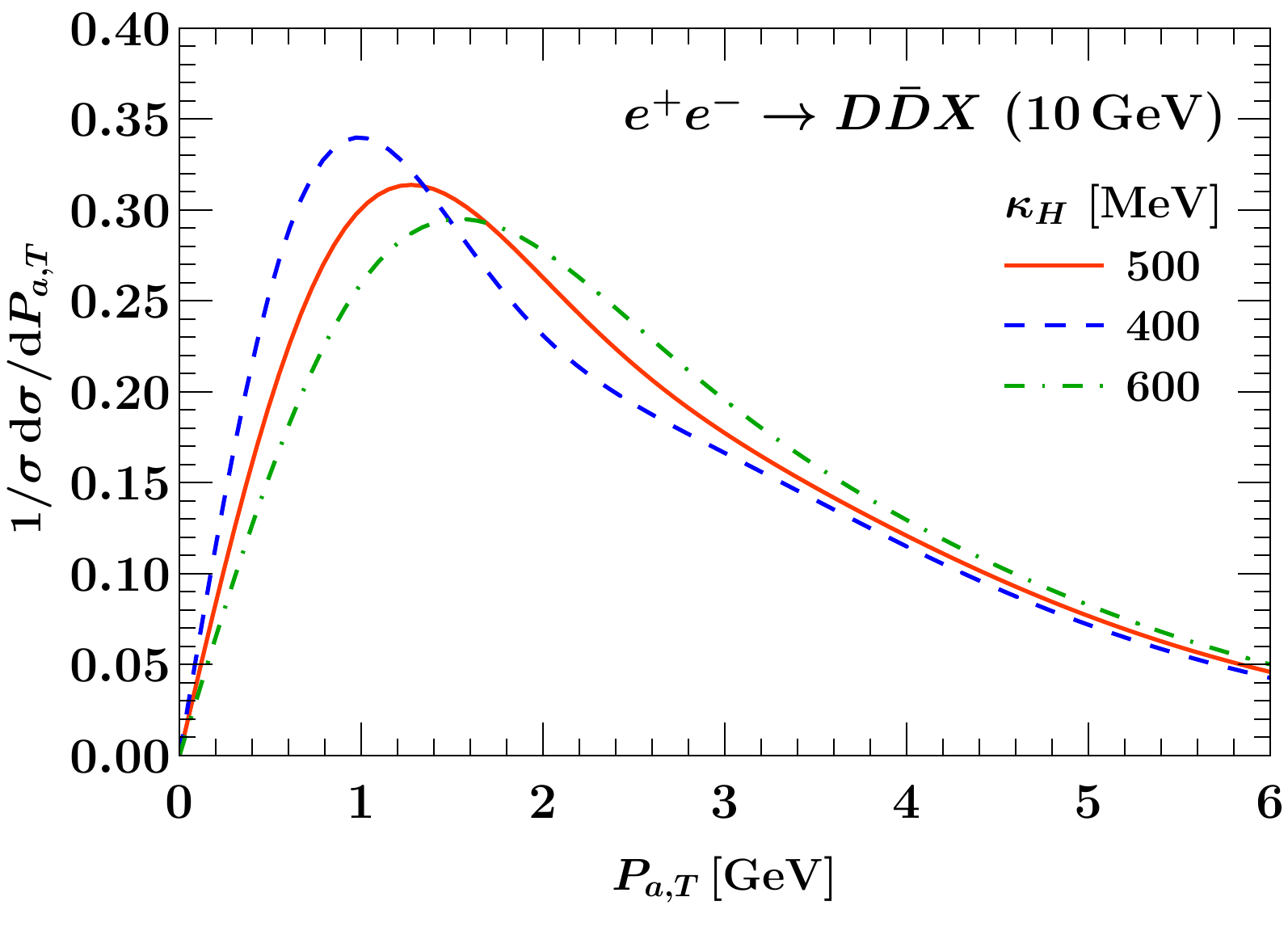}%
\hfill%
\includegraphics[width=\WidthTwoSubfigs]{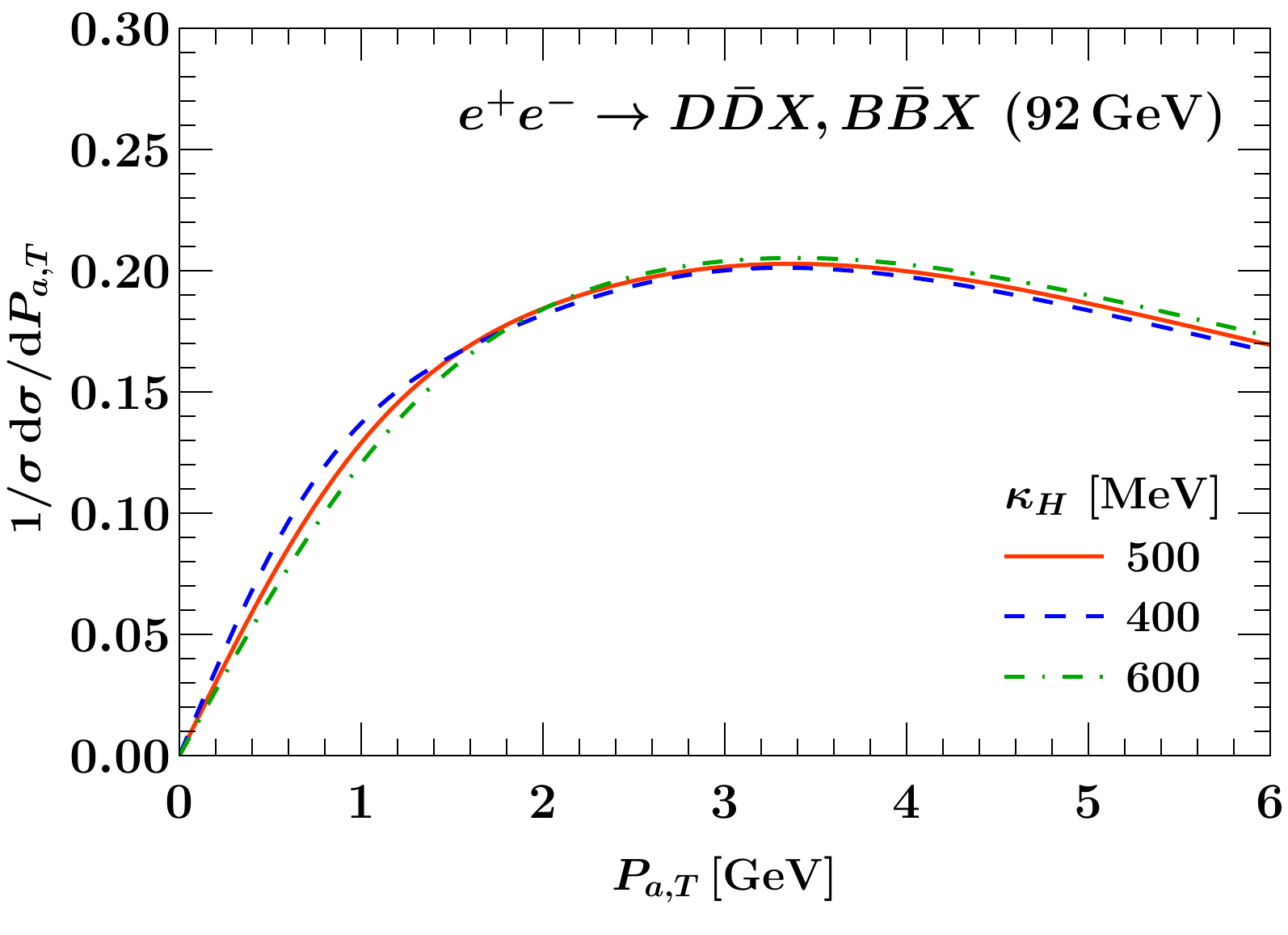}%
\\
\includegraphics[width=\WidthTwoSubfigs]{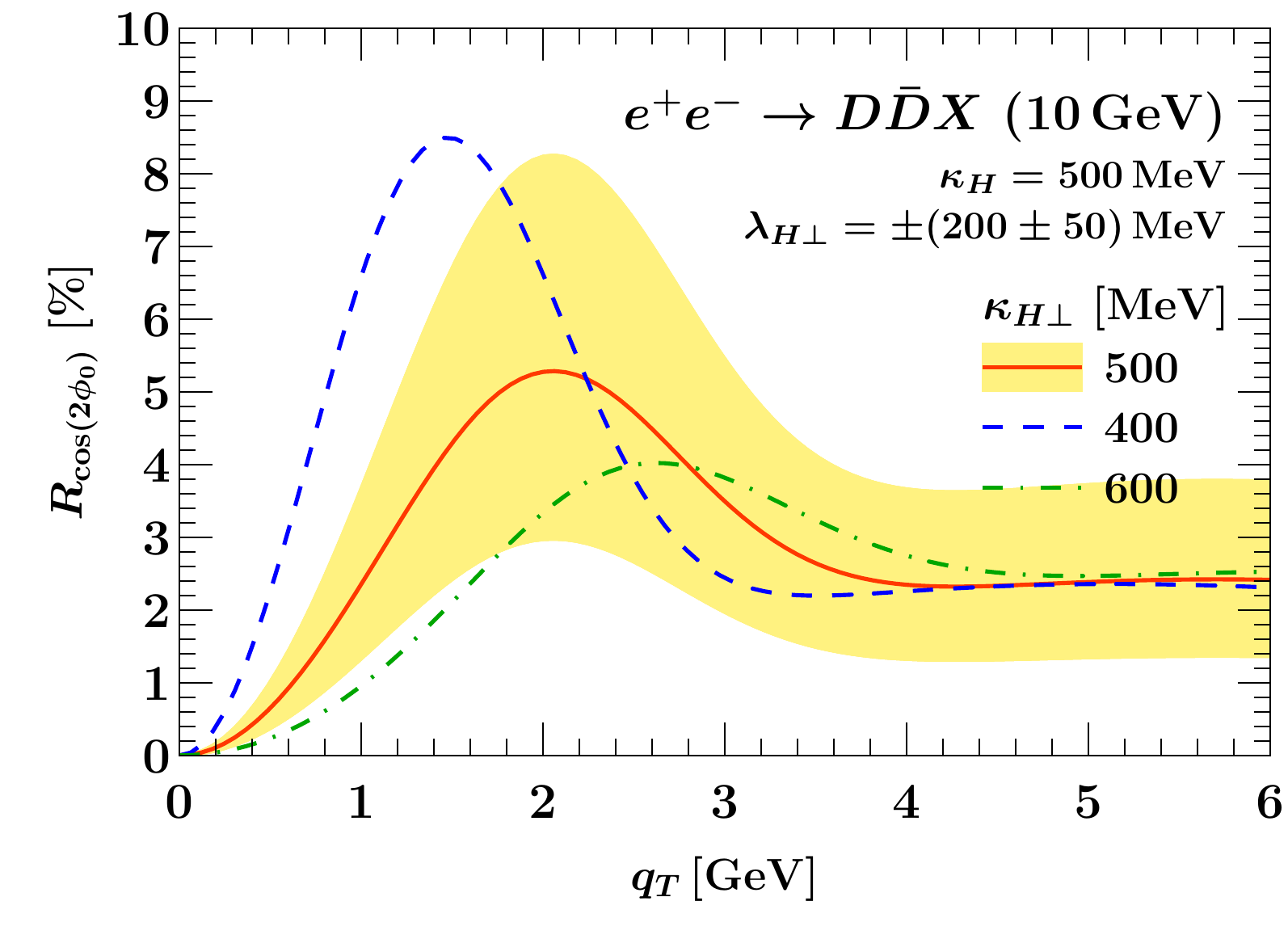}%
\hfill%
\includegraphics[width=\WidthTwoSubfigs]{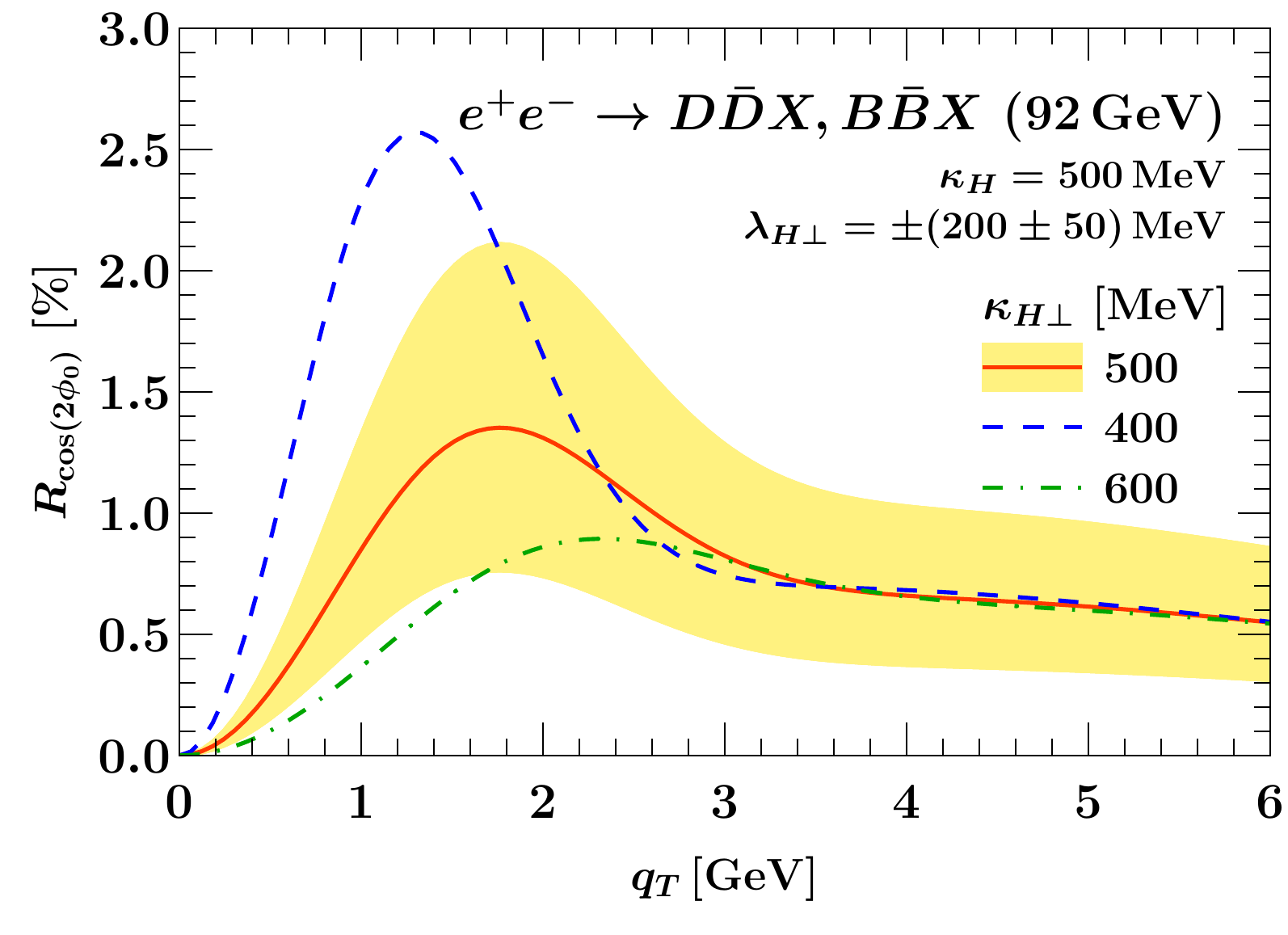}%
\caption{
Normalized TMD cross sections (top)
and Collins effect strengths (bottom)
for charm quarks at $Q = 10 \GeV$ (left)
and charm and bottom quarks at $Q = m_Z$ (right)
as a function of $P_{aT}$ and integrated over $z_H$.
The yellow bands in the case of the Collins effect
correspond to the indicated variations of the sign and magnitude of $\lambda_{H\perp}$.
}
\label{fig:ee_hadrons}
\end{figure*}

In \fig{ee_hadrons} we show the predicted $e^{+}e^{-} \to D \bar D X$ or $B \bar B X$ cross sections
as a function of hadron transverse momentum $P_{a,T}$,
and the Collins effect strength $R_{\cos(2\phi_0)}$
as a function of $q_{T}$.
The universality for charm and bottom quarks follows along the same lines as for \fig{tmd_ffs},
and holds as long as the center-of-mass energy is sufficient to produce the quark-antiquark pair in a boosted state.
This is the case for charm mesons at typical continuum center-of-mass energies at existing $B$ factories,
and for both charm and bottom mesons at higher values of $Q$ such as at the $Z$ pole.
The Collins effect is smaller at higher center-of-mass energies
because $\chi_{1,H}^{\perp}$ is linearly suppressed in $b_{T}$ compared to the unpolarized,
which means it predominantly contributes at larger values of $b_T$
where the Sudakov suppression at higher energies tends to be stronger.

We show the results of varying
$\kappa_{H}$ ($\kappa_{H}^{\perp}$)
for the unpolarized (Collins) TMD FF,
and illustrate the variation of $\lambda_{H\perp}$
by the yellow band, exactly as in \fig{tmd_ffs}.
Note that the information about the absolute sign of the Collins function is lost in $e^{+}e^{-}$ collisions,
i.e., for two charge-conjugate hadrons
we end up with a positive effect strength for any value of $\lambda_{H\perp} = \lambda_{\bar H \perp}$
since the effect is proportional to the square of the Collins function.
One may nevertheless extract the relative factor
between e.g.\ the $D$ and $D^*$ Collins function,
which heavy-quark spin symmetry predicts to be exactly minus one, see \eq{spin_symmetry_relations_h1perphq},
by measuring the Collins effect separately for $e^{+}e^{-} \to D \bar D X$ and $e^{+}e^{-} \to D^{*} \bar D X$.
Explicitly, our prediction from heavy-quark spin symmetry reads
\begin{align}
R_{\cos(2\phi_0)}^{D \bar{D}}
= - \frac{1}{3} R_{\cos(2\phi_0)}^{D^* \bar{D}}
= - \frac{1}{3} R_{\cos(2\phi_0)}^{D \bar{D}^*}
= + \frac{1}{9} R_{\cos(2\phi_0)}^{D^* \bar{D}^*}
\end{align}
We point out that for generic $\ord{\lqcd}$ model parameters,
the Collins effect strength reaches the several-percent level
for continuum open charm production at existing $B$ factories,
in line with our expectation of an effect strength
that is comparable to the light quark case,
making a future dedicated measurement (or search) appear very feasible.

\subsubsection{Comment on claims regarding a mass suppression of the Collins effect}
\label{sec:collins_comment}

In $e^+ e^-$ collisions,
the ``intrinsic'' heavy-quark Collins effect we analyzed above has been disregarded so far.
Note that this effect is in general distinct from the large background contribution
of $D\bar D$ weak decays to e.g.\ a measurement of the Collins effect on a $K\bar K$ sample.
This contribution is indeed considered
in experimental analyses \cite{BaBar:2013jdt, BaBar:2015mcn, Belle:2005dmx, Belle:2008fdv}
and subtracted as a background using Monte-Carlo simulations and heavy-quark enriched samples,
but cannot be immediately interpreted as a sign of a (nonperturbative) Collins effect
since the progenitor $D\bar D$ pair in this case is not constrained
to be near the back-to-back limit by the measurement,
meaning that e.g.\ perturbative gluon emissions can also
induce azimuthal correlations on the $D\bar D$ pair and thus their weak decay products.

\Refcite{BaBar:2013jdt} mentions that it would be possible
to look for the intrinsic heavy-quark Collins effect
with some further improvements to their analysis,
but also incorrectly expects that the Collins effect
should be parametrically suppressed by the mass of heavy quarks.
The argument sketched in that reference (see beginning of their section~IV)
is that helicity flips should wash out the spin correlation
between the heavy quark and the antiquark.
This is not the case, as we have argued above:
The quarks are approximately massless at the scale $\mu \sim Q$ at which
they are produced, and thus are produced with fully entangled spin states,
such that there is no suppression by the mass from physics at this scale.
Similarly, in our detailed analysis of the Collins FF at the scale $\mu \leq k_T \leq m$,
we find \emph{no suppression} of the effect by the mass,
and the Collins effect in particular is fully allowed by heavy-quark symmetry
when accounting for the presence of lightlike Wilson lines.
Note that this is not contradictory to the fact that we do find a suppression of the Collins effect
by $\lqcd/k_T$ at large $k_T$,
since this suppression is exactly commensurate
with the twist suppression of the two Collins functions in the light-quark case,
which has been mapped out extensively~\cite{Belle:2005dmx, Belle:2008fdv, BaBar:2013jdt, BaBar:2015mcn, Belle:2019nve}.
We conclude that the prospects for a measurement of the intrinsic,
nonperturbative heavy-quark Collins effect at $B$ factories
are even better than anticipated in \refcite{BaBar:2013jdt}.

\subsection{Accessing heavy-quark TMDs at the future EIC}
\label{sec:tmds_sidis}

TMD fragmentation functions may also be accessed
from single-inclusive measurements with one identified hadron
in electron-nucleon collisions, $e^-(\ell) + N(P) \to e^-(\ell') + H(P_H) + X$,
where the scattering is mediated by an off-shell photon
with momentum $q = \ell - \ell'$ (and $Q^2 \equiv -q^2 > 0$).
The fully differential cross section for this process in the TMD regime
reads~\cite{Ji:2004wu, Ji:2004xq, Bacchetta:2006tn, Collins:1350496, Ebert:2021jhy}
\begin{align} \label{eq:tmd_factorization_sidis}
\frac{\df \sigma_{e N \to eH X}}{\df x \, \df y \, \df z_H \, \df^2 \vec{P}_{H,T}}
&= \sigma_0 \Bigl\{ W_{UU,T}(Q^2, x, z_H, \vec{P}_{H,T}/z_H)
\nn \\ & \quad
+ \lambda_e S_L \, \sqrt{1 - \eps^2} \, W_{LL}(Q^2, x, z_H, \vec{P}_{H,T}/z_H)
\nn \\ & \quad
+ S_L \, \eps \, \sin(2 \phi_H) \,  \, W_{UL}^{\sin(2 \phi_H)}(Q^2, x, z_H, \vec{P}_{H,T}/z_H)
\Bigr\}
\,.\end{align}
On the left-hand side, $x = Q^2/(2 P \cdot q)$, $y = (P \cdot q )/ (P \cdot \ell)$,
$z_H = (P \cdot P_H)/(P \cdot q)$, and $\vec{P}_{H,T}$ is
the outgoing hadron's transverse momentum relative to $\vec{q}$ in the Breit frame.
On the right-hand side,
\begin{alignat}{3}
\sigma_0 &= \frac{\aem^2 \pi y}{z_H Q^2 (1- \eps)}
\,, \qquad &
\eps = \frac{1 - y}{1 - y + y^2/2}
\,,\end{alignat}
up to power corrections in $x M_N/Q$, $x M_H M_N/(z_H Q^2)$,
or $x P_{H,T} M_N/(z_H Q^2)$, all of which are small in the TMD regime of interest,
and $\phi_H$ is the azimuthal angle of the hadron transverse momentum
in the Trento (photon) frame~\cite{Bacchetta:2006tn}.
The beam polarization information is encoded
in the lepton beam helicity $\lambda_e$
and the covariant nucleon spin vector $S^\mu = (0, S_T \cos \phi_S, S_T \sin \phi_S, -S_L)$
as decomposed in the Trento frame.
We have dropped terms proportional to $S_T$, which cannot be populated
by leading-power heavy-quark TMD PDFs, see \sec{polarized_heavy_quark_tmd_pdfs}.
We have also dropped terms proportional to the Boer-Mulders function,
whose twist-2 matching in the heavy-quark case is suppressed
by at least one additional power of $\as$.
The hadronic structure functions factorize in terms of one TMD PDF and one TMD FF each,
\begin{align} \label{eq:structure_functions_sidis}
W_{UU,T}(Q^2, x, z_H, \vec{q}_T)
&= \cF_{eN} \Bigl[
   \cH \, {f}_{1} \, {D}_{1}
\Bigr]
\,, \nn \\
W_{LL}(Q^2, x, z_H, \vec{q}_T)
&= \cF_{eN} \Bigl[
   \cH \, {g}_{1L} \, {D}_{1}
\Bigr]
\,, \nn \\
W_{UL}^{\sin(2 \phi_H)}(Q^2, x, z_H, \vec{q}_T)
&= -\cF_{eN} \Bigl[
   \cH \, {h}_{1L}^{\perp(1)} \, {H}_{1}^{\perp(1)}
\Bigr]
\,,\end{align}
where the convolution in transverse momentum may be written in position space as~\cite{Boer:2011xd}
\begin{align} \label{def_f_sidis}
\cF_{e N}\Bigl[ \cH \, {g}^{(n)} \, {D}^{(m)} \Bigr]
&= 2z_H \int_0^\infty \! \frac{\df b_T\,b_T}{2\pi}
(M_N b_T)^n (- M_H b_T)^m
J_{n+m}(b_T q_T)
\nn \\ & \quad \times
\sum_{i} \cH_{ei \to ei}(Q^2, \mu) \,
{g}_{i/N}(x, b_T, \mu, Q^2) \,
{D}_{H/i}(z_H, b_T, \mu, Q^2)
\,,\end{align}
and the hard function for scattering a quark off a virtual photon is
\begin{align}
\cH_{ei \to ei}(Q^2, \mu) = \abs{e_i}^2 + \ord{\as}
\,.\end{align}
As for $e^+ e^-$ collisions, the TMD factorization theorems
in \eq{structure_functions_sidis}
only assume that the hard scale $Q \sim zQ$ is large compared to all low scales,
and thus hold for both light and heavy hadron production without modification.
Again, the heavy quark is approximately massless at the hard scale
such that helicity is conserved during the hard scattering.
This means that while the production mechanisms for longitudinally
or transversely polarized heavy quarks from an incoming nucleon are different from light quarks
(and are fully perturbative),
the way they imprint on the distribution of final-state hadrons is the same,
leaving nonzero spin asymmetries
\begin{align}\label{eq:sidis_ssa}
A_{LL}(Q^2, x, q_T)
&= \frac{
   \int \! \df z_H \,
   W_{LL}(Q^2, x, z_H, \vec{q}_T)
}{
   \int \! \df z_H \,
   W_{UU,T}(Q^2, x, z_H, \vec{q}_T)
}
\,, \nn \\[0.4em]
A_{UL}^{\sin(2 \phi_H)}(Q^2, x, q_T)
&= \frac{
   \int \! \df z_H \,
   W_{UL}^{\sin(2 \phi_H)}(Q^2, x, z_H, \vec{q}_T)
}{
   \int \! \df z_H \,
   W_{UU,T}(Q^2, x, z_H, \vec{q}_T)
}
\,.\end{align}
In particular, the $\sin(2\phi_H)$ modulation induced
by a nucleon beam polarization flip gives direct access
to the heavy-quark Collins function including its sign,
which is not accessible in $e^+ e^-$ collisions.

\begin{table}
\renewcommand{\arraystretch}{1.2}
\centering
\begin{tabular}{c|c|c|c|c}
\hline\hline
$\sigma(eN \to eHX)~[\!\pb]$    & $c$, $x > 0.01$ & $c$, $x > 0.1$ & $b$, $x > 0.01$ & $b$, $x > 0.1$ \\
\hline
$q_T < 2 \GeV, Q > 4 \GeV$ & 84 & 3.47 & 18 & 0.65 \\
$q_T < 4 \GeV, Q > 10 \GeV$ & 16 & 1.45 & 4.9 & 0.42 \\
\hline\hline
\end{tabular}%
\caption{
Total cross sections in picobarn for producing charm (left two columns) or bottom-quark hadrons (right two columns)
in the TMD region at the future $18 \times 275 \GeV^2$ EIC for different cuts
on $x > x_\mathrm{min}$, $Q > Q_\cut$, $q_T = P_{H,T}/z < q_T^\cut$.
See the text for further details on the acceptance cuts we consider.
}
\label{tab:eic_sample_sizes}
\end{table}

To assess the statistical power of the future EIC to constrain charm and bottom quark TMD dynamics,
we first estimate the expected sample size of heavy hadrons in electron-proton collisions.
To do so, we consider the total cross section for producing
a heavy quark in the TMD region summed over beam polarizations,
\begin{align} \label{eq:total_xsec_sidis}
\sigma_{e N \to eH X}(Q_\cut, q_T^\cut)
&= \int \! \df x \, \df y \, \df z_H \,
\df^2 \vec{P}_{H,T} \, \frac{\df \sigma_{e N \to eH X}}{\df x \, \df y \, \df z_H \, \df^2 \vec{P}_{H,T}}
\\ & \quad \times
\Theta(\qTcut - P_{H,T}/z_H) \, \Theta(Q - Q_\cut) \, \Theta_\mathrm{DIS}(x, y)
\nn \\
&= \sigma_0 \int \! \df x \, \df y \, \df z_H \, \Theta(Q - Q_\cut) \, \Theta_\mathrm{DIS}(x, y) \,
2z_H^3 \int_0^\infty \! \df b_T \, \qTcut \,
J_1(b_T \qTcut)
\nn \\ & \quad \times
\sum_{i} \cH_{ei \to ei}(Q^2, \mu) \,
{f}_{1\,i/N}(x, b_T, \mu, Q^2) \,
{D}_{1\,H/i}(z_H, b_T, \mu, Q^2)
\nn \,,\end{align}
where $\Theta_\mathrm{DIS}(x, y)$ denotes DIS acceptance cuts given by
\begin{align} \label{eq:dis_cuts}
x &> x_\mathrm{min}
\,, \quad
&0.01 &< y < 0.95
\,, \quad
&W^2 &= \Bigl( \frac{1}{x} - 1 \Bigr) Q^2 > 100 \GeV^2
\,.\end{align}
We consider the EIC at beam energies $E_e = 18 \GeV$ and $E_N = 275 \GeV$.
Any experimental cuts on $z_H > z_\cut$
and the additional prefactor of $z_H^3$ in \eq{total_xsec_sidis}
are irrelevant at our working order
because the heavy quark carries all the energy in all regimes,
i.e., $z_H = 1$ either at leading power in the heavy-quark expansion
or at the leading perturbative order,
see the comments below \eq{tmd_ff_unpol_model}.
For this estimate we set $\kappa_H = 0$ in the unpolarized heavy-quark TMD FF,
since the total integral of the cross section up to $\qTcut \gg \lqcd$
is independent of it up to corrections of $\ord{\lqcd^2/\qTcut}$~\cite{Ebert:2022cku},
and sum over all heavy hadrons containing the heavy quark, exploiting $\sum_H \chi_H = 1$.
This means that the total rate at which heavy quarks are produced
is predicted fully perturbatively, as expected.

Our results for the expected total charm and bottom-quark TMD
cross sections are given in \tab{eic_sample_sizes}
for $Q_\cut = 4 \GeV$ and $Q_\cut = 10 \GeV$, where higher $Q_\cut$ allows for mapping out
the TMD region to higher $q_T$ before encountering power corrections,
but at the cost of much lower rates.
(We have also adjusted $q_T^\cut$ accordingly in each case.)
Scaled to an integrated luminosity of $10 \fb^{-1}$,
we expect a total charm quark sample of $35 \times 10^3$
in the TMD region for the loose cut on $Q$
and in the region $x > 0.1$ where polarization effects
are expected to be most pronounced, see \fig{tmd_pdfs},
and where a measurement of the $\sin (2\phi_H)$ asymmetry is the most promising.
This suggests that even with this limited integrated luminosity,
percent-level asymmetries should be statistically resolvable.

\begin{figure*}
\includegraphics[width=\WidthTwoSubfigs]{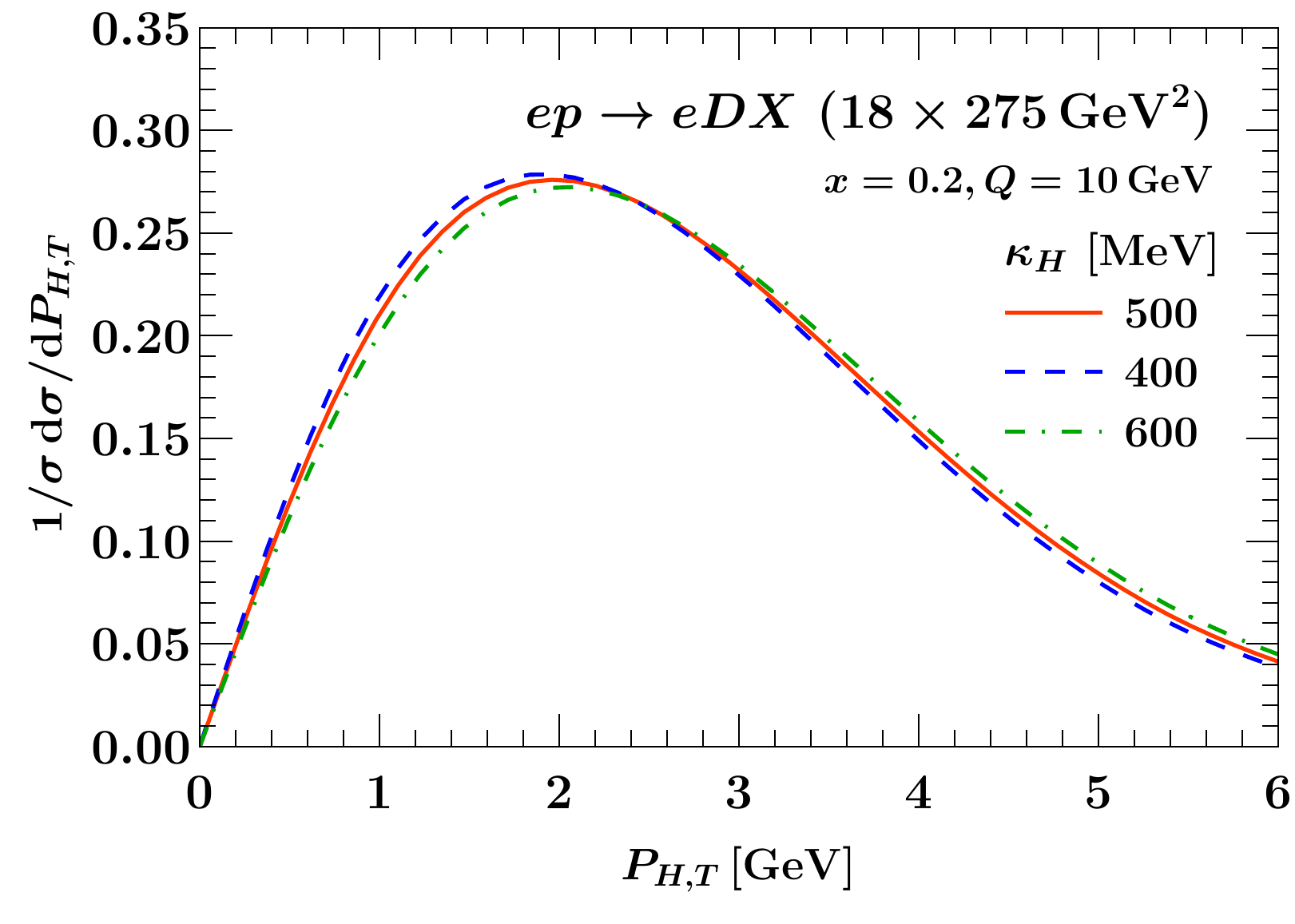}%
\hfill%
\includegraphics[width=\WidthTwoSubfigs]{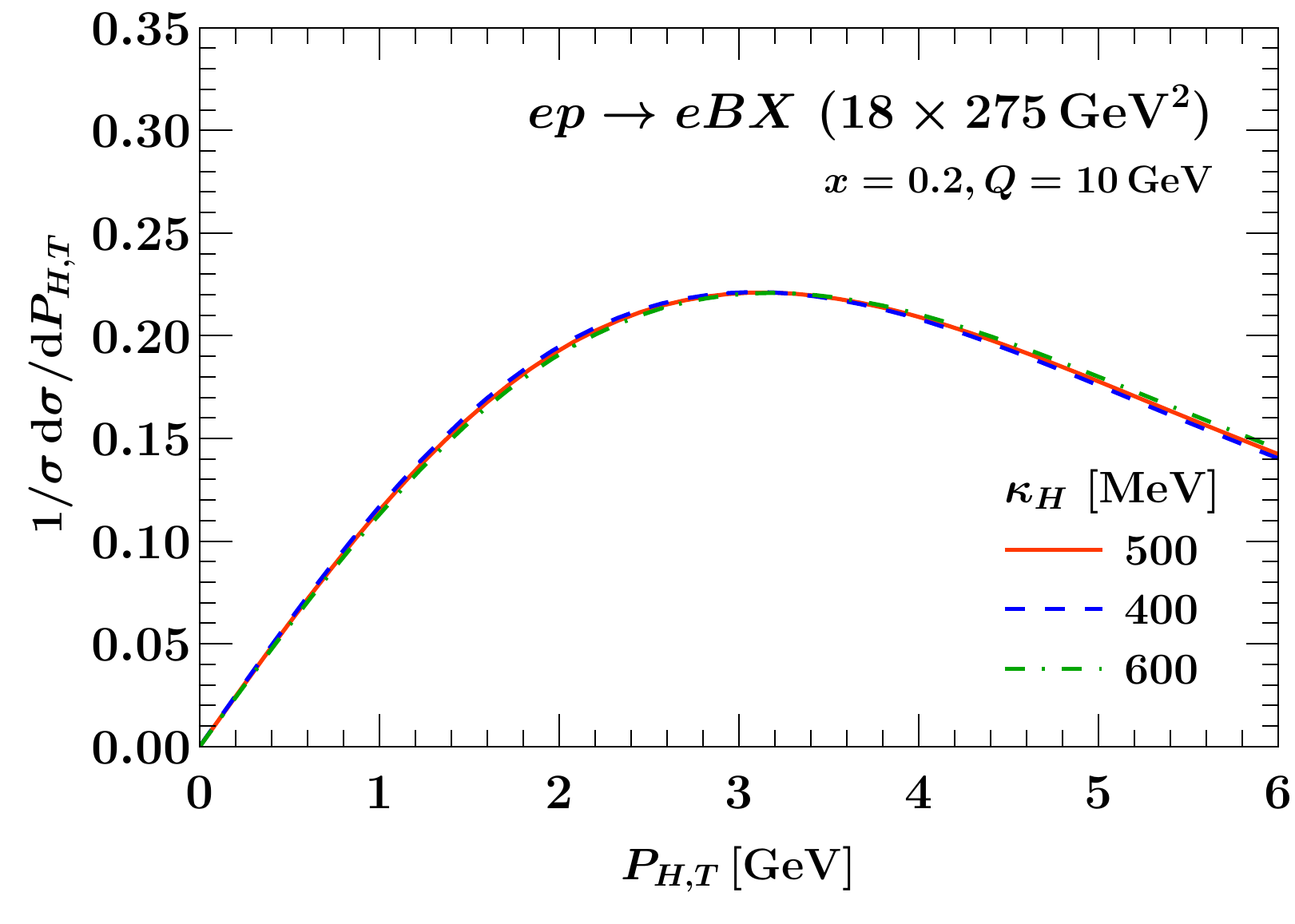}%
\\
\includegraphics[width=\WidthTwoSubfigs]{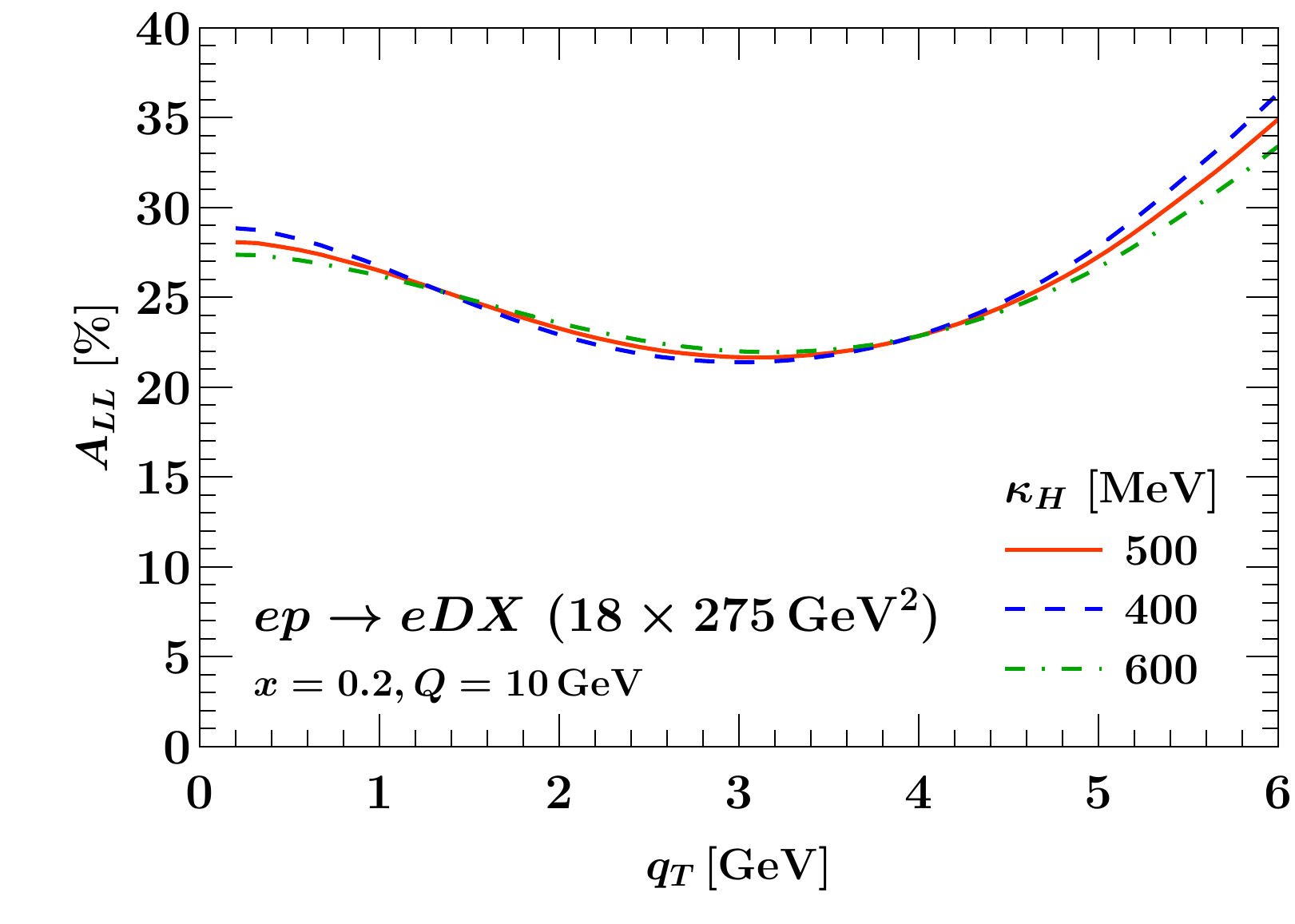}%
\hfill%
\includegraphics[width=\WidthTwoSubfigs]{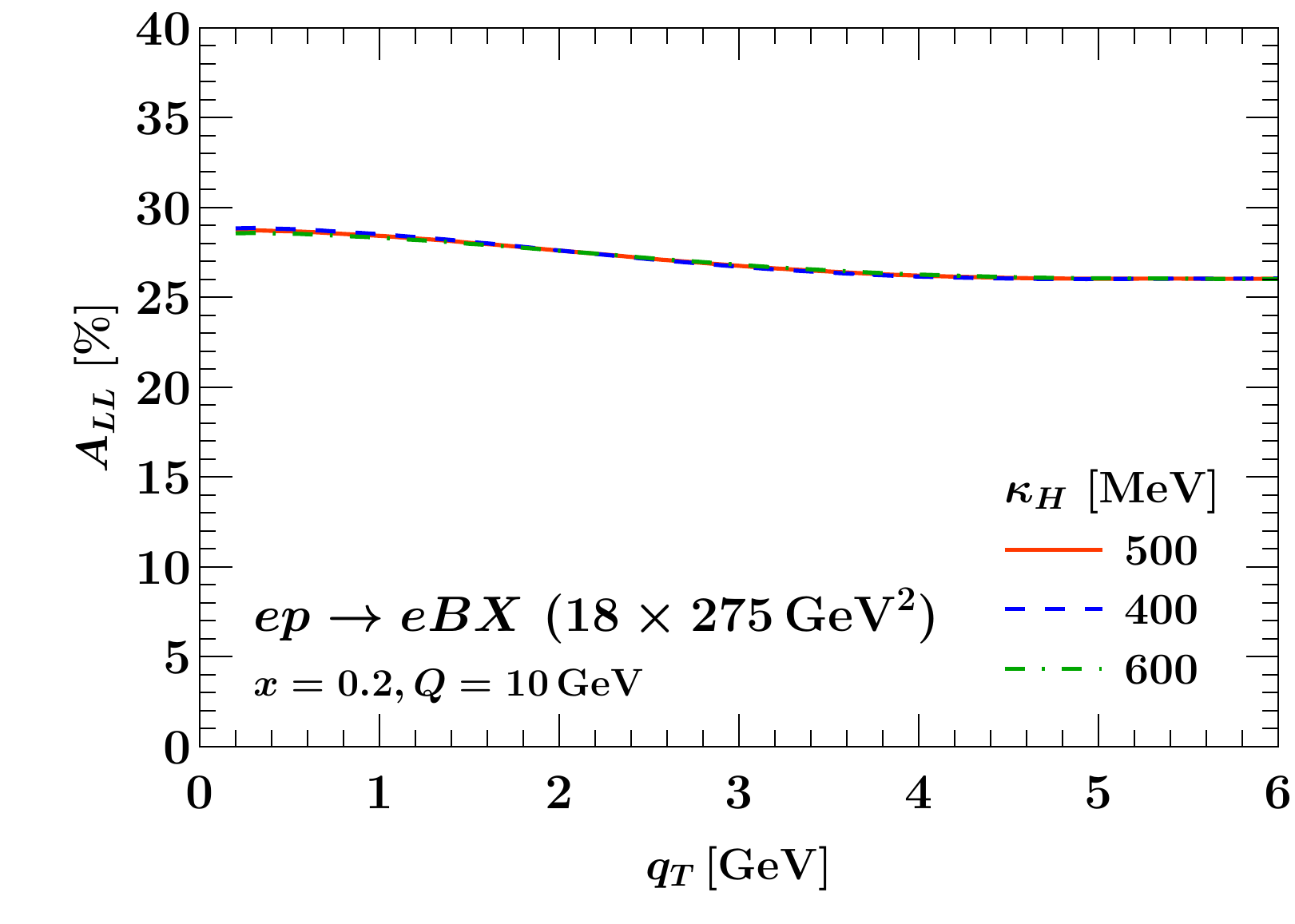}%
\\
\includegraphics[width=\WidthTwoSubfigs]{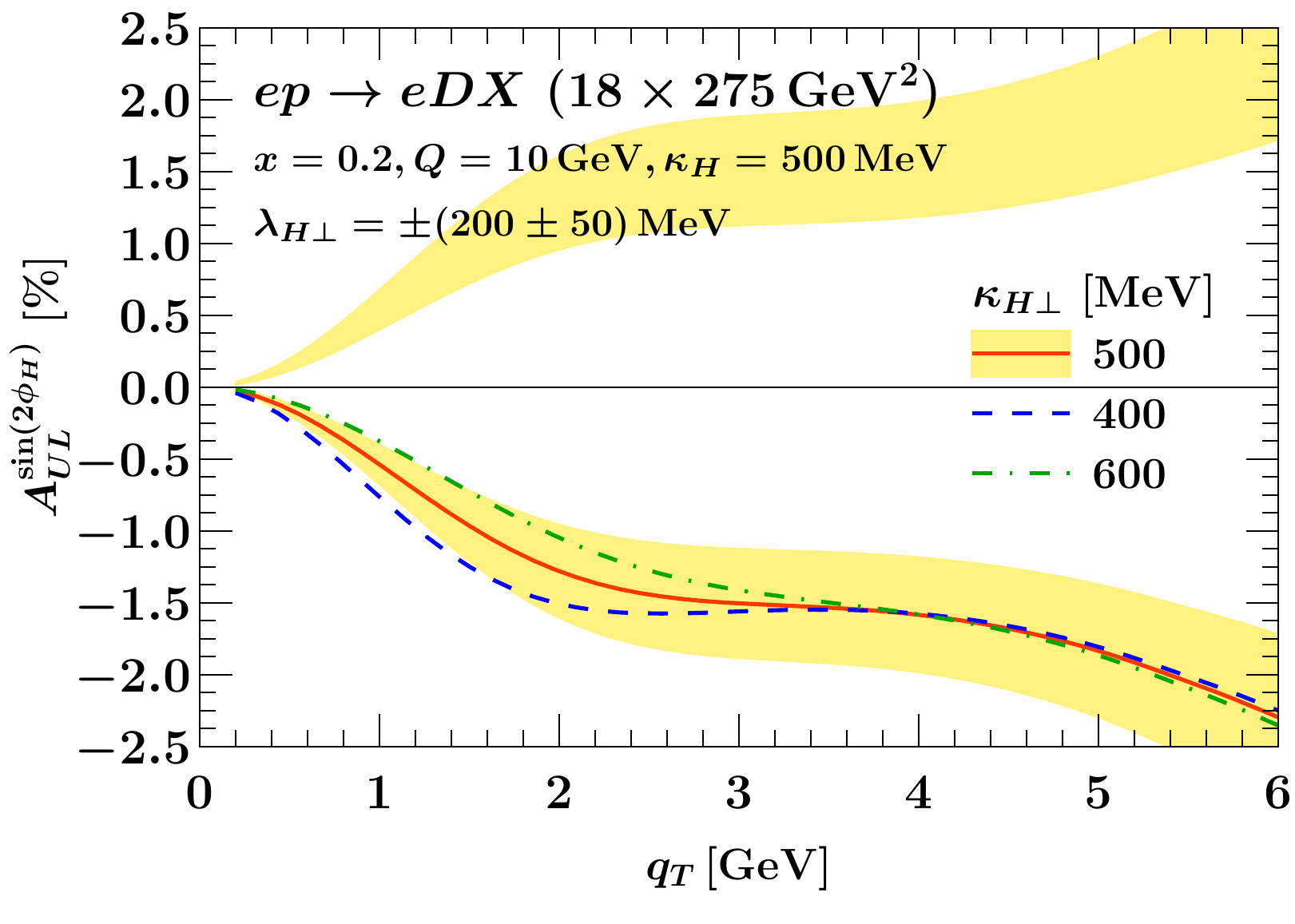}%
\hfill%
\includegraphics[width=\WidthTwoSubfigs]{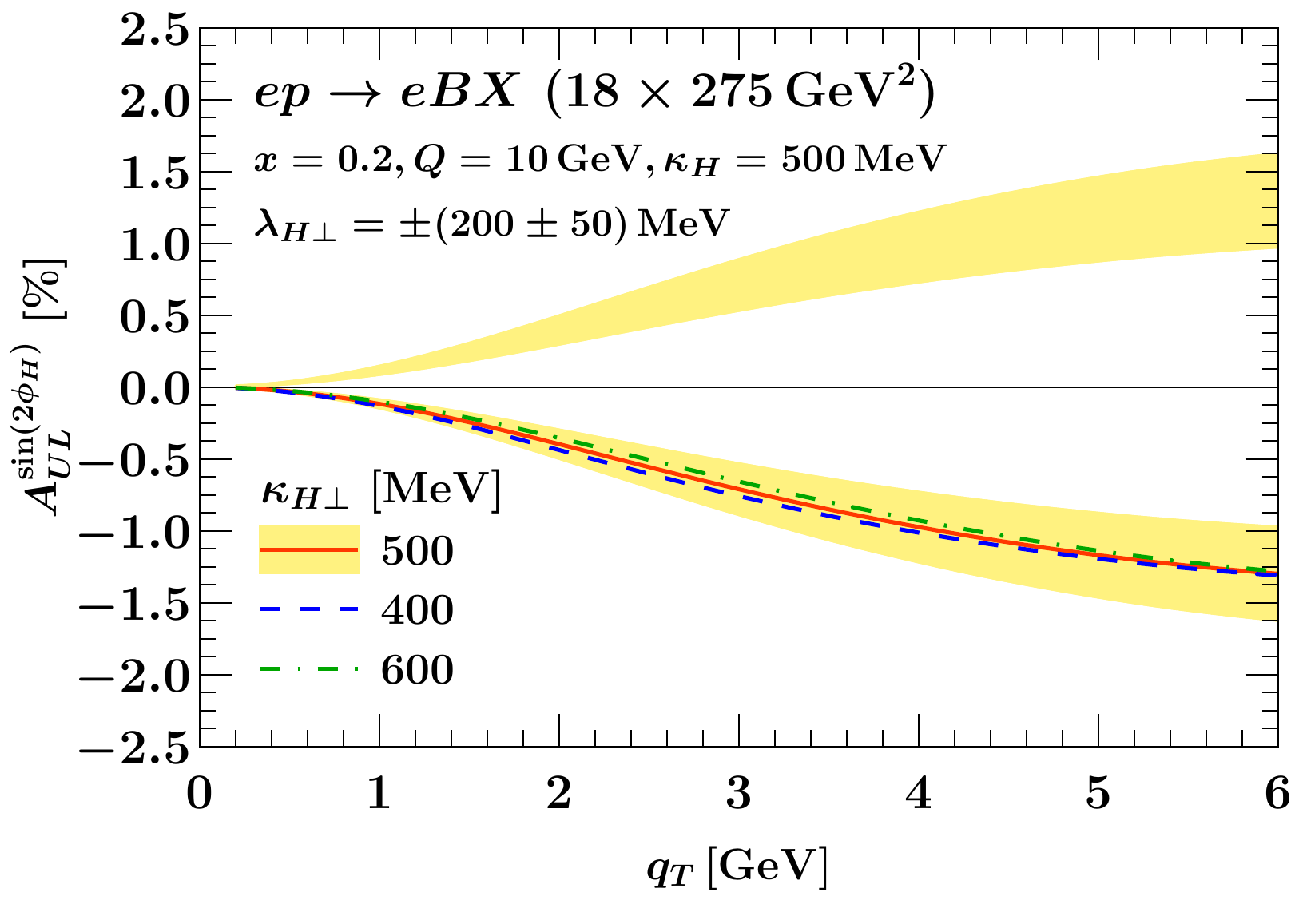}%
\caption{
Normalized unpolarized TMD cross sections (top),
longitudinal spin asymmetry (center),
and $\sin(2\phi_H)$ spin asymmetry (right)
for charm quarks (left) and bottom quarks (right)
at the future EIC.
The yellow bands correspond
to the indicated variations of the sign and magnitude of $\lambda_{H\perp}$,
i.e., the heavy-quark Collins function.
}
\label{fig:sidis}
\end{figure*}

In \fig{sidis} we show the results for the unpolarized SIDIS cross section
with a $D$ ($B$) meson in the final state,
and for the two spin asymmetries defined in \eq{sidis_ssa}.
Note that the effect of different $\kappa_{H}$
in the unpolarized TMD fragmentation function
is negligible in the cross section and the $A_{LL}$ asymmetry,
which as expected are dominated by perturbative physics.
The $A_{LL}$ asymmetry is very sizable at $\sim 30\%$ at the chosen value of $x = 0.2$.
On the other hand, the $A_{UL}$ asymmetry
is substantially smaller ($1-2\%$) for the generic $\ord{\lqcd}$ parameters we picked here
due to the smaller value of both $h_{1L}^{\perp}$ compared to $g_{1L}$
and $H_{1}^{\perp}$ compared to $D_{1}$
in most of the contributing TMD region,
see \figs{tmd_ffs}{tmd_pdfs} and the surrounding discussion.
The numerically smaller value of $h_{1L}^{\perp}$ for bottom quarks
discussed around \fig{tmd_pdfs} is likewise reflected in the size
of the asymmetry for bottom compared to charm quarks.
We emphasize that a measurement of $A_{UL}$,
compared to the Collins effect in $e^+ e^-$ collisions,
has the unique benefit of accessing
the absolute sign of the heavy-quark Collins function.
Resolving this sign should well be possible
within the expected statistics at the future EIC.
While we leave the study of systematic effects
(such as luminosity uncertainties) to future work, we note that the
requirements that the established heavy-flavor/gluon distribution program of the EIC
places on instrumentation
have already been analyzed in depth in \refcite{AbdulKhalek:2021gbh}.
Among these requirements are secondary vertex reconstruction capabilities
and the momentum resolution on soft pions from $D$ decays,
all of which will also benefit the kind of
differential measurements of semi-inclusive heavy-quark fragmentation
that we propose here.

\section{Summary of main results and outlook}
\label{sec:conclusions}

In this paper, we have studied the transverse momentum-dependent (TMD)
dynamics of bottom or charm quarks
with mass $m \equiv m_c, m_b \gg \lqcd$
fragmenting into heavy hadrons for the first time.
We considered two parametric regimes for the transverse momentum $k_T$,
(a)~$\lqcd \lesssim k_{T} \ll m$,
where the hadron transverse momentum $k_{T}$
is determined by nonperturbative soft radiation into the final state,
and (b)~$\lqcd \ll m \lesssim k_{T}$,
where $k_{T}$ is set by perturbative emissions.
We assumed throughout that the heavy quark is produced
at a hard scale $Q \gg m, k_T$, i.e., it is boosted
in the frame of the hard scattering,
such that standard TMD factorization applies at the scale $Q$
and only the low-energy TMD matrix elements are modified by the heavy quark dynamics.
In both regimes, the dynamics at scales below the heavy quark mass
are constrained by heavy-quark symmetry
and encoded in novel low-energy matrix elements in boosted Heavy-Quark Effective Theory (bHQET):
\begin{itemize}
   \item
   We showed that in regime (a), the unpolarized and Collins TMD fragmentation functions (FF)
   match onto new, universal nonperturbative bHQET matrix elements
   $\chi_{1,H}(k_T)$ and $\chi_{1,H}^{\perp}(k_T)$,
   which we dubbed TMD fragmentation factors.
   \item
   In regime (b), we made use of the twist expansion for light-quark TMD FFs
   and combined it with the matching collinear FFs onto bHQET
   to identify the relevant leading or subleading bHQET matrix elements.
   \item
   An important new ingredient in this analysis is
   the unpolarized partonic heavy quark TMD FF $d_{1 \, Q/Q}$,
   a perturbative Wilson coefficient
   that appears in our analysis for the first time
   and that we expect to appear also in other contexts
   like flavor-tagged energy-energy correlators in the back-to-back limit.
   \item
   We find that the Collins TMD FF scales as $\lqcd/k_T$
   at $k_T \gg \lqcd$ --- but is \emph{not} suppressed by the quark mass
   --- and identified the coefficient
   as a new subleading bHQET matrix element probing gluon correlations
   within the fragmentation process.
   \item
   We used heavy-quark spin symmetry
   to express the TMD fragmentation factors
   in terms of the underlying spin density matrix of the light hadron constituents.
   \item
   For the unpolarized TMD FF $D_{1\,H/Q}(z_H, k_T, \mu, \zeta)$, this allowed us to prove
   the following relations between renormalized TMD FFs
   within spin symmetry multiplets:
   \begin{align}
   D_{1\,H/Q} = \frac{1}{3} D_{1\,H^*/Q}
   \,, \qquad
   D_{1\,\Sigma_Q/Q} = \frac{1}{2} D_{1\,\Sigma_Q^*/Q}
   \,, \qquad
   D_{1\,H_1/Q} = \frac{3}{5} D_{1\,H_2^*/Q}
   \end{align}
   These relations are a powerful generalization of known results
   for inclusive heavy-quark fragmentation,
   demonstrating for the first time that they hold
   point by point in transverse momentum.
   \item
   We showed that the Collins function arises from correlations between
   hadronic radiation into the final state
   and the transverse polarization of the light constituents,
   which in turn is correlated with the heavy-quark spin
   by the experimental reconstruction of e.g.\ $D$ vs.\ $D^*$ mesons.
   This new picture of spin correlations in the heavy-quark limit
   allowed us to prove the following novel sum rule
   for the renormalized heavy-quark Collins function,
   which holds up to corrections of $\ord{\lqcd/m}$
   and up to radiative corrections in $\alpha_s$ at the scale $\mu \sim m$, and likely beyond,
   \begin{align}
      \sum_{H \in M_\ell} H_{1\,H/Q}^\perp(z_H, k_T, \mu, \zeta) = 0
   \,,\end{align}
   where $M_\ell = \{\Lambda_Q\}, \{H, H^*\},
   \{\Sigma_Q, \Sigma_Q^*\}, \{H_1, H_2^*\}, \dots$
   is a heavy-quark spin symmetry multiplet
   specified by the total angular momentum
   and flavor content of the light hadron constituents.
\end{itemize}

\noindent
To extend our analysis to the possible phenomenology
at the future Electron-Ion Collider (EIC),
we also considered the production of polarized heavy quarks
from a polarized nucleon,
which is encoded in all-order matching relations
between heavy-quark TMD PDFs and twist-2 collinear light-parton PDFs:
\begin{itemize}
   \item
   We find that terms proportional to the transverse nucleon polarization
   vanish for heavy quarks at twist-2 to all orders in the strong coupling
   due to chirality and flavor conservation in the light-quark sector.
   \item
   In contrast to the light-quark case, transverse quark polarization states
   \emph{are} populated from unpolarized and linearly polarized nucleons
   because the quark mass breaks chirality.
   \item
   We find nontrivial matching coefficients at $\ord{\as}$ for
   the heavy-quark worm-gear $L$ and helicity TMD PDFs
   onto the gluon helicity collinear PDF,
   both of which we computed explicitly for the first time.
   We anticipate that the heavy-quark Boer-Mulders function
   will receive a contribution from the twist-2 collinear gluon PDF
   starting at $\ord{\as^2}$, where it becomes allowed by time-reversal invariance.
\end{itemize}

\noindent
Combining the standard TMD factorization theorems
for $e^{+}e^{-}$ to hadrons and SIDIS
with simple numerical models
for the new nonperturbative functions we identified,
we provided predictions for unpolarized heavy-quark TMD cross sections,
the Collins effect strength for heavy quarks at $e^{+}e^{-}$ colliders
(and in particular for $c\bar c$ continuum production at current $B$ factories),
as well as for the relevant spin asymmetries at the future EIC:
\begin{itemize}
   \item
   We find that a measurement of the intrinsic heavy-quark Collins effect
   is well within reach of existing $B$ factories,
   and is motivated by the rich nonperturbative structure
   of the heavy-quark Collins function that our analysis revealed.
   \item
   The fact that transversely polarized heavy quarks are produced from linearly polarized nucleons
   at a significant rate, as encoded in the worm-gear $L$ matching coefficient,
   in addition provides a clean avenue for probing
   the heavy-quark Collins functions
   in heavy-quark SIDIS at the future EIC,
   including its absolute sign.
\end{itemize}

\noindent
The theoretical framework we developed in this paper
paves the way for many promising future applications:
\begin{itemize}
   \item
   While we only considered the case
   of unpolarized heavy hadrons in this work,
   an immediate next application of our framework
   are polarized vector mesons or baryons containing heavy quarks.
   This gives access to a larger set of
   transverse-momentum dependent \emph{polarized}
   fragmentation functions~\cite{Boer:1997mf, Callos:2020qtu, Kang:2021kpt}
   which in the heavy-quark case
   resolve the light spin density matrix in even greater detail
   and obey additional sum rules.
   \item
   Another promising prospect is to consider
   heavy-quark TMD fragmentation within jets,
   which makes its rich physics accessible in hadron collisions.
   This extension is in fact straightforward
   because our results for the heavy-quark TMD FFs
   hold independent of the factorization theorem they appear in.
   This makes it possible to insert them into the hadron-in-jet frameworks
   of \refscite{Kang:2017glf, Kang:2020xyq} in a plug-and-play fashion
   as long as $Q \sim p_T^\mathrm{jet} R \gg m, k_T$.
   Yet another possibility, which could mitigate
   the effect of nonglobal logarithms that can become nonperturbative
   in our regime of interest, would be to apply grooming to the jet
   and study the hadron transverse momentum spectrum with respect
   to the groomed jet axis~\cite{Makris:2017arq, Makris:2018npl},
   see also \ftn{groomed_heavy_jets}.
   We look forward to the attendant phenomenology,
   which may in addition serve as a vacuum baseline for TMD interactions
   of open charm and bottom quarks with the quark-gluon plasma in heavy-ion collisions.
   \item
   Other natural extensions are higher-order calculations
   of the various new partonic matching coefficients we introduced in this paper,
   which will reduce the perturbative uncertainties on the lowest-order
   theory predictions we provided here. This will also involve analyzing
   the renormalon structure and optimizing the choice of quark mass scheme.
   In addition, one could consider the matching onto subleading
   bHQET fragmentation matrix elements (for TMD FFs)
   or onto twist-3 collinear PDFs (for TMD PDFs,
   extending the work of \refcite{Scimemi:2018mmi} to the massive case),
   which would make it possible to interpret phenomenological extractions
   in terms of higher-point correlation functions.
   Higher-order resummed predictions for heavy-quark TMD spectra
   then immediately follow from our factorization results
   by solving the attendant renormalization group equations,
   and will serve as powerful, highly differential benchmarks
   of the heavy-quark physics encoded in present and future
   parton showers, including their interface with hadronization models,
   on which our field-theory analysis of the nonperturbative
   dynamics places rigorous constraints.
\end{itemize}

\noindent
In conclusion, our analysis reveals that a wealth of information
on the all-order and nonperturbative structure of QCD
resides in the transverse momentum dependence of heavy-quark fragmentation.
An experimental exploration of this new subfield of TMD physics
is in immediate reach of existing $B$ factories
and will be an exciting addition to the planned heavy-flavor
physics program of the future EIC.

\acknowledgments

We thank Kyle Lee and Iain Stewart for insightful discussions
and their comments on the manuscript, and Markus Diehl
for pointing out an error regarding the all-order twist-2 matching
of the heavy-quark Boer-Mulders function
in an earlier version of this manuscript.
We thank Frank Tackmann for his leading role and many contributions
in developing the \texttt{SCETlib} numerical library.
This work was supported in part by the Office of Nuclear Physics of the U.S.\
Department of Energy under Contract No.\ DE-SC0011090,
and within the framework of the TMD Topical Collaboration, by the European Research Council (ERC) 
under the European Union's Horizon 2020 research and innovation programme 
(Grant agreement No.\ 101002090 COLORFREE) and the PIER Hamburg Seed Project PHM-2019-01.
Z.S. was also supported by a fellowship from the MIT Department of Physics.
R.v.K. gratefully acknowledges financial support for this project by the Fulbright Foreign Student Program, which is sponsored by the U.S. Department of State.
(The contents of this publication are solely the responsibility of the authors
and do not necessarily represent the official views of the Fulbright Program and the Government of the United States.)
R.v.K. thanks the MIT Center for Theoretical Physics for its hospitality.

\addcontentsline{toc}{section}{References}
\bibliographystyle{jhep}
\bibliography{../paper/refs}

\end{document}